%% file: arxiv_quantumsurvey.tex
\pdfoutput=1

\documentclass[10pt]{article}

\usepackage{authblk}                % for \affil
\usepackage[square,numbers]{natbib} % needed for the references, when using ACM-Reference-Format outside an

\usepackage{amsmath,amssymb}                % for math operations

\input{common_settings.tex}

\input{header.tex}

\input{generated_files/collection_statistics.tex}

\input{generated_files/RQ_results.tex}

\usepackage{geometry}
\title{
\input{title.tex}
}

\author{Man Zhang}
\author{Yuechen Li}
\author{Tao Yue\thanks{Corresponding author}}
\author{Kai-Yuan Cai}

\affil{Beihang University}
\date{}

\begin{document}

\maketitle

\begin{abstract}
\input{abstract}
\end{abstract}

{\bf Keywords}: \input{keywords.tex}

\input{content.tex}

%%%%%%%%%%%%%%%%%%%%%%%%%%%%%%%%%%%%%%%%%%%%%%%%%%%%%%%%%%%%%%%%%%%%%%%%%%%%

\bibliographystyle{ACM-Reference-Format} % this requires ACM-Reference-Format.bst in same folder
%\bibliographystyle{acm}  % this is ancient from 80s, which does not support URL and DOI

%%https://arxiv.org/help/submit_tex#latex
%
% IMPORTANT: For final version arXiv, use generated bbl.
%            In such case, do not use compile.sh to build final pdf, but rather call pdflatex directly, eg
%            pdflatex arxiv

\input{arxiv_quantumsurvey.bbl}

%\bibliography{quantum_op_emp}

%If needed
\newpage
\input{appendix.tex}

\end{document}

%% file: common_settings.tex
\usepackage{listings}    % for snippets of code using \begin{lstlisting}.
\usepackage{graphicx}    % for figures, eg, \includegraphics
\usepackage{subcaption}  % for subfigures, ie figures with internal figures
\usepackage{booktabs}    % for better tables, eg using \toprule
\usepackage{hyperref}    % to handle URL links with \url
\usepackage{xurl}        % to handle very long URLs inside \url{} (done automatically, nothing to set manually)
\usepackage[htt]{hyphenat} % for line breaks in texttt, using \-
\usepackage{microtype} % ligatures are extremely annoying, especially when copy&paste text from PDF
\DisableLigatures{}    % https://tex.stackexchange.com/questions/439651/how-do-i-disable-ligatures

\usepackage{caption}
\usepackage{setspace}
\usepackage{multirow}
\usepackage{enumerate}
\usepackage{pdflscape}
\usepackage{pgfplots} % bar lib

\usepackage{xcolor}
\definecolor{codegreen}{rgb}{0.25,0.5,0.35}
\definecolor{codegray}{rgb}{0.5,0.5,0.5}
\definecolor{codepurple}{rgb}{0.6,0,0}
\definecolor{backcolour}{rgb}{0.95,0.95,0.92}
\definecolor{colorstring}{rgb}{0.5,0,0.35}
\definecolor{rltred}{rgb}{0.5,0,0}
\definecolor{rltgreen}{rgb}{0,0.5,0}
\definecolor{rltblue}{rgb}{0,0,0.5}
\definecolor{DarkGreen}{rgb}{0.00,0.60,0.00}
\definecolor{ScarletRed}{rgb}{0.80,0.00,0.00}
\definecolor{blizzardblue}{rgb}{0.67, 0.9, 0.93}
\definecolor{green-yellow}{rgb}{0.68, 1.0, 0.18}
\definecolor{dkgreen}{rgb}{0,0.6,0}
\definecolor{gray}{rgb}{0.5,0.5,0.5}
\definecolor{mauve}{rgb}{0.58,0,0.82}
\definecolor{lightgrey}{rgb}{0.90,0.90,0.90}
\definecolor{grey}{gray}{0.75}
\definecolor{light-gray}{gray}{0.80}

\usepackage{colortbl}

\lstdefinestyle{mystyle}{
    escapechar=©, %  use ©\label{}© when needing \label pointing to line numbers
    language=Python,
	backgroundcolor=\color{backcolour},
    basicstyle=\footnotesize\ttfamily,
   	identifierstyle=\footnotesize\ttfamily,
	commentstyle=\color{codegreen},
	keywordstyle=\color{colorstring}\bfseries,
	morekeywords={OR, AND},
	numberstyle=\ttfamily\color{codegray},
	stringstyle=\ttfamily\color{DarkGreen},
	breakatwhitespace=false,
	breaklines=true,
	captionpos=b,
	keepspaces=true,
	numbers=left, % possible values are (none, left, right)
	numbersep=2pt,
	showspaces=false,
	showstringspaces=false,
	showtabs=false,
	tabsize=2
}
\usepackage{braket}		% Quantum notifications
\usepackage{mathtools}	% Math-related signs

\usepackage{pifont} 
\usepackage{xspace}

\usepackage{tcolorbox}
\newtcolorbox{resultsbox}[1][]
{
	colframe=gray!100, 
	colback=white!100, 
	coltitle=white,
	title=#1 
}

\newenvironment{results}[1][]{
	\begin{resultsbox}[#1]
	}{
	\end{resultsbox}
}

\usepackage{ifthen}
\newboolean{showcomments}
\setboolean{showcomments}{true} % comment this line to deactivate comments

\ifthenelse{\boolean{showcomments}}{
	\newcommand{\nbc}[3]{
		{\colorbox{#3}{\bfseries\sffamily\scriptsize\textcolor{white}{#1}}}
		{\textcolor{#3}{\sf\small$\langle$\textit{#2}$\rangle$}}}
	
}{
	\newcommand{\nbc}[3]{}

}

\newcommand\tao[1]{\nbc{Tao}{#1}{red}}
\newcommand\man[1]{\nbc{Man}{#1}{blue}}
\newcommand\yc[1]{\nbc{Yuechen}{#1}{orange}}

\usepackage[ruled,linesnumbered]{algorithm2e}
\usepackage{algpseudocode}

%% file: header.tex
\newcommand{\rqExperimentDesign}{\textbf{RQ1:} How was the evaluation designed in the primary studies?\xspace}
\newcommand{\rqHyperparameterSetting}{\textbf{RQ2:} What hyperparameters have been configured in the primary studies?\xspace}
\newcommand{\rqExperimentSetting}{\textbf{RQ3:} What experimental settings do the primary studies consider?\xspace}
\newcommand{\rqMetrics}{\textbf{RQ4:} What evaluation metrics are employed in the primary studies?\xspace}
\newcommand{\rqBenchmarks}{\textbf{RQ5:} What are the case studies used in the primary studies?\xspace}
\newcommand{\rqBaselines}{\textbf{RQ6:} What are the baselines used in the primary studies?\xspace}
\newcommand{\rqTooling}{\textbf{RQ7:} Which tools have been made publicly available?\xspace}

\newcommand{\totalNumRQs}{7\xspace}

%% file: generated_files/collection_statistics.tex
\newcommand{\dbIEEE}{IEEE\emph{Xplore}\xspace}
\newcommand{\dbACM}{ACM\xspace}
\newcommand{\dbScopus}{Scopus\xspace}
\newcommand{\dbWiley}{Wiley\xspace}
\newcommand{\dbWoS}{Web of Science\xspace}
\newcommand{\dbSpringer}{Springer Nature Link\xspace}

\newcommand{\finalselected}{76\xspace}
\newcommand{\finalSLRSelected}{77\xspace}

\newcommand{\paperfirst}{jhaveri2023cloning}
\newcommand{\papersecond}{ammermann2024quantum}
\newcommand{\paperthird}{gan2024research}
\newcommand{\paperfourth}{ren2024dynamic}
\newcommand{\paperfifth}{schonberger2023digitalannealing}
\newcommand{\papersixth}{barletta2024quantum}
\newcommand{\paperseventh}{groppe2021optimizing}
\newcommand{\papertenth}{bajaj2022test}
\newcommand{\papereleventh}{mandal2024evaluating}
\newcommand{\papertwelfth}{deepalakshmi2022optimized}
\newcommand{\paperthirteenth}{wang2024quantum}
\newcommand{\paperfourteenth}{dornala2023quantum}
\newcommand{\papersixteenth}{zhang2017test}
\newcommand{\papertwentieth}{guo2023effective}
\newcommand{\papertwentysecond}{chen2020quantum}
\newcommand{\papertwentythird}{rani2023novel}
\newcommand{\papertwentyfourth}{kiruthiga2019enriched}
\newcommand{\papertwentyfifth}{serrano2024minimizing}
\newcommand{\papertwentysixth}{wu2020hybrid}
\newcommand{\papertwentyseventh}{bettonte2022quantum}
\newcommand{\papertwentyeighth}{kumari2016comparing}
\newcommand{\papertwentyninth}{liu2024quantum}
\newcommand{\paperthirtieth}{jin2021cross}
\newcommand{\paperthirtyfirst}{zhang2021new}
\newcommand{\paperthirtysecond}{jin2016parameter}
\newcommand{\paperthirtythird}{bajaj2022improved}
\newcommand{\paperthirtyfourth}{guo2022synergic}
\newcommand{\paperthirtyfifth}{tong2023can}
\newcommand{\paperthirtysixth}{xiong2016optimized}
\newcommand{\paperthirtyseventh}{hussein2021quantum}
\newcommand{\paperthirtyeighth}{bajaj2022Sensorstest}
\newcommand{\paperthirtyninth}{xiong2016virtual}
\newcommand{\paperfortieth}{fi13100260}
\newcommand{\paperfortyfirst}{jain2023quantum}
\newcommand{\paperfortysecond}{deng2022research}
\newcommand{\paperfortythird}{muniswamy2024joint}
\newcommand{\paperfortyfourth}{pathak2023opposition}
\newcommand{\paperfortyfifth}{bhatia2020quantum}
\newcommand{\paperfortyeighth}{ou2023quantum}
\newcommand{\paperfiftieth}{saxena2024constrained}
\newcommand{\paperfiftysecond}{uotila2025left}
\newcommand{\paperfiftythird}{trummer2015multiple}
\newcommand{\paperfiftyfourth}{nayak2023constructing}
\newcommand{\paperfiftyfifth}{barletta2022quantum}
\newcommand{\paperfiftysixth}{caivano2022quantum}
\newcommand{\paperfiftyeighth}{fankhauser2023multiple}
\newcommand{\papersixtieth}{trummer2024leveraging}
\newcommand{\papersixtyfirst}{schonberger2023quantum}
\newcommand{\papersixtythird}{trovato2025reformulating}
\newcommand{\papersixtyfourth}{wang2024test}
\newcommand{\papersixtyfifth}{miranskyy2022using}
\newcommand{\papersixtysixth}{wagner2019quantum}
\newcommand{\papersixtyseventh}{lou2018failure}
\newcommand{\papersixtyeighth}{blanco2024qiss}
\newcommand{\papersixtyninth}{barletta2023quantum}
\newcommand{\paperseventieth}{zhang2020intelligent}
\newcommand{\paperseventyfirst}{jin2015prediction}
\newcommand{\paperseventysecond}{balicki2021many}
\newcommand{\paperseventythird}{shen2021failure}
\newcommand{\paperseventyfourth}{kumar2025optimizing}
\newcommand{\paperseventyfifth}{guo2023tolerance}
\newcommand{\paperseventysixth}{guo2023memetic}
\newcommand{\paperseventyseventh}{hussein2020quantum}
\newcommand{\paperseventyninth}{roy2024quantum}
\newcommand{\papereightieth}{qiao2020novel}
\newcommand{\papereightysecond}{naik2025energy}
\newcommand{\papereightythird}{pathak2024cooperative}
\newcommand{\papereightyfourth}{urgelles2022multi}
\newcommand{\papereightyfifth}{liu2022hpcp}
\newcommand{\papereightysixth}{ahanger2022quantum}
\newcommand{\papereightyseventh}{emu2022resource}
\newcommand{\papereightyeighth}{emu2024warm}
\newcommand{\papereightyninth}{krishna2023optimal}
\newcommand{\paperninetieth}{niu2024performance}
\newcommand{\paperninetyfirst}{bhatia2019quantum}
\newcommand{\paperninetysecond}{li2019quantum}
\newcommand{\paperninetythird}{schonberger2023ready}

%% file: generated_files/RQ_results.tex
\newcommand{\RQfirstNumOfSettings}{73\xspace}

\newcommand{\RQfirstNumOfCaseStudies}{76\xspace}
\newcommand{\RQfirstPercentOfCaseStudies}{100.00\%\xspace}
\newcommand{\RQfirstNumOfBaselines}{71\xspace}
\newcommand{\RQfirstPercentOfBaselines}{93.42\%\xspace}
\newcommand{\RQfirstNumOfEffectiveness}{76\xspace}
\newcommand{\RQfirstPercentOfEffectiveness}{100.00\%\xspace}
\newcommand{\RQfirstNumOfEfficiency}{1\xspace}
\newcommand{\RQfirstPercentOfEfficiency}{1.32\%\xspace}
\newcommand{\RQfirstNumOfComplexity}{22\xspace}
\newcommand{\RQfirstPercentOfComplexity}{28.95\%\xspace}
\newcommand{\RQfirstNumOfCost}{38\xspace}
\newcommand{\RQfirstPercentOfCost}{50.00\%\xspace}

%% file: title.tex
% use \\ to break line, if needed
Empirical Studies on Quantum Optimization for Software Engineering: A Systematic Analysis

%% file: abstract.tex
In recent years, quantum, quantum-inspired, and hybrid algorithms are increasingly showing promise for solving software engineering optimization problems. However, best-intended practices for conducting empirical studies have not yet well established. 
In this paper, based on 
%77 
the primary studies identified from the latest systematic literature review on quantum optimization for software engineering problems, we conducted a systematic analysis on these studies from various aspects including experimental designs, hyperparameter settings, case studies, baselines, tooling, and metrics.
We identify key gaps in the current practices such as limited reporting of the number of repetitions, number of shots, and inadequate consideration of noise handling, as well as a lack of standardized evaluation protocols such as the adoption of quality metrics, especially quantum-specific metrics. Based on our analysis, we provide insights
% and recommendations\man{need to discuss if we can provide recommendations}
for designing empirical studies and highlight the need for more real-world and open case studies to assess cost-effectiveness and practical utility of the three types of approaches: quantum-inspired, quantum, and hybrid.
This study is intended to offer an overview of current practices and serve as an initial reference for designing and conducting empirical studies on evaluating and comparing quantum, quantum-inspired, and hybrid algorithms in solving optimization problems in software engineering.

%% file: keywords.tex
%Quantum Optimization, Quantum-inspired Algorithm, Classical Software Engineering, Systematic Literature Review
Quantum Optimization, Quantum-inspired Algorithm, Software Engineering Optimization Problem, Empirical Study

%% file: content.tex
%%%%%%%%%%%%%%%%%%%%%%%%%%%%%%%%%%%%%%%%%%%%%%%%%%%%%%%%%%%%%%%%%%%%%%%%%%%%

\input{intro}

%%%%%%%%%%%%%%%%%%%%%%%%%%%%%%%%%%%%%%%%%%%%%%%%%%%%%%%%%%%%%%%%%%%%%%%%%%%%%%%%%%%%%%%%%%%%%%%%%%%%%%%%%%%%%%%%%%%%
\section{Background} 
\label{sec:background}

\input{background}

%%%%%%%%%%%%%%%%%%%%%%%%%%%%%%%%%%%%%%%%%%%%%%%%%%%%%%%%%%%%%%%%%%%%%%%%%%%%

\input{researchmethod}

%%%%%%%%%%%%%%%%%%%%%%%%%%%%%%%%%%%%%%%%%%%%%%%%%%%%%%%%%%%%%%%%%%%%%%%%%%%%

\input{empiricalstudy}

%%%%%%%%%%%%%%%%%%%%%%%%%%%%%%%%%%%%%%%%%%%%%%%%%%%%%%%%%%%%%%%%%%%%%%%%%%%%
\section{Discussions and Insights}
\label{sec:discussions}

From the perspective of evaluating quantum optimization algorithms, Abbas et al.~\cite{abbas2024challenges} classifies metrics for enabling fair comparison of quantum and classical optimization solutions into two main categories, i.e.,
\textit{resource cost} and
%\textit{run time}, 
\textit{quality},
but unfortunately did not provide an overview of concrete metrics. 
Considering that these metrics were proposed to evaluate quantum algorithms and to compare them with quantum, quantum-inspired, and classical baselines, we consider them suitable for evaluating all types of optimization approaches. 
Then, we examine results of our study and align them with the categories in Sections~\ref{subsec:resource} and~\ref{subsec:quality}, aiming to identify common gaps. 
In addition, we discuss aspects such as experimental setup and hyperparameters (Section~\ref{subsec:parameters}), case studies and baseline selections (Section~\ref{subsec:casestudies}), and availability of proposed approaches (Section~\ref{subsec:transparent}).

\subsection{Resource cost}
\label{subsec:resource}
Resource cost refers to the computational resources required by an optimization solution.
However, in the context of quantum optimization, these resources may differ across computing paradigms and execution environment. 
Distinct from the execution platform, the optimization approaches can be classified into three types: \textit{quantum-inspired} executed on classical computers, \textit{quantum} executed on (simulated) quantum computers, and \textit{hybrid} combining classical and (simulated) quantum computing resources.

\begin{figure}
	\centering
	\includegraphics[width=0.98\textwidth]{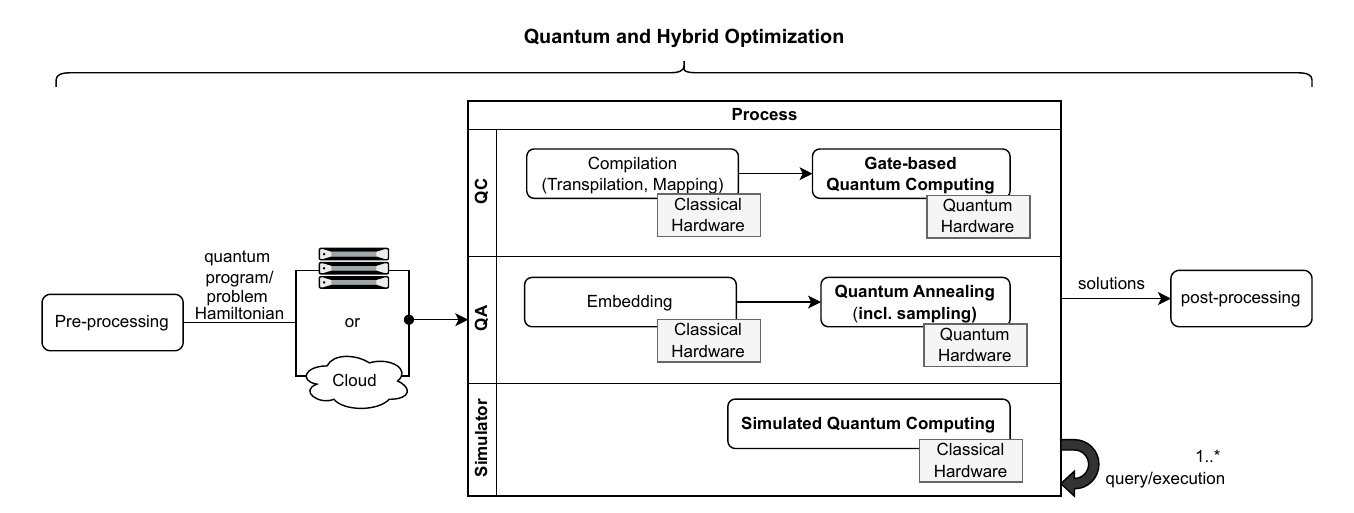}
	\caption{Typical execution phases of quantum and hybrid optimization on various platforms}
	\label{fig:runtime}
\end{figure}

Quantum-inspired optimizations are performed entirely on classical hardware, and classical computing resources are commonly satisfiable. In contrast, quantum resources are limited. With current techniques, quantum computing could consist of several phases that can span both classical and quantum hardware as illustrated in Figure~\ref{fig:runtime}. 
It is therefore crucial to further distinguish resource cost associated with each phase and any communication involved, when feasible.
More specifically, the pre-processing phase is typically performed on classical hardware that carries out the tasks such as data encoding, then outputs quantum program or encoded problem (e.g., Hamiltonian) that can be processed by quantum computers or simulators.
In addition, depending on the quantum computing paradigm, there are also various steps performed on classical hardware before the quantum computation.
For instance, in gate-based quantum computers, compilation converts a high-level quantum program into hardware-executable instructions, while transpilation, a subset of compilation, transforms logical quantum circuits into optimized, hardware-executable gate representations.
In quantum annealers, embedding is a specialized transpilation step that assigns logical qubits to physical ones, considering quantum hardware topology.
Furthermore, post-processing involves classical steps after quantum execution, such as measurements and error mitigation, when evaluating a quantum optimization algorithm.
Moreover, quantum computing is commonly offered via cloud platforms, allowing users to execute programs or encoded problems on remote quantum hardware or simulators.
Such cloud-based access introduces additional cost, e.g., communication overhead and extra expenses, into the overall optimization process.
Therefore, the feasible strategies with which an optimization solution can be configured on the available platform also affects the resources required.
Overall, in the context of quantum optimization, resource costs mainly refer to hardware requirements, expenses for third-party services, and run time (a commonly used resource metric based on time).

Regarding hardware requirements, quantum computing follows fundamentally different principles from classical computing. 
Its resources are measured not by traditional metrics such as CPU cores, GPUs, or RAM, but by quantum-specific properties, including qubit count, coherence time, and gate fidelity.
However, based on our analysis results (see Section~\ref{subsec:numberQubits}), fewer than one-third of the primary studies employing quantum or hybrid algorithms explicitly report qubit count used in their empirical evaluations. 
We also found that current empirical studies are limited by available computational hardware, with most experiments using 4–32 logical qubits on simulators and 4–247 qubits on quantum devices (Section~\ref{subsec:numberQubits}). 
As hardware improves, future studies can address larger and more realistic SE problems. 
Furthermore, as noted in Section~\ref{subsec:EffectivenessMetrics}, quantum-specific evaluation metrics (e.g., circuit depth and width, coherence time, and the number of logical qubits) appeared in only one of the primary studies. This omission is significant: in the NISQ era, where hardware constraints are severe, transparent reporting of quantum resource costs, especially qubit requirements, is essential for evaluating the practical feasibility of proposed approaches~\cite{yue2023QRE}. 
We strongly encourage researchers to report backend characteristics and limitations, such as qubit count, and connectivity, to enhance transparency, reproducibility, and comparability. 

Time serves as a common and fair metric for comparing solutions across various computing paradigms and execution environments~\cite{abbas2024challenges} .
In our analyses (Section~\ref{subsubsec:cost}), we observed that time is the most commonly reported cost metric.
However, only 38 out of a total \finalselected primary studies provided this info which highlights a gap in empirical research.
In terms of optimization types, a higher proportion of quantum (60\%) and hybrid optimization (87.5\%) studies report time costs compared to quantum-inspired approaches (42.5\%) that may indicate growing awareness among researchers of the importance of resource costs in optimizations with quantum computing.
%In many practical scenarios, 
Regarding metrics of run time,
for quantum-inspired algorithms, they are typically constrained by a time budget or by the number of fitness function evaluations, as in GA and other evolutionary algorithms (Section~\ref{subsec: TerminationCriteriaNoise}). These constraints affect the overall execution time, so fair comparisons across algorithms should account for them by reporting total runtime, number of evaluations, or using consistent stopping criteria.
For quantum and hybrid optimization, as we discussed, understanding the time required for each phase is crucial for practical applications.
Unfortunately, as reported in Section~\ref{subsubsec:cost}, the majority of primary studies only reported the total execution time, with only a few providing a breakdown of the time usage. 
We look forward to seeing the community adopt more standardized practices for reporting time in future studies. 
To ensure a fair comparison with classical and quantum-inspired algorithms, the runtime for classical computing should report the total execution time and, if possible, provide a breakdown of individual phases, e.g., pre-processing, classical computation (using general-purpose hardware such as GPUs or specialized hardware such as digital annealers) and post-processing.

Regarding  expenses for third-party services, 24 primary studies employed cloud-based quantum hardware, but only one primary study reported \textit{economic cost}~\cite{ \papertwentyfifth}, and none included communication overhead.

Our analysis shows that reporting is limited across all aspects of resource costs, including hardware requirements, runtime, and expenses for third-party services.
Standardized reporting practices are needed to build cumulative knowledge across the community.
Based on the computing paradigm, we recommend the following:
\begin{itemize}
	\item For quantum and hybrid optimization, it is necessary to break down the phases when feasible to better assess resource costs.
	\item When reporting quantum computational resources, it is recommended to include backend characteristics and limitations, such as qubit count and connectivity.
	\item To enable fair comparisons across computing paradigms, runtime including communication overhead must be reported. 
	Specifically, for quantum and hybrid optimization, runtime should be reported for each phase.
	\item For any third-party service used in an approach, associated expenses should be also reported.
\end{itemize}

\subsection{Quality metric}
\label{subsec:quality}

Quality refers to how the optimization is effective to tackle the problems of various complexity.
In our study, we first analyzed the problems in terms of their complexity by examining the problem size, the density of variable connections, and constraint types, measuring the practical difficulty of the problems.
%\emph{Problem complexity} measures the practical difficulty of the problem, typically in terms of its size, the density of variable connections, or the nature of constraints. 
In Section~\ref{subsec:ProblemComplexity}, we grouped problem-specific complexity metrics into two main mechanisms: size and dependencies/constraints. We observed that most of the primary studies relied primarily on size as the measure of problem complexity, while comparatively fewer considered dependencies or constraints. These metrics are independent of the computation paradigm, as they pertain to the problem domain itself, which reflects the current practice in empirical studies. However, relying solely on size may be limiting, since it does not capture structural hardness or inter-variable relationships that can significantly affect the difficulty of solving real-world software engineering problems. As a general guideline, empirical studies should always aim to measure problem complexity, such that the scalability and robustness of proposed solutions can be sufficiently evaluated.

Regarding quality metrics, we consider problem-specific metrics, general effectiveness metrics, quantum-specific metrics and diversity (see Section~\ref{subsec:EffectivenessMetrics}). 
Our classification substantially extends \emph{solution quality} reported in~\cite{abbas2024challenges}, since our scope focuses on SE problems where a large number of problem-specific metrics are employed, such as test suite size reduction in test minimization and response time in resource scheduling. We regard these metrics as essential, as they directly capture the quality of solutions from the perspective of addressing real-world SE problems. We also observed the use of more general metrics such as recall, accuracy, RMSE, and AR, which provide common ground for comparing different approaches across tasks. Moreover, some quantum-specific metrics, notably the energy value, are employed to characterize the solution landscape and are particularly useful for analyzing and optimizing quantum algorithms. Taken together, we argue that the quality of solutions should be assessed in multiple dimensions: problem-specific metrics are essential for capturing the practical value of solving real-world SE problems, general metrics enable comparability across studies, and quantum-specific metrics offer insights into the unique behavior of quantum optimization methods. Finally, we call for empirical studies that systematically investigate which metrics are most suitable for fairly comparing algorithms across different computing paradigms, to foster consistency and fairness in evaluation practices.

There are several quantum-specific metrics that are commonly used to evaluate algorithm performance in quantum computing. For example, the probability of finding the ground state measures how often the algorithm returns the optimal solution across repeated runs, reflecting solution reliability; the energy expectation value is the average energy of the solutions obtained, indicating how close the algorithm comes to the optimal solution on average; AR quantifies the quality of the solutions relative to the optimum, providing a normalized measure of performance  \cite{willsch2020benchmarking, lubinski2024optimization}. While these metrics are widely used in quantum computing research, our study shows that their adoption in SE applications of quantum and hybrid algorithms is inadequately considered. Specifically, as discussed in Section~\ref{subsec:EffectivenessMetrics}, energy value is adopted by only 10 primary studies, while AR was adopted in only three primary studies. This highlights a clear gap between the evaluation practices in quantum computing and the current empirical studies in SE. We recommend that future SE studies systematically incorporate such metrics.

\subsection{Experimental setup and hyperparameters}
\label{subsec:parameters}
%repetition and shots of experiments
%\textbf{Repetition. }

Surprisingly, we found that only about half of the primary studies report repeated experiments (see Section~\ref{subsec:Repetition}). This is unexpected, as repeated runs are a well-established practice in evaluating heuristic and metaheuristic algorithms, particularly when assessing quantum and hybrid approaches. Due to the inherent uncertainty in these algorithms, including probabilistic outcomes and, in the case of quantum algorithms, the limited number of shots per execution, repeated measurements are essential to obtain reliable and statistically meaningful observations. Following practices from both SBSE and quantum computing research, we recommend performing multiple independent runs, typically 30, or 10 when the reported experimental cost is high, unless new evidence suggests otherwise, to ensure robust and reproducible results.

%\textbf{Shots.} 
Due to the inherent randomness and noise in quantum computers, it is important to perform multiple shots to obtain more statistically reliable results based on the outcome distribution.
However, as discussed in Section~\ref{subsec:shots}, only 15 out of 29 relevant primary studies reported the number of shots. It is below the expectation, as in quantum optimization, the number of shots is a key factor affecting the accuracy of measured probabilities, energy values and ARs. Typically, it is set between 100 and 10000 (Figure~\ref{fig:RQ2_hasShots_summary_bar}) to balance requirements on statistical confidence with execution time and hardware cost. Smaller and low-variance problems may require fewer shots, while larger or more sensitive problems often need more to obtain reproducible outcomes. 

In terms of noise handling, it is a critical factor in the NISQ era, as it directly affects the performance of quantum algorithms. However, in our study, we found that noise is almost entirely neglected: only two of the primary studies explicitly addressed it (Section~\ref{subsec: TerminationCriteriaNoise}). Metrics commonly used to quantify noise such as quantum gate fidelity and quantum process fidelity were not reported in any primary study. This gap is partly due to the difficulty researchers face in obtaining such information, as not all quantum hardware platforms provide access to detailed noise or fidelity data. The lack of reporting underscores a significant challenge in fully evaluating quantum and hybrid optimization algorithms for SE problems. 
Though various efforts (e.g., \cite{muqeet2024mitigating, sato2025bug}) have been made to address noise challenges (e.g., quantum software testing and bug localization) in SE, further research is still needed. We also emphasize the importance of communication between the quantum computing and SE communities, as solutions developed in quantum computing (e.g., Mitiq proposed in \cite{larose2022mitiq}) may be applicable to SE. We look forward to continued collaboration and cross-fertilization of ideas between these fields.

\begin{comment}

\tao{Man: I not sure whether it is ok to place the below in this subsection. }
%\textbf{Scalability.} 
Current empirical studies are limited by available computational hardware, with most experiments using 4–32 logical qubits on simulators and 4–247 qubits on quantum devices (Section~\ref{subsec:numberQubits}). As hardware improves, future studies can address larger and more realistic SE problems. 
\yc{Q: This shows overlap with the section for metrics.} 
We strongly encourage researchers to report backend characteristics and limitations, such as noise, qubit count, and connectivity, to enhance transparency, reproducibility, and comparability. Standardized reporting practices are needed to build cumulative knowledge across the community.
\end{comment}

Regarding hyperparmaters, for quantum-inspired algorithms, which are often derived from existing evolutionary algorithms such as PSO or GA, key parameters including population size and mutation or crossover settings (Section~\ref{subsubsec: quantum-inspired hyperparameter}), where applicable, should be explicitly reported and, when possible, empirically justified. Since these algorithms inherit many design choices from their classical counterparts, the settings of the base algorithms should also be documented, and parameter tuning is ideally required to ensure fair comparison and reliable performance assessment. In addition, quantum-specific parameters should be configured properly such as initial qubit states. 

Setting objective weights to convert a multi-objective problem into a single-objective one is a common practice in SBSE. How to determine these weights is not specific to quantum or classical approaches; rather, weights are typically established based on empirical data, user preferences, or problem requirements, and in some cases, equal weighting is adopted as a simple solution. In any case, we strongly suggest to provide a clear rationale for how these weights are determined to ensure transparency and reproducibility.

\subsection{Case studies and baselines}
\label{subsec:casestudies}
%benchmark vs. real-world SE problems
%\textbf{Case studies.} 
Compared to benchmark problems like MaxCut, SE problems are far more heterogeneous, constraint-rich, and domain-specific (see Table~\ref{tab:rq3_problem_metrics}). We observe that quantum or hybrid solutions effective on benchmarks may not be feasible or cost-effective for solving SE problems. Real-world applications require careful consideration of hardware availability and performance (e.g., noise, qubit count) as well as the ability to encode complex dependencies into QUBO or Ising models, for instance. Though some real-world case studies exist (see Table~\ref{tab:rq3_problem_metrics}), we still see the need to call for more empirical investigations on applying quantum, quantum-inspired and hybrid approaches to SE problems, and when possible make them openly available. 

%Baselines
%\textbf{Baselines.} 
Selecting appropriate baselines is critical in evaluating quantum, quantum-inspired, and hybrid algorithms for solving SE problems. Without meaningful baselines, it is difficult to assess whether the observed performance gains are due to the inherent advantages of quantum approaches or simply to experimental conditions. In particular, when addressing real SE problems, it is essential to compare against strong classical counterparts, such as state-of-the-art heuristics, metaheuristics, or exact solvers, which often perform competitively in practice. Such comparisons not only ensure fairness but also provide valuable insights into whether quantum solutions offer genuine advantages in terms of scalability, quality, or resource efficiency. The results of our study (see Section~\ref{sec:baselines}) show that this is indeed the case, as many primary studies evaluate quantum or hybrid algorithms against classical methods (e.g., SA, GA, PSO). Also, we recommend including quantum or hybrid baselines for approaches upon the same quantum optimization algorithms, such as exploring different encoding approaches and solvers against the proposed one.

Moreover, only a small proportion of the primary studies made their benchmarks and toolings publicly available on open-source platforms like GitHub and Zenodo. Since the artifact availability is a trending requirement in the SE community, we encourage future studies to ensure that their artifacts can be archived in the long term and easily accessible. Apart from that, it is still valuable to have artifacts functional and reusable, according to the criteria for ACM artifact review and badging.\footnote{\url{https://www.acm.org/publications/policies/artifact-review-badging}} For example, the released artifacts are suggested to include complete functionalities for reproducing all the relevant empirical results and provide readable documentation for possible reuse in future studies.

\subsection{Availability of artifacts}
\label{subsec:transparent}

Availability is a standard expectation in the SE community. 
In SE venues, public availability of tools or datasets is often mandatory.
However, in our analysis, we found that among the 76 primary studies, only 15 employed open-source case studies, and just 12 made their tools publicly accessible.
This indicates a significant gap between the SE community’s expectations and current practice in quantum optimization research for SE.

To improve transparency and reproducibility, we recommend that future studies ensure public access to code, tools, and datasets, which is essential for the development of this research community.
Establishing dedicated venues in this area could further promote artifact availability. 
In addition, adopting common platforms or repositories for sharing quantum optimization approaches could facilitate broader collaboration and encourage the creation and sharing of benchmarks for fair comparison. 
Future work could also investigate barriers to open-source adoption in this area, such as hardware dependency, or deployment complexity, and propose strategies to overcome them.

%%%%%%%%%%%%%%%%%%%%%%%%%%%%%%%%%%%%%%%%%%%%%%%%%%%%%%%%%%%%%%%%%%%%%%%%%%%%
\section{Threats To Validity}
\label{sec:threats}
Several threats may affect the validity of our survey. Selection bias could arise from our choice of databases and inclusion/exclusion criteria; although, in our previous work~\cite{zhang2025quantum}, we systematically collected 77 primary studies, some relevant work might have been omitted. Data extraction and interpretation bias may occur due to subjective judgment when coding study attributes or answering the seven RQs, though we mitigated this by using an extraction protocol (which was optimized iteratively) and cross-validation among multiple authors. Finally, generalizability of our findings may be limited, as the survey reflects the current state of research in quantum, quantum-inspired, and hybrid algorithms applied to SE, which is still an emerging area with limited empirical studies.

%%%%%%%%%%%%%%%%%%%%%%%%%%%%%%%%%%%%%%%%%%%%%%%%%%%%%%%%%%%%%%%%%%%%%%%%%%%%
\section{Related Work} 
\label{sec:relatedwork}

\input{related}

%%%%%%%%%%%%%%%%%%%%%%%%%%%%%%%%%%%%%%%%%%%%%%%%%%%%%%%%%%%%%%%%%%%%%%%%%%%%
\section{Conclusion}
\label{sec:conclusions}

With the increasing complexity of software, quantum optimization has become a promising solution to address challenges in software engineering (SE) across development, operations, and maintenance. 
To better understand how evaluations are conducted at the intersection of quantum computing and SE, this study provides a systematic analysis of primary research on quantum optimization for SE problems, offering an overview of current evaluation practices and their implications.

Following evaluation practices of SE, we examined the primary studies from seven perspectives: evaluation design, hyperparameter configurations, experimental settings, evaluation metrics, case studies, baseline selection, and the availability of proposed approaches.
Reporting of resource costs is essential, especially for optimization with quantum computers.
However, our analysis shows that current practices remain limited across all aspects of resource cost reporting, including hardware requirements, runtime, and expenses for third-party services.
In quantum optimization, randomness is inherent for all three types of approach.
Proper handling of randomness, such as through repeated executions or multiple shots, is important for ensuring statistical relivable conclusion and results.
Moreover, while transparency is well established in the SE community, it remains underdeveloped in quantum optimization for SE.
As an emerging intersection of quantum computing and SE,  this research area still lacks standardized protocols for conducing evaluation.
Future work should focus on developing standardized guidelines for resource cost reporting, experimental settings, hyperparameter configurations, and criteria for selecting  evaluation metrics, benchmark problems and baselines. 
In addition, building open-source platforms and repositories could further allow fair, transparent, and reproducible comparisons across quantum, hybrid, and quantum-inspired optimization approaches.

%%%%%%%%%%%%%%%%%%%%%%%%%%%%%%%%%%%%%%%%%%%%%%%%%%%%%%%%%%%%%%%%%%%%%%%%%%%%
\section*{Acknowledgement}
This work is supported by the State Key Laboratory of Complex \& Critical Software Environment (SKLCCSE, grant No. CCSE-2024ZX-01).

%% file: intro.tex
%%%%%%%%%%%%%%%%%%%%%%%%%%%%%%%%%%%%%%%%%%%%%%%%%%%%%%%%%%%%%%%%%%%%%%%%%%%%
\section{Introduction}
Quantum computing is attracting increasing attention, with recent breakthroughs enabling practical applications across fields such as drug discovery, logistics, and finance~\cite{hidary2021quantum}. In the last few years, researchers have recognized its potential for tackling Software Engineering (SE) problems and started investigating the application of quantum optimization approaches, achieving promising results~\cite{wang2024test, mandal2024evaluating, ammermann2024quantum}. Quantum-inspired algorithms (e.g., quantum-behaved particle swarm optimization (QPSO)~\cite{sun2004particle}) have also been applied in SE, for example in software requirements selection~\cite{kumari2016comparing} and task scheduling in software development projects~\cite{ou2023quantum}, although they have not become mainstream within Search-Based Software Engineering (SBSE)~\cite{harman2012search} in particular, and SE in general. Despite these advances, when considering the range of problems traditionally addressed by SBSE, there remains significant potential for quantum, quantum-inspired, and hybrid algorithms to outperform classical search approaches, particularly on real-world SE tasks that involve complex constraints, dependencies, and scalability challenges. 

Heuristic algorithms (e.g., hill climbing) and metaheuristics (e.g., Genetic Algorithms (GA), Particle Swarm Optimization (PSO)), and especially quantum, quantum-inspired, and hybrid algorithms, are inherently non-deterministic. Conducting empirical studies to evaluate their cost-effectiveness and compare them in the context of specific SE optimization problems is therefore essential. In the SBSE community, a body of knowledge has been accumulated over more than a decade, including guidelines for designing empirical studies~\cite{arcuri2013parameter}, selecting quality indicators for evaluating multi-objective search algorithms~\cite{ali2020quality}, determining repetitions to achieve statistical confidence~\cite{arcuri2014hitchhiker}, etc. However, this knowledge needs to be updated and extended to accommodate emerging applications of quantum, quantum-inspired, and hybrid algorithms. Before doing so, it is necessary to systematically examine the existing literature to understand what has already been studied and identify gaps for future research.

To this end, we first checked all of \finalSLRSelected primary studies collected through the Systematic Literature Review (SLR)~\cite{zhang2025quantum} to examine whether they included an evaluation. 
One study~\cite{\papersixtyfirst} did not, and was therefore excluded.
Subsequently, we analyze the \finalselected primary studies comprehensively from various aspects, including experimental design, hyperparameter settings, experimental settings, evaluation metrics, case studies, baselines, and tooling, with the aim of answering seven research questions (RQs) concerning the current state of empirical studies on applying quantum, quantum-inspired, and hybrid optimization algorithms in SE. 

From our analysis, we discover the following key findings: 1) most primary studies include empirical evaluations, which provide a rich source of information for this work; 2) common settings of hyperparameters include population size in quantum-inspired algorithms and penalty coefficients in Quadratic Unconstrained Binary Optimization (QUBO) formulations, etc.; 3) the most common numbers of repetitions are 30 and 10; however, only about half of the studies report the number of repetitions or shots; 4) reported logical qubit counts range from 4-32 on simulators to 4-247 on quantum devices; 5) convergence indicators are the most frequently used termination criteria; 6) noise is rarely considered; 7) size is the dominant measure of problem complexity, as expected; 8) problem-specific metrics are very diverse as expected, while general metrics (e.g., recall, precision) are also adopted and quantum-specific metrics (e.g., energy value) though their adoption remains limited; 9) cost metrics are primarily reported in terms of time; 10) real-world case studies are more common than benchmarks or synthetic ones; 11) classical approaches are the most widely used comparison baselines for evaluating quantum, hybrid and quantum-inspired approaches; and 12) some tools developed in the primary studies have been made publicly available.

Based on these findings, we identify some gaps in the current body of knowledge. For instance, reporting practices remain inconsistent, as many studies do not disclose repetitions, number of shots, etc.; there is no consensus on which general or quantum-specific metrics are most appropriate for fair cross-paradigm comparisons; more case studies are expected to be open-sourced. Together, these gaps highlight the need for developing standards and best practices for conducting empirical studies in quantum, quantum-inspired, and hybrid .
We hope this study provides an overview of existing practices, highlights current gaps, and serves as an initial reference for designing and conducting rigorous empirical studies in this emerging area. For enabling replicability and reproducability, we make the artifact of our analysis available at \url{https://github.com/WSE-Lab/QuantumOpt4SE-EmpiricalStudies}.

Structure of the paper: Section~\ref{sec:background} provides the background; The research method is discussed in Section~\ref{sec:researchmethod}, followed by presenting findings that answer each RQ (Section~\ref{sec:results}). %(Section~\ref{sec:RQ1}-Section~\ref{sec:tooling}). 
In Section~\ref{sec:discussions}, we provide insights we obtained from our findings, followed by the threats to validity (Section~\ref{sec:threats}) and the related work (Section~\ref{sec:relatedwork}). We conclude the paper in Section~\ref{sec:conclusions}.

%% file: background.tex
Quantum computing is proposed based on the four fundamental postulates of quantum mechanics, i.e., the postulates for state space, time evolution, quantum measurement, and composite systems~\cite{nielsen2010quantum}. From the perspective of the four postulates, we will compare classical computing and quantum computing in the following parts of this section. Also, we will discuss how quantum-inspired and quantum algorithms are associated with the four postulates, particularly in terms of their implementations. 

\textbf{State space.} 
% One of the significant distinctions between quantum computing and classical computing lies in the basic computational units. 
Unlike classical computing using bits, quantum computing employs qubits (a.k.a. quantum bits)\, which can exist not only in the computational basis states $\ket{0}$ and $\ket{1}$, but also in any superposition of these states. This is because pure quantum states can be mathematically denoted as state vectors in the Hilbert space. Hence, a single-qubit pure state can be denoted in the form of a linear combination, i.e., $\ket{\psi}=\alpha\ket{0}+\beta\ket{1}$, where $\alpha,\beta$ are complex numbers representing the amplitudes corresponding to $\ket{0}$ and $\ket{1}$, respectively, and $\left\vert\alpha\right\vert^2+\left\vert\beta\right\vert^2=1$ holds. Quantum optimization algorithms (e.g., Quantum Approximate Optimization Algorithm (QAOA)~\cite{farhi2014quantum}]) leverage superposition and interference to represent and manipulate candidate solutions simultaneously. Meanwhile, quantum-inspired optimization algorithms (e.g., Quantum-Inspired Evolutionary Algorithms~\cite{han2002quantum}) often borrow concepts such as superposition to guide and diversify the search space~\cite{zhang2025quantum}. For instance, individuals of the population (a set of candidate solutions maintained and evolved by a quantum-inspired algorithm) can be encoded as qubits associated with state vectors like $[\alpha,\beta]^{\top}$.  

\textbf{Time evolution.} Quantum algorithms are carried out through the evolution of qubits, where the qubit evolution of a closed system follows the Schrödinger equation~\cite{griffiths2018introduction}. Gate-based quantum systems, for instance, introduce quantum gates to perform operations on qubits, analogous to logical gates on bits in classical computing. The requirement for a closed system, indicating no interaction with the outer environment ideally, guarantees quantum algorithms to proceed as expected. This suggests that, in the design and implementation of purely quantum algorithms, intermediate states within a quantum subroutine cannot be directly inspected or modified once the corresponding quantum system has been constructed. Taking an example of a famous pure quantum algorithm, Grover Search~\cite{grover1996fast}, the number of iterations of the quantum subroutine must be determined prior to executing the quantum circuit. This contrasts with quantum-inspired optimization algorithms, where the termination necessity of a loop can often be adaptively determined based on intermediate available variables.

\textbf{Quantum measurement.} Quantum measurement is the only way to convert inaccessible quantum information into classical information observable in the classical world. In many quantum algorithms, the operations of quantum measurement are merely placed at the end of a quantum circuit~\cite{Abhijith2022Quantum}. However, quantum measurement breaks the assumption of a closed system and irreversibly alters the original states by projecting them into one measurement basis. Besides, the inherently probabilistic nature of quantum computing can be noticed by quantum measurement. The measurement outcomes of a quantum state are distributed according to the squared magnitudes of the amplitudes of the corresponding basis states, e.g., the probability of obtaining 1 upon measuring $\frac{\sqrt{3}}{3}\ket{0}+\frac{\sqrt{6}}{3}e^{\frac{\pi i}{4}}\ket{1}$ as $\frac{2}{3}$, where $e^{\frac{\pi i}{4}}$, as a phase factor, does not change the probability, but only the relative phase between states. To this end, executing measurements with a relatively sufficient number of times enables outputs for both quantum-inspired and quantum optimization algorithms to be valid with statistical significance. 

\textbf{Composite systems.} The state spaces of separable quantum subsystems (e.g., multiple qubits) combine via the tensor product, which empowers an $n$-qubit state with the capacity to exist in a superposition of $2^n$ computational basis states. This exponential representational capacity allows quantum-inspired and quantum optimization algorithms to represent the search space more compactly, compared to quantum-irrelevant classical approaches whose complexity could scale linearly with the number of candidate solutions. For instance, with 3 binary variables and no constraints, the classical search space has \(2^3=8\) valid solutions, each of which is represented explicitly
%~\yc{Typo: the classical search space has \(2^3=8\) valid solutions, each of which is represented explicitly}. 
In quantum representation, 3 qubits span the \(2^3\) computational basis states \(|000\rangle,\dots,|111\rangle\) and a single quantum state can encode all these solutions simultaneously in superposition. 
It is also important to notice that from a scalability perspective, simulating large-scale quantum systems on classical hardware is costly because the state space of an $n$-qubit system grows exponentially, requiring storage of $2^n$ complex amplitudes. As a result, classical simulators are typically limited to relatively small problem sizes.

%% file: researchmethod.tex
\section{Research Method}
\label{sec:researchmethod}

In this section, we discuss the research method we applied for conducting the study, including RQs, the primary studies, data extraction and analysis.

%----------------------------------------------------------
\subsection{Research Questions}
\label{subsec:rq}

In this study, we aim to answer the following \totalNumRQs research questions (RQs):
	\begin{itemize}
		\item \rqExperimentDesign
		\item \rqHyperparameterSetting
		\item \rqExperimentSetting
		\item \rqMetrics
		\item \rqBenchmarks
		\item \rqBaselines
		\item \rqTooling		
	\end{itemize}

These RQs comprehensively examine various aspects of the empirical studies of the primary studies. This includes identifying which parameters have been set and how (e.g., the number of repetitions and shots), which case studies have been employed (e.g., benchmarks vs. real-world case studies), what baselines have been used (classical, quantum, or quantum-inspired), and which tools have been made available. Through such analyses, we aim to provide a comprehensive review of this field and offer insights to guide the future conduct of empirical studies

%----------------------------------------------------------
\subsection{Included Primary Studies}
\label{subsec:primaryStudies}             

Our analysis was carried out on primary studies identified from a SLR~\cite{zhang2025quantum}
%~\yc{an SLR?} \man{as S refers to systematic, we may better keep them consistent with `a`}
on quantum techniques for SE optimization problems published in the past decade, i.e., from 2014 to 2025.
The SLR was conducted by following a standard process during which six digital databases (i.e., \dbACM, \dbIEEE, \dbScopus, \dbSpringer, \dbWoS, and \dbWiley) were systematically searched with carefully crafted search queries.
By following the pre-defined paper selection procedure and conducting the snowballing, the SLR ends up with \finalSLRSelected primary studies. 
To investigate the presence of empirical studies on quantum optimization for SE, we first examined whether each of the 77 primary studies included an empirical evaluation.
During this initial check, we found that the primary study~\cite{\papersixtyfirst} did not contain any evaluation and thus was excluded from our analysis. 
Consequently, our systematic analysis was conducted on the remaining \textbf{\finalselected primary studies}.

%----------------------------------------------------------
\subsection{Data Extraction}
\label{subsec:extraction}

We first identified the information to be extracted for studying the current presence of how the evaluation was conducted, including details of the reported experiment design (such as the problems addressed, hyperparameters, and experiment settings), key aspects of the evaluation (such as baselines, case studies, and employed metrics), and the availability of the tools.
We then reviewed the \finalselected primary studies based on the pre-defined criteria for data extraction.

\begin{table}
	\small
	\centering
	\caption{Information extracted for answering RQs}
	\label{tab:data_extraction}
	%\resizebox{.99\textwidth}{!}{
		\input{tables/data-extraction.tex}
		%}
\end{table}

Table~\ref{tab:data_extraction} provides the detailed information extracted to answer the RQs.
More specifically, we examined each paper for reported experiment design details (e.g., baselines, settings) and for whether the evaluation assessed an approach in terms of cost, effectiveness, etc. (RQ1).
To address RQ2, we collected hyperparameters used in the current research.
For experiment settings (RQ3), we collected general settings (e.g., the number of repetitions and termination criteria) which influence the validity of conclusions in the context of classical SE evaluation.
We also considered quantum-specific settings, including the number of shots and qubits configured or used, and noise of the quantum hardware or simulator.
Regarding evaluation metrics (RQ4), we collected not only metrics used to assess the proposed approach but also metrics used to measure the problem complexity.
For case studies (RQ5), we categorized them into three types: benchmarks, artificial and real-world case studies, and also recorded their accessibility to support further research in the field.
To answer RQ6, we collected information about baselines employed in each paper, and classified them into three categories: classical, quantum and hybrid approaches.
To promote transparency and reproducibility, we collected the accessibility of these artifacts for each paper to answer RQ7.

%% file: tables/data-extraction.tex
\begin{tabular}{ l  p{.8\textwidth}}
	\toprule 	
	RQs     & Information being extracted \\
	\midrule
	RQ1		& 1) For each paper, whether the experimental design reported details such as baselines, experiment settings, problems, and case studies \\ 
				& 2) For each paper, whether the evaluation assessed the approach in terms of cost, effectiveness, efficiency, and complexity\\ 
	RQ2		& Hyperparameter configurations used in the evaluation of each approach in each paper \\	
	RQ3		& 1) Repetitions configured in the evaluation of each paper \\
				& 2) Termination criteria defined in the evaluation of each paper \\
				& 3) Shot, Qubit, and Noise settings considered in the evaluation of each paper \\
	RQ4		& 1) Complexity of the SE problem tackled by the proposed approach in the evaluation of each paper\\
				& 2) Metrics of cost, effectiveness, efficiency, and complexity applied to evaluate the approach in each paper\\
	RQ5		& 1) Types of case studies used in the evaluation of each paper, such as Benchmark, Artificial, and Real-world \\
				& 2) Number of case studies used in the evaluation of each paper \\
				& 3) Accessibility of each case study in the paper \\
				& 4) Details of each accessible case study, including SE optimization problem, type, description, and URLs \\
	RQ6		& 1) Type of baselines involved in the evaluation of each paper, such as Classical, Quantum, Hybrid\\
				& 2) Number of baselines compared in the evaluation of each paper\\ 
	RQ7		& Accessibility of proposed quantum optimization approach and details (e.g., approach type, URL) in each paper \\

	\bottomrule 
\end{tabular}

%% file: empiricalstudy.tex
%%-------------------------------------------------------------------------%%
\section{Result Analysis}
\label{sec:results}

In this section, we discuss our findings and answers to each RQ.

\subsection{\rqExperimentDesign}
\label{sec:RQ1}

\begin{figure}
	\centering
	\begin{subfigure}{0.48\textwidth}
		\centering
		\includegraphics[width=\textwidth]{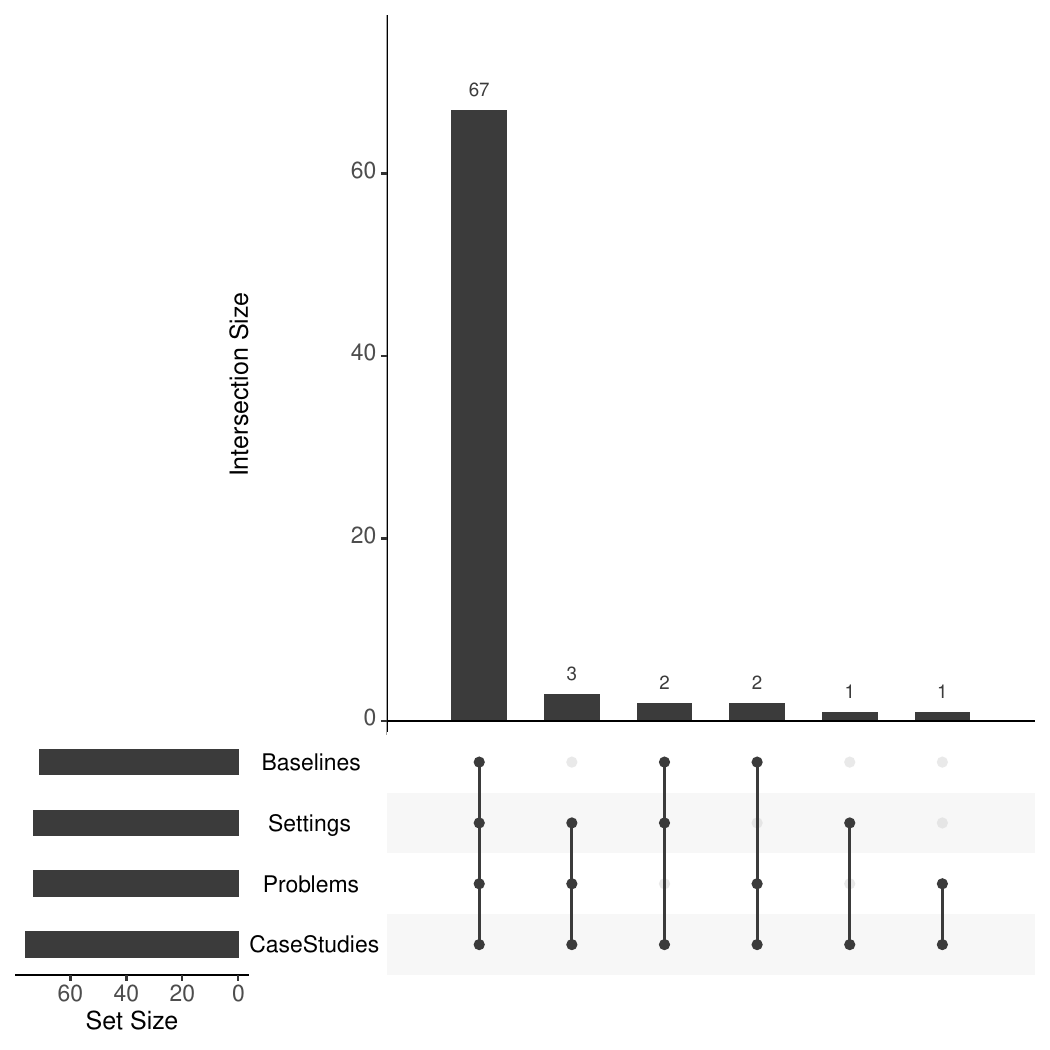}
		\caption{Experiment design}
		\label{fig:RQ1_exp_setup_upset}
	\end{subfigure}
	%\hfill
	\begin{subfigure}{0.48\textwidth}
		\centering
		\includegraphics[width=\textwidth]{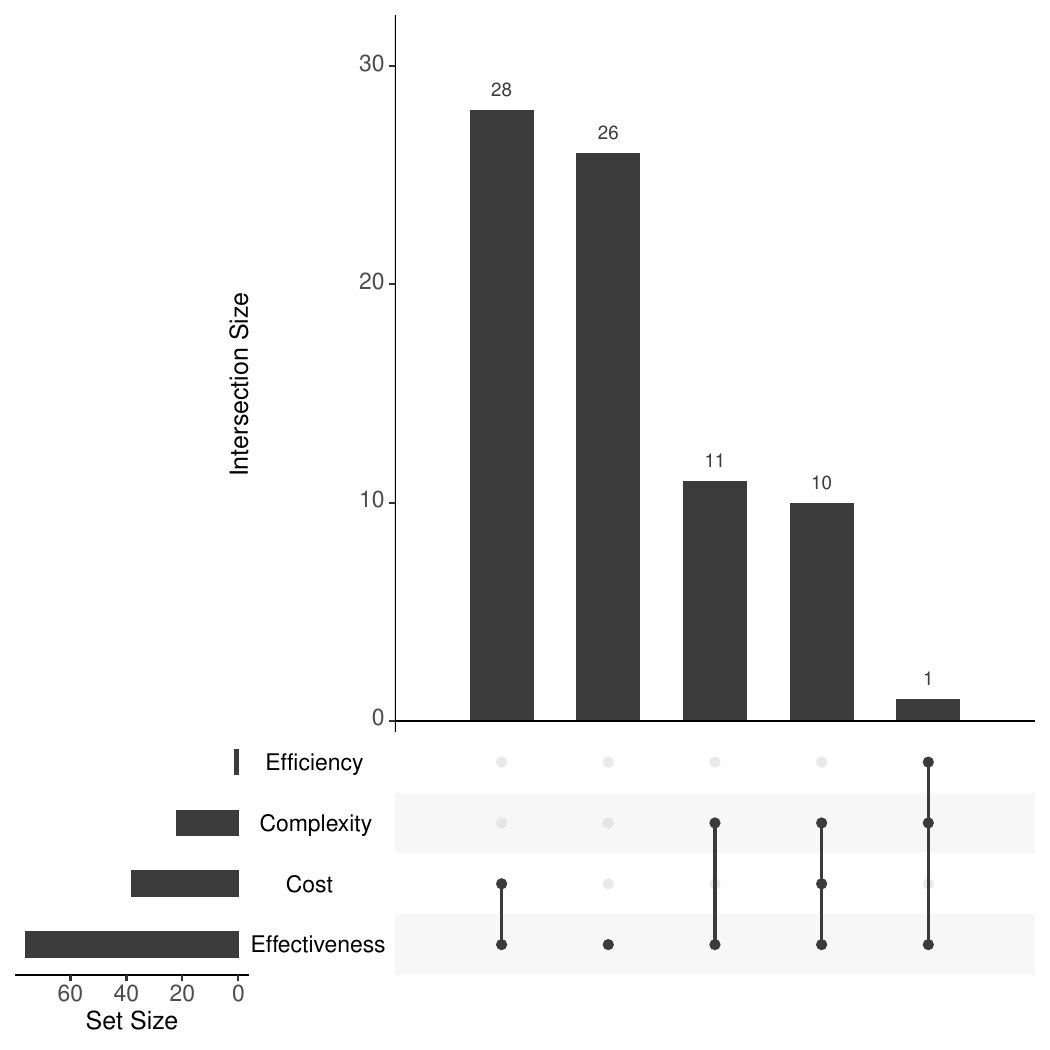}
		\caption{Evaluation metrics}
		\label{fig:RQ1_metrics_setup_upset}
	\end{subfigure}
	\caption{Overview of evaluation design (a) and evaluation metrics (b) - RQ1}
	\label{fig:overall_figure}
\end{figure}

With RQ1, we aim to gain an overview of the empirical studies presented in the primary studies, in terms of the experiment settings, metrics employed, case studies and baselines used, and results are reported in Figure~\ref{fig:overall_figure}.
As shown in Figure~\ref{fig:RQ1_exp_setup_upset}, 
%\RQfirstNumOfProblems (\RQfirstPercentOfProblems) primary studies characterize the problems being handled with problem-specific metrics; 
\RQfirstNumOfSettings (67+3+2+1, 96.05\%) primary studies present their experiment settings; \RQfirstNumOfBaselines (67+2+2, \RQfirstPercentOfBaselines) and \RQfirstNumOfCaseStudies (67+3+2+2+1+1, \RQfirstPercentOfCaseStudies) primary studies employ at least one baseline and case study in their evaluations, respectively. When looking at the vertical bars of the figure, we can easily see that the empirical studies of 67 out of \finalselected primary studies concern all the four aspects. This observation emphasizes the necessity of our analysis, as the majority of the primary studies are empirical, which provides a rich body of information to examine. This is mainly because the field of quantum optimization, especially in the Noisy Intermediate-Scale Quantum (NISQ) era, heavily relies on empirical studies; current quantum hardware suffers from noise and decoherence, which as a result, theoretical predictions often fail to match experimental outcomes, and algorithms must be tested, tuned, and validated on real devices, which makes empirical evaluation essential. Moreover, many quantum and quantum-inspired algorithms are heuristic in nature, and as a result, their evaluation in terms of cost-effectiveness naturally relies on empirical studies. 

%\begin{figure}%[htbp]
%	\centering
%	\includegraphics[width=0.8\textwidth]{generated_files/RQ1_exp_setup_upset.pdf}
%	\caption{Overview of the Experiment Design - RQ1}
%	\label{fig:RQ1_exp_setup_upset}
%\end{figure}

%Metrics 
When particularly looking at metrics, as shown in Figure~\ref{fig:RQ1_metrics_setup_upset}, \RQfirstNumOfEffectiveness (28+26+11+10+1, \RQfirstPercentOfEffectiveness) primary studies have their approaches evaluated from the aspect of effectiveness, while \RQfirstNumOfCost (28+10, \RQfirstPercentOfCost), \RQfirstNumOfComplexity (11+10+1, \RQfirstPercentOfComplexity), and \RQfirstNumOfEfficiency (\RQfirstPercentOfEfficiency) primary studies examine their approaches in terms of approach's cost, problem complexity and approach's efficiency, respectively. Efficiency here is about how effectively resources (such as time) are used to produce a desired outcome; hence it differs from cost.
%~\yc{Should we explain a little bit why efficiency is separate from cost, e.g., opposed purposes for maximization or minimization? For example, in the following sections, time is attributed to cost instead of efficiency.} 
This is understandable because quantum optimization is an emerging domain, hence the first priority is to determine whether an approach can solve a problem well, as a gatekeeper, before considering cost, scalability and efficiency. Near half of the primary studies employ one or more cost metrics, which is encouraging, as consistent measurement of cost, especially time cost, is complicated by hardware dependencies, hybrid quantum algorithm structures, etc. More details about the metrics will be provided in Section~\ref{subsec:EffectivenessMetrics}. 
When looking at the figure vertically, we can observe that 26 of \finalselected primary studies only concern effectiveness, while 28 concern both cost and effectiveness. This suggests that although cost is recognized as an important factor, it has not yet become a standard evaluation metric, even though ideally it should be considered in all primary studies.

%\yc{This indication looks somehow irrelevant, according to two values for concerning effectiveness only and concerning both.}\tao{I want to say that ideally cost is supposed to be considered in all primary studies. }. 

%There is no single primary study that concerns all four aspects: effectiveness, cost, problem complexity and efficiency, which implies that quantum optimization, as an emerging field, has insufficient infrastructure (e.g., case studies, mature hardware) to facilitate comprehensive evaluation of quantum optimization approaches across all dimensions. \man{this needs to be updated. complexity indicates complexity metric for assessing approach. we also remove scalability.}\yc{I agree with Man to update this.}

%\begin{figure}[htbp]
%	\centering
%	\includegraphics[width=0.9\textwidth]{generated_files/RQ1_metrics_setup_upset.pdf}
%	\caption{Overview of metrics employed - RQ1 \tao{Man: we remove Scalability from the figure. }}
%	\label{fig:RQ1_metrics_setup_upset}
%\end{figure}

\begin{results}[Findings of Experiment Designs]
Most of the \finalselected primary studies present empirical research that explicitly addresses experimental settings, problem characteristics, baseline methods, and case studies, which offers comprehensive data for our analysis. All the involved primary studies evaluate their approaches in terms of effectiveness, and near half employ at least one cost metrics. 
\end{results}

\subsection{\rqHyperparameterSetting}
In Table~\ref{tab:rq2_hyperparameter}, we present the hyperparameters that were configured by multiple approaches in the primary studies. We also classify them into two categories according to the types of quantum optimization approaches. 
%For hybrid approaches, most concerns are about quantum-specific parameters \tao{Yuechen, for hybrid approaches, do we have parameters of classical components? }~\yc{Still, some general parameters are involved in hybrid approaches, such as the dataset split ratio for learning-based tasks and the weight of each objective for a wide range of reformulations that tranform muti-objective into single-objective expressions.}.

\begin{table}
	\small
	\centering
	\caption{Summary of hyperparameter settings. Column \textit{Category} indicates if a parameter is classical, quantum relevant (\textit{quantum}), or applicable to both classical and quantum (\textit{both}). }
	\label{tab:rq2_hyperparameter}
	\resizebox{.9\textwidth}{!}{
		\input{generated_files/RQ2_hyperparameters.tex}
	}
\end{table}

%below are the parameters inherited from the base algorithms such as GA and PSO
%population size
\subsubsection{Quantum-inspired approaches}
\label{subsubsec: quantum-inspired hyperparameter}
When working with quantum-inspired algorithms, population size is the most critical hyperparameter to tune. As shown in Table~\ref{tab:rq2_hyperparameter}, 31 primary studies explicitly configure this parameter, which is more than any other hyperparameter. This is mainly because the majority of quantum-inspired approaches were built on classical population-based metaheuristics such as PSO \cite{guo2023effective, wu2020hybrid} and GAs (\cite{xiong2016optimized, qiao2020novel}), which inherently require the specification of \textit{population size} to govern the scale and diversity of the search process.
%mutation and crossover settings
Similarly, these classical evolutionary algorithms on which many quantum-inspired approaches are built require the configuration of \textit{mutation and crossover} mechanisms (e.g., mutation and crossover rates or probabilities). 
%\tao{Yuechen: they should mutation rate/probability or crossover rate/probability, right? If yes, then we should name them precisely}\yc{Response: Most papers are related to mutation rate and crossover rate. However, some incoporare rotation angles for the mutation and crossover phase. I am not confident that they can regared as ``rate''.}. 
For instance, in \cite{qiao2020novel}, the authors proposed a GA-based quantum-inspired approach, which naturally inherits these parameters to tune both mutation and crossover rates to balance exploration and exploitation.

In addition to the parameters inherited from the base algorithms, some quantum-inspired approaches also configure three parameters that are quantum inspired: initial qubit states, the parameter for updating quantum rotation angles, and contraction-expansion coefficient. 
%\yc{I am not sure whether to regard contraction-expansion as quantum-relevant. ChatGPT said no.}\tao{Yuechen: you labelled it :) i think it is a quantum-inspired parameters. I changed the wording to quantum inspired, instead of quantum relevant. }
%initial qubit state 
Especially the \textit{initial qubit states} is the second most frequently configured hyperparameter, as 12 of the primary studies explicitly initialize the qubit states (Table~\ref{tab:rq2_hyperparameter}). 
Initializing qubits is analogous to determining individual start positions in evolutionary algorithms. Moreover, we observe that many papers configure the initial qubit state as $\frac{1}{\sqrt{2}}\ket{0}+\frac{1}{\sqrt{2}}\ket{1}$ with equal amplitudes mimicking Hadarmard gates on the default state, indicating that each case to be searched is assigned with identical probability.
%parameter for updating quantum rotation angles
The hyperparameter of \textit{updating quantum rotation angles} $\Delta\theta$ controls the incremental update of rotation angles $\theta_m$, which controls the probability amplitudes of qubit-like states encoding candidate solutions. Specifically, each angle $\theta_m$ defines the quantum-inspired state: \[|\psi_m\rangle = \cos(\theta_m) |0\rangle + \sin(\theta_m) |1\rangle,\] where $\cos(\theta_m)$ and $\sin(\theta_m)$ are the probability amplitudes associated with $|0\rangle$ and $|1\rangle$, respectively. The probability of measuring qubit $m$ as $1$ can then be: the squared magnitude of its amplitude:  
\[p_m = \left| \langle 1 | \psi_m \rangle \right|^2 = \sin^2(\theta_m).\] By tuning $\theta_m$ via $\Delta\theta$, the algorithm gradually increases the amplitude of high-fitness solutions, which hence enhances their measurement probability. The whole idea is to mimic the amplitude amplification in quantum algorithms such as Grover's search. 
%contration-expansion coefficient
The \textit{contraction-expansion coefficient} $\alpha$ is a hyperparameter of Quantum-inspired Particle Swarm Optimization (QPSO), which is analogous to the inertia weight $w$ of Particle Swarm Optimization (PSO) (modulating velocity dynamics of particles), for balancing exploration and exploitation during the search process. 

\subsubsection{Quantum and hybrid approaches}
For quantum or hybrid approaches, as shown in Table~\ref{tab:rq2_hyperparameter}, \textit{penalty coefficient }$\lambda$ has been configured by 13 approaches, which all use either QAOA or QA to solve QUBO problems with different quantum hardware. Since the standard QUBO formulation has no constraints, to handle real-world problems (often having constraints) violations of such constraints are penalized by adding them as quadratic penalty terms to the objective function being optimized. The penalty coefficient controls how strongly violations of constraints are discouraged. In other words, actually this parameter is used to enforce hard constraints by transforming them into soft penalties in the objective function. Properly tuning $\lambda$ is critical as too small of it leads to infeasible solutions, while too large of it possibly dramatically changes the objective landscape and ends up in suboptimal regions. Although this parameter is critical, not all the primary studies that rely on QUBO formulations configure this parameter; Instead of explicitly configuring this parameter, within the 15 papers that discuss penalty coefficients, only two adopt fixed default values, such as~\cite{\paperfirst} with $\lambda=2.0$, while the remained 13 explore coefficients adaptive to specific arguments, like the primary study ~\cite{\papersecond} setting $\lambda$ as 1.5 times of the sum of all feature costs.  
% Fixed: ID-1 (i.e.,2.0), ID-84 (i.e.,27)
% Unfixed: ID-2, ID-5, ID-25, ID-27, ID-52, ID-53, ID-54, ID-58, ID-60, ID-61, ID-63, ID-68, ID-90
% Total number of papers: 14 for quantum/hybrid and 1 for quantum-inspired

% Optimizer of variational algorithms
The \textit{selection of classical optimizers for variational algorithms} was made in seven primary studies, in which classical optimizers (e.g., COBYLA in \cite{barletta2023quantum}) were selected to tune parameters in variational quantum circuits, such as in Variational Quantum Eigensolver (VQE) (e.g., \cite{nayak2023constructing}) and QAOA (e.g., \cite{ammermann2024quantum}). These optimizers are for updating parameters based on evaluation of cost functions defined in the variational algorithms, where each cost value is estimated through repeated measurements of the quantum state, which consequently drives the parameterized quantum circuit to evolve toward an optimal parameter configuration.

%QAOA layers
\textit{The number of QAOA layers} (denoted as $p$) determines the depth of quantum circuits and thus the algorithm's ability to refine its search for an optimal solution. Each layer introduces a pair of tunable parameters that guide the evolution of the quantum state. While increasing the number of layers potentially improves solution quality, it also deepens the circuit, which actually amplifies the impact of noise. As a result, the number of layers in QAOA is a critical hyperparameter that must be carefully chosen to balance accuracy against the practical limitations of quantum hardware in the NISQ era. Our study indicates that all seven primary studies using QAOA configured this parameter (Table~\ref{tab:rq2_hyperparameter}). 
%\tao{Yuechen, can you double check whether all approaches using QAOA configured this parameter? }\yc{Yes, for all of seven papers.}

\subsubsection{Parameters common across all types of approaches}

% for both categories
Two parameters are commonly configured across the three types of quantum optimization methods: dataset split ratio and weight of each objective. 

%Dataset split ratio
For learning based tasks (e.g., failure prediction~\cite{shen2021failure}), the \textit{dataset split ratio} (e.g., proportions allocated for training, validation, and test sets) is a critical hyperparameter, as its configuration significantly influences model generalization, evaluation reliability, etc. As shown in Table~\ref{tab:rq2_hyperparameter}, there are, in total, six primary studies explicitly configure this parameter: two employing quantum-inspired approaches and four utilizing quantum or hybrid approaches. We want to point it out that regardless of whether the learning framework is classical, quantum or hybrid, the dataset split ratio remains an essential hyperparameter, as it fundamentally shapes the model’s ability to generalize and be reliably evaluated. 
%\tao{Yuechen, can you check these two sets of primary studies and comment a bit here on whether there are some differences in setting the dataset split ratio across the two categories? The thing is that obviously we have less data in quantum and hybrid. It is better to provide concrete data such as the commonly used setting of the ratio is...}
%\yc{The limited number of samples hardly supports to conclude such differences.}
Specifically, among the six papers involving the dataset split ratio, we discover that none includes the validation set. A half of papers~\cite{\papereleventh, \paperfiftysixth, \papersixtyninth} adopt the proportion of training and testing sets as 80\% and 20\%, and one paper~\cite{\papersixth} configures a similar split ratio, i.e., 70\% and 30\%.
% Train/Validate/Test
% Quantum/hybrid: ID-6(7/0/3), ID-11(8/0/2), ID-56(8/0/2), ID-69(8/0/2)
% Quantum-inspired: ID-71(3/0/2), ID-73(2/0/1)

%Weight of each objective
Multi-objective optimization problems are often converted into single-objective formulations before being solved, which is achieved typically via weighted summation or other aggregation methods, because the relative importance of competing objectives fundamentally influences solution tradeoffs, convergence behaviour, and final performance. This holds true across paradigms, i.e., quantum-inspired, quantum, and hybrid approaches. Due to algorithmic simplicity, hardware constraints, or compatibility with existing solvers, weighted aggregation remains prevalent across all paradigms, often at the cost of reduced exploration of tradeoffs among competing objectives. 
As shown in Table~\ref{tab:rq2_hyperparameter}, there are, in total, 11 primary studies using weights to convert multi-objective optimization problems into single-objective formulations.
% \tao{Yuechen: can you check paper ID-11, which was labelled as Single. All the rest are multi-objective}\yc{This paper assigns weights to the linear and quadratic terms in its QUBO formulation. Actually, the two terms cannot be regarded as two objectives, but the hyperparameters are indeed ``weights''.}. \tao{If this is the case, we should remove paper ID-11} 
%Notably, equal weighting is the most commonly adopted strategy and is often applied without justification, which is also the case in classical optimization\tao{Man: this is true right? If yes, we need a reference}\man{I am not sure. If there is no clear prior objective, it might be a case.}. \tao{Yuechen: can you check these 11 papers and conclude whether they all use equal weighting. If not what else they used.}
Five papers~\cite{\paperthirteenth, \papersixtythird, \papersixtyfourth, \paperfortyfirst, \paperseventieth} merely take into account equal weights for each objective, while the others incorporate different weights for objectives owing to particular requirements and purposes. For example,~\cite{\papersixtyeighth} discusses four groups of weights to balance response time and sustainability (i.e., (0.0, 1.0), (0.2, 0.8), (0.5, 0.5), and (0.8, 0.2)), where the weights are determined by various prioritization requirements.
% 11 papers using weights
% Quantum/hybrid: 	ID-11(0.98/0.02), ID-13(equal), ID-63(equal), ID-64(equal), ID-68(0.0/1.0, 0.2/0.8, equal, 0.8/0.2)
% Quantum-inspired: ID-40(both equal and 0.2/0.8), ID-41(equal), ID-70(equal), ID-79(non-equal), ID-82(0.4/0.4/0.2), ID-92(0.5/0.1/0.1/0.1)

\begin{results}[Findings of Hyperparameters]
	For quantum-inspired algorithms, population size, mutation, and crossover rates are commonly tuned as they derive from classical metaheuristics such as PSO, and additional quantum-inspired parameters (e.g., initial qubit states, rotation update rates, and contraction-expansion coefficients) are also required to control search dynamics; For quantum or hybrid approaches, the penalty coefficient is often tuned when formulating problems as QUBO; variational algorithms require careful configuration of optimizers, while for QAOA specifically, the number of layers is a critical hyperparameter that directly impacts solution quality and convergence; Across quantum-inspired, quantum, and hybrid approaches, the dataset split ratio is configured to ensure reliable model evaluation in learning tasks, while objective weights are set to convert multi-objective problems into single-objective formulations. 	
\end{results}

%%-------------------------------------------------------------------------%%
\subsection{\rqExperimentSetting}

\begin{figure}
	\centering
	\includegraphics[width=.6\textwidth]{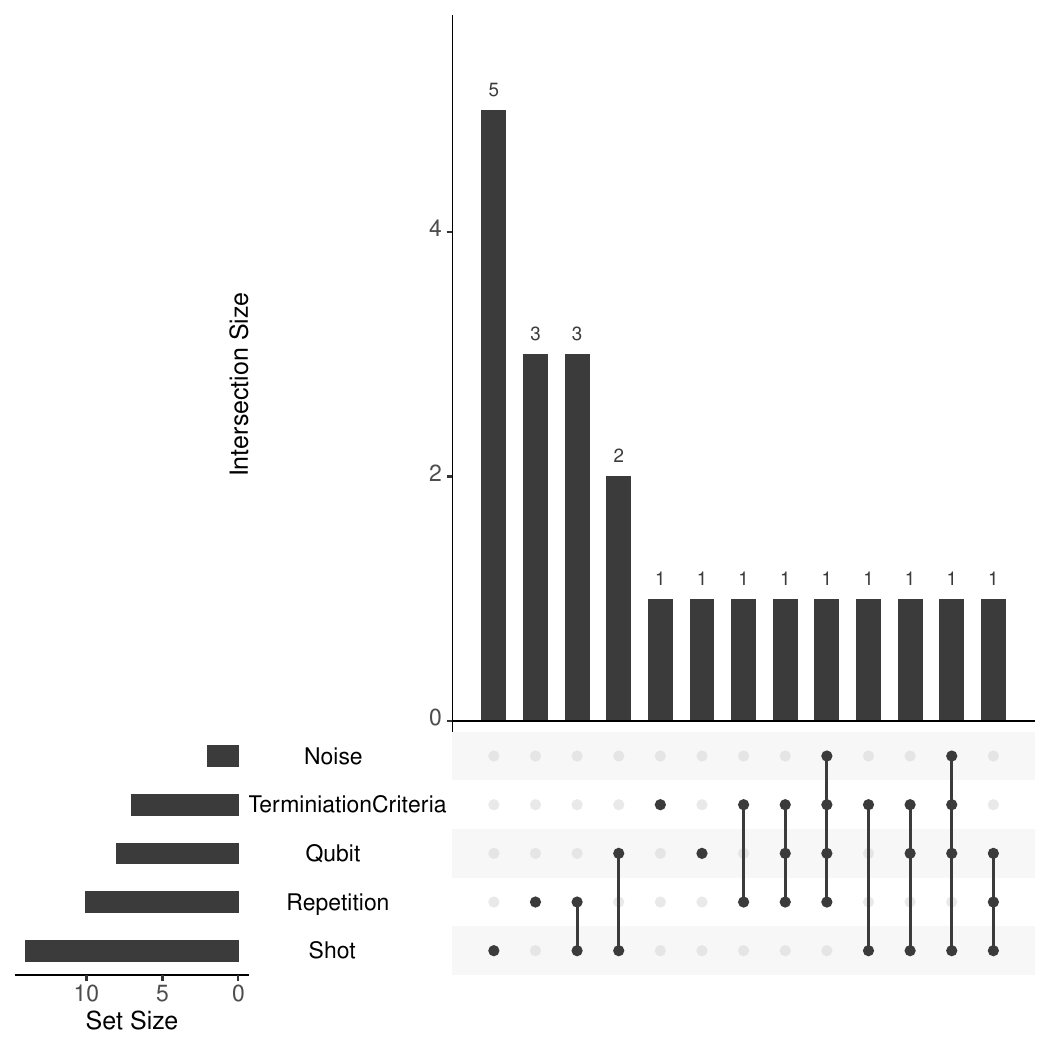}
	\caption{Overview of parameter settings - RQ3}
	\label{fig:RQ2_quantum_approach_settings_upset}
\end{figure}

%overall
The parameter settings of the approaches are of great importance in empirical studies because they directly influence the performance, confidence, and reproducibility of experimental results. In Figure~\ref{fig:RQ2_quantum_approach_settings_upset}, we present an overview. From the figure, we can observe that among the five concerned aspects (i.e., the number of shots, termination criteria, the number of repetitions, the number of qubits, and noise), the number of shots is the most concerned parameter, while noise is the least concerned parameter. This is reasonable because quantum measurements are probabilistic and hence a quantum circuit needs to be executed many times (i.e., taking many "shots") to estimate the probability distributions over all possible outcomes. In other words, more shots imply better estimation of output probabilities. In addition, it has impacts on the performance of quantum algorithms and has been considered as a universal parameter to be tuned for balancing accuracy and cost. On the other hand, noise is often ignored due to the fact that modelling realistic noise is not only complex but also platform-dependent. Hence, many studies actually only use ideal simulators, although it is widely accepted that considering noise is crucial for evaluating real-world performance of quantum solutions in the NISQ era. Termination criteria are important because they determine when the optimization process should stop, balancing solution quality against computational cost.

When examining Figure~\ref{fig:RQ2_quantum_approach_settings_upset} vertically, we observe that, aside from noise, there are cases where two or more of the other four parameters (the number of shots, termination criteria, the number of repetitions, and the number of qubits) were configured together. However, the number of such cases is very small (e.g., 1–3 instances per combination). As such, it is not possible to draw solid conclusions. This suggests that parameter settings were often made as ad hoc methodological choices rather than as part of established and systematic best practices. 

\subsubsection{Number of Repetitions}
\label{subsec:Repetition}

\begin{figure}
	\centering
	\begin{subfigure}{0.45\textwidth}
		\centering
		\includegraphics[width=\textwidth]{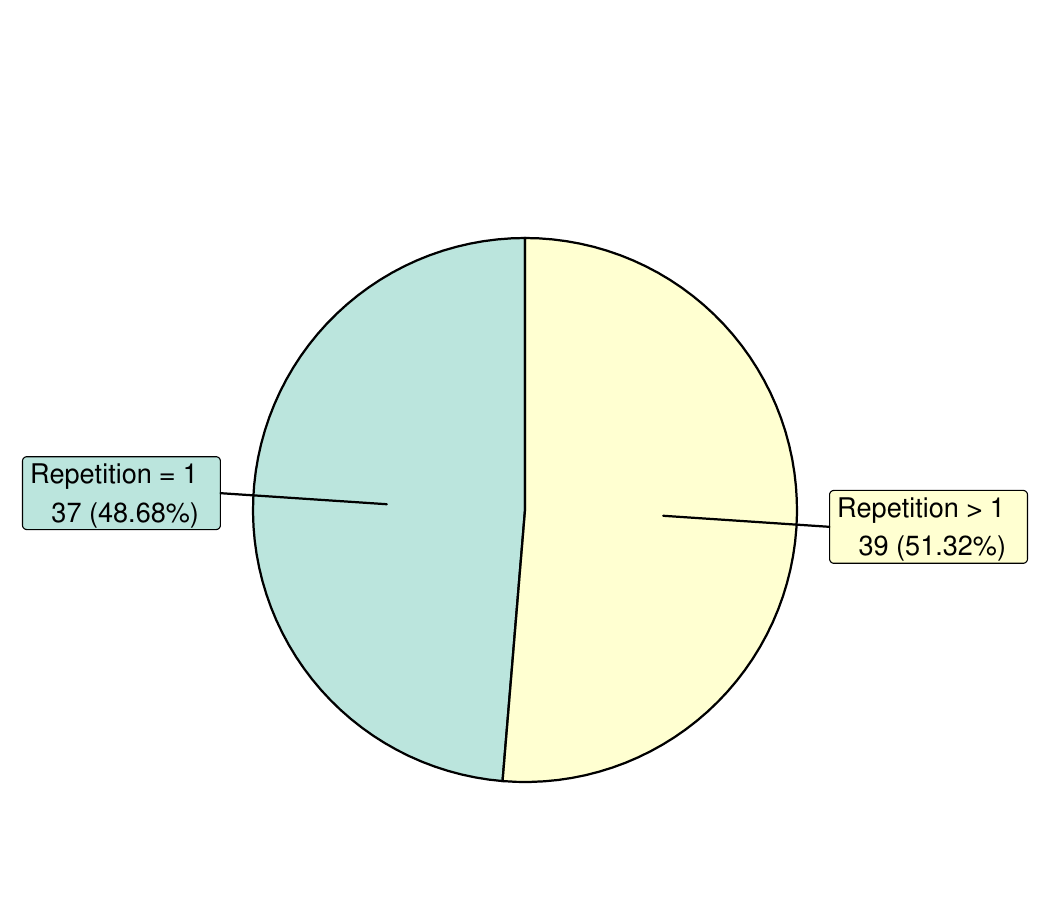}
		\caption{The ratio of primary studies with and without multiple repetitions}
		\label{fig:RQ2_hasRepetition_summary_pie}
	\end{subfigure}
	\hfill
	\begin{subfigure}{0.45\textwidth}
		\centering
		\includegraphics[width=\textwidth]{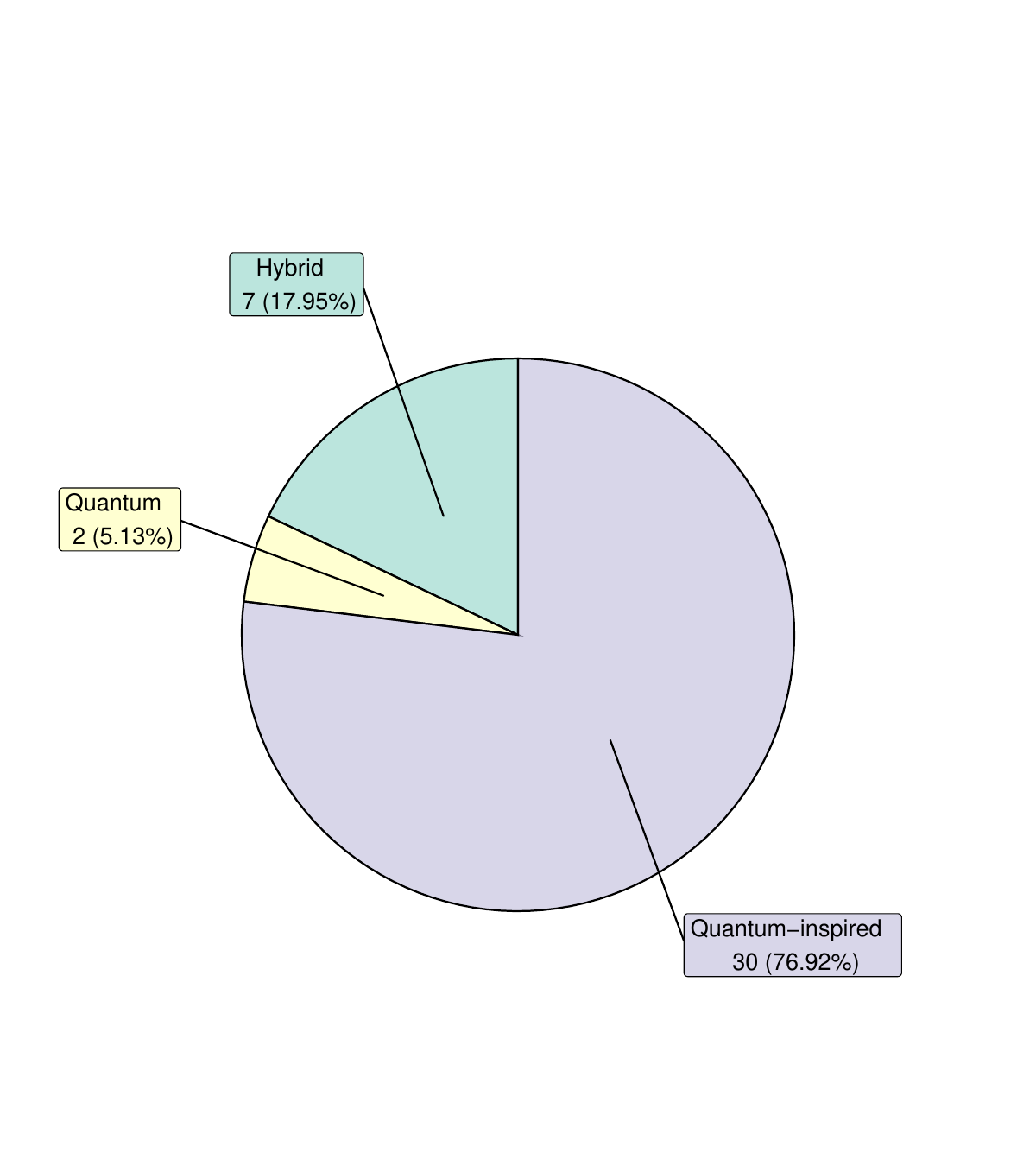}
		\caption{The ratio of primary studies with multiple repetitions among the three optimization types}
		\label{fig:RQ2_repetition_byQOpType_summary}
	\end{subfigure}
	\hfill
	\begin{subfigure}{0.8\textwidth}
		\centering
		\includegraphics[width=0.9\textwidth]{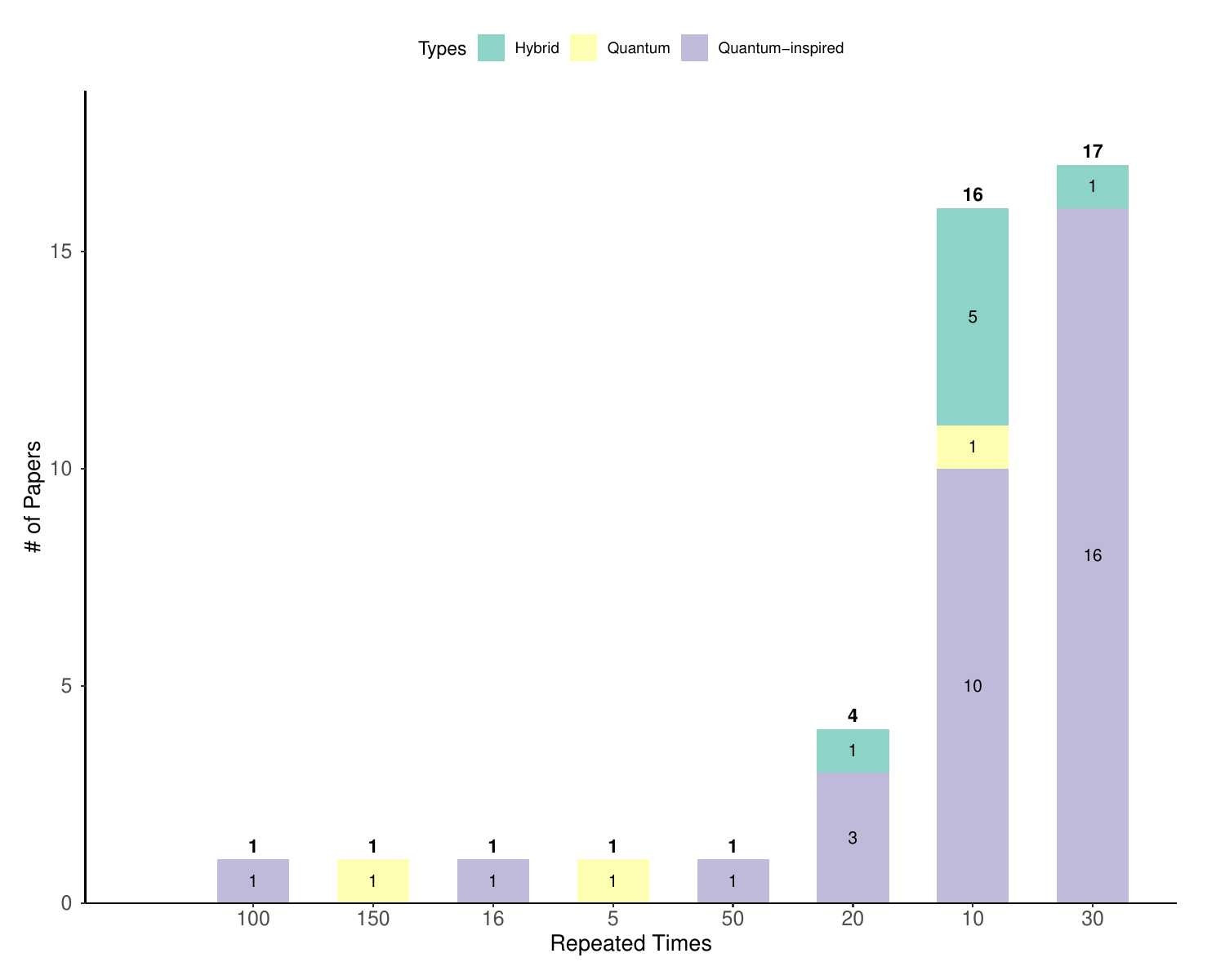}
		\caption{The number of repetitions across the three optimization types}
		\label{fig:RQ2_repetition_byQOpType_groupedbar}
	\end{subfigure}
	\caption{Overview of settings of repetitions - RQ3}
	\label{fig:overall_repetition_figure}
\end{figure}

One typical experimental setting is the number of repetitions (or runs, trials) for characterizing the variability of an entire approach, which can be either classical or quantum. The variability might be caused by randomness, heuristic nature of the approach, noise from quantum hardware, etc. As shown in Figure~\ref{fig:RQ2_hasRepetition_summary_pie}, only 39 out of \finalselected primary studies (51.32\%) have their experiments repeated (i.e., the number of repetitions is greater than one). This clearly shows a notable limitation in the current empirical research on quantum optimization experiments reported in the primary studies, because repeating experiments is crucial for ensuring the statistical confidence of results, especially in quantum optimization, where outcomes can be influenced by noise, hardware variability, and probabilistic nature of quantum systems. 

When further classifying the primary studies that have their experiments repeated according to the quantum optimization types, as shown in Figure~\ref{fig:RQ2_repetition_byQOpType_summary} observe that quantum-inspired optimization algorithms dominate these studies (76.92\%), followed by hybrid approaches (17.95\%), and then pure quantum approaches (5.13\%). Currently, this is understandable because quantum-inspired optimization algorithms are classical ones, which can benefit from a mature body of knowledge in experimental design, statistical validation, and best practices for reproducibility. In addition, they run on classical computing infrastructure, which offers abundant, accessible computational resources to enable experiment repetitions. In contrast, pure quantum and hybrid approaches rely on NISQ-era quantum hardware, which remains expensive and limited in accessibility; even when simulated, these algorithms often face scalability challenges due to the exponential cost of representing quantum states classically.

We further present the numbers of repetitions across the three different quantum optimization types in Figure~\ref{fig:RQ2_repetition_byQOpType_groupedbar}. First of all, we can see that the number of repetitions ranges from 5 to 1000, demonstrating a large variability. Second, the numbers of 30 and 10 are the top two choices of 17 and 16 primary studies, respectively. The other seven settings were used in the empirical studies presented in fewer than 5 primary studies. 
This is because the study conducted by~\cite{arcuri2011practical,arcuri2014hitchhiker} suggests that at least 30 runs are needed to analyze randomized algorithms, and the use of 30 has been widely adopted in evolutionary computation~\cite{shi2014multipopulation}. The number 10 is also the choice of 16 primary studies' experiments, which we think it is a choice made to strike a balance between statistical rigour and computational cost, resource limitations and even feasibility.
We also observe that more quantum and hybrid approaches being evaluated with experiments with 10 repetitions than 30 (6 vs. 1). This is again because quantum and hybrid optimization experiments often rely on access to quantum hardware and/or resource-intensive quantum simulators, both of which are computationally expensive and time-constrained.
%Surprisingly, there are one hybrid and one quantum approaches opt for the number 1000.
% \tao{Yuechen, we need to check the paper and understand why they chose 1000. We should also check the studies with numbers of 50, 100, 150 as well. 16 is an interesting number :D}\yc{Unfortunately, no paper explains the choice of special numbers for repeats. They might mention the reason for including repeats.}\yc{I am bothered that several works do not clarify the hyperparameters (i.e., algorithm steps, independent repeats, and number of measurement) very well}. 
Nevertheless, there is no standardized or systematic approach for setting the number of repetitions.

% ID-34 | 50 | for parameter tuning
% ID-74 | 100 | no explanation
% ID-77 | 50 | no explanation
% ID-90 | 5; 150 | The final results are evaluated on the test set, with all reported metrics representing the average of 5 runs with random initialization.; To further explore this, we conducted additional experiments using the Hybrid method, selecting 135 features for the Item-KNN with both “all" and “Indiv" approaches, repeating the process 150 times.
% -------------------------------------------------------------------
% Disccusion board
% Paper |  Repeats  |  Evidence  |  YC's remarks
% ID-23 |  16  | In this study, the entire simulation is run sixteen times, beginning at the anchor position (0,0) coordinates and ending at the last anchor point (15,0) coordinates. | I suspect that ``sixteen times'' does not indicate the independent repeats.

%\begin{figure}[htbp]
%	\centering
%	\includegraphics[width=0.9\textwidth]{generated_files/RQ2_repetition_byQOpType_groupedbar.pdf}
%	\caption{Descriptive statistics of the number of repetitions across the three optimization types - RQ3}
%	\label{fig:RQ2_repetition_byQOpType_groupedbar}
%\end{figure}

\begin{results}[Findings of Number of Repetitions]
Around half of the \finalselected primary studies report repeated experiments, the majority of which (76.92\%) are quantum-inspired approaches, followed by hybrid and quantum ones (17.95\% and 5.13\% respectively); The numbers 30 and 10 are the most used settings for the number of repetitions.
\end{results}

\subsubsection{Number of Shots} 
\label{subsec:shots}
The number of shots (denoted as $s$) is a critical quantum-specific setting for quantum, hybrid and Digital Annealing (DA) approaches. In quantum and hybrid approaches, multiple shots are required because quantum measurements are inherently probabilistic, and repeated circuit executions are needed to estimate the output distribution reliably. Although DA is implemented on classical hardware, it emulates the sampling behavior of QA via stochastic processes where repeated runs serve a similar purpose: collecting statistical data to approximate the solution distribution. Note that the number of shots corresponds to $s$ repetitions of the same evolution and measurement in one call of a quantum system. 
%
%\tao{Yuechen: We perhaps can provide a psedo code to illustrate the case when repetitions and shots are both needed. For instance, for paper [136], the number of shots is 100 while the repetition is 10.  A simple illustration might be good here. }\yc{OK. I add Algorithm~\ref{alg:repeats_Shots}} \tao{Yuechen: we need a hybrid approach to illustrate this, I suppose. }
%
In Algorithm~\ref{alg:repeats_Shots}, we show the procedure of an empirical evaluation on solving one problem by a quantum optimization algorithm, where we should employ a quantum system (e.g., a quantum circuit or a quantum annealer) to represent this problem. The pseudocode demonstrates that $s$ shots are performed during one call of the backend. Then, we can gain $s$ measurements and then calculate a corresponding experimental metric $m$. For statistical significance, the final metric to evaluate the approach's performance should not rely on one derived metric. Instead, we suggest repeating calling the backend (e.g., $r$ repeats) to obtain a reasonable number of measurements.
Note that in Quantum Annealing (QA) and DA, considering that an annealer executes a single, physical process in one repetition and samples from the final quantum state, the entire process is to sample the output distributions. Hence, the number of repetitions is often called the number of \textit{reads} or \textit{samples}, which are equivalent to the number of shots, in our study. 

\begin{algorithm}[!t]
	\small
	\caption{Empirically evaluating a quantum system with independent repetitions} 
	\label{alg:repeats_Shots}
	\input{psedo_code/repeats_vs_shots.tex}
\end{algorithm}

\begin{figure}
	\centering
	\begin{subfigure}{0.4\textwidth}
		\centering
		\includegraphics[width=\textwidth]{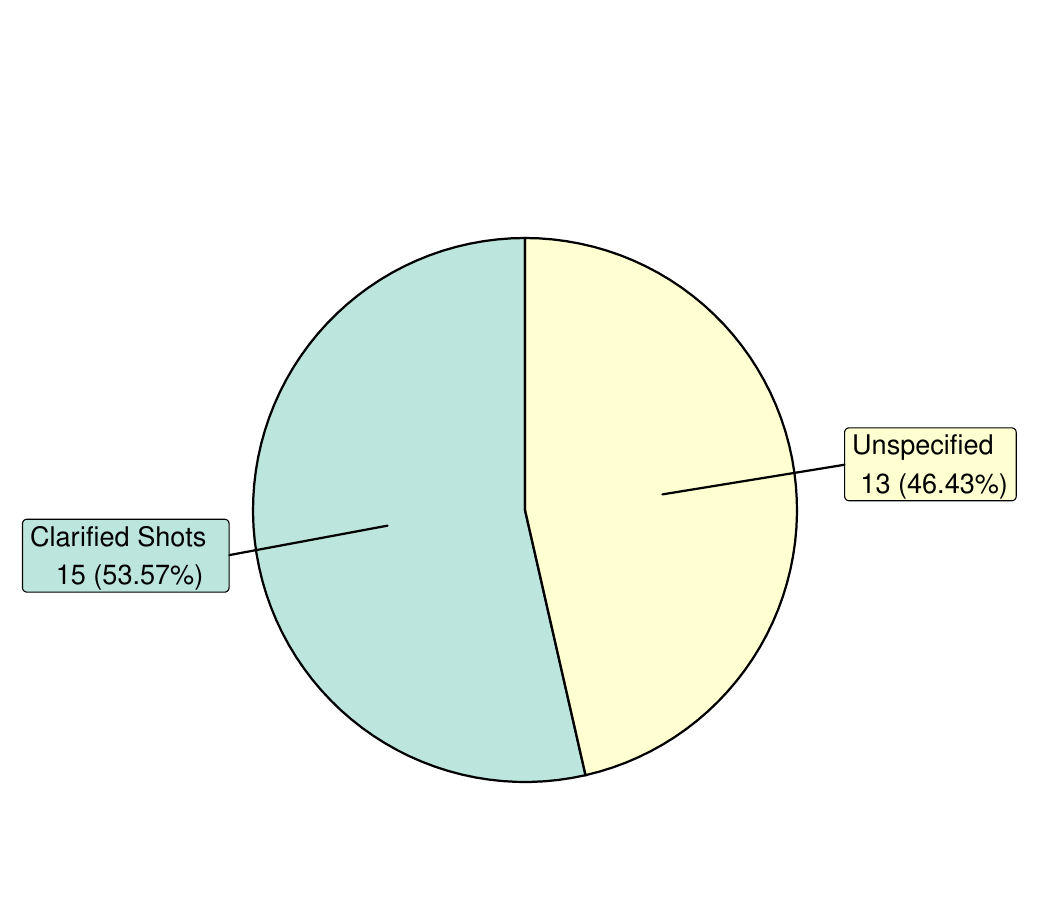}
		\caption{Overview of settings of the number of shots in quantum, hybrid and DA approaches}
		\label{fig:RQ2_hasShots_summary_pie}
	\end{subfigure}
	\hfill
	\begin{subfigure}{0.55\textwidth}
		\centering
			\includegraphics[width=\textwidth]{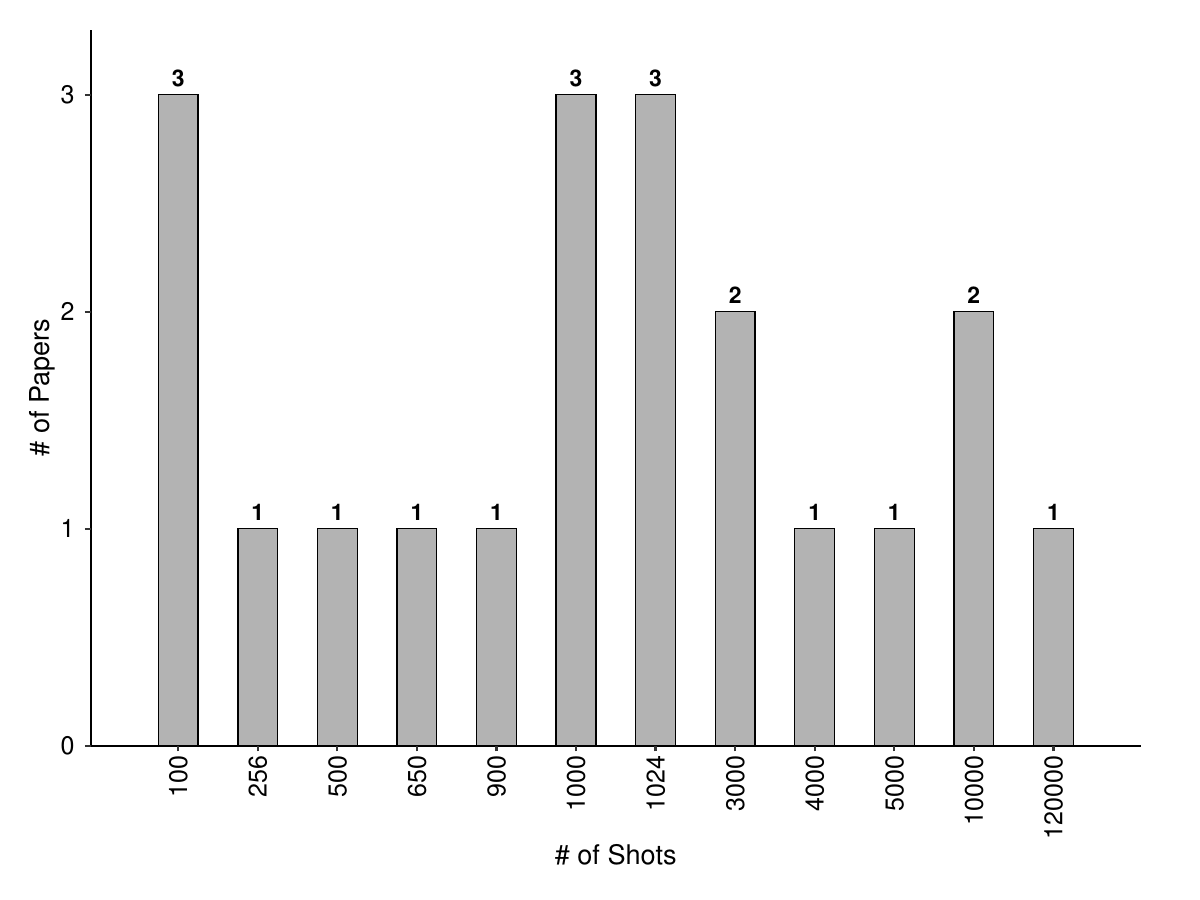}
		\caption{Descriptive statistics of settings of the number of shots}
		\label{fig:RQ2_hasShots_summary_bar}
	\end{subfigure}

	\caption{Overview of settings of shots - RQ3}
	\label{fig:overall_shots_figure}
\end{figure}

%Overview
Overall, as illustrated in Figure~\ref{fig:RQ2_hasShots_summary_pie}, only 15 primary studies (53.57\%) report the setting of number of shots, indicating a critical gap in current practices for quantum optimization studies, as the number of shots directly affects the accuracy and statistical confidence of quantum computation outcomes.

%\begin{figure}
%	\centering
%	\includegraphics[width=0.6\textwidth]{generated_files/RQ2_hasShots_summary_pie.pdf}
%	\caption{Overview of settings of the number of shots in quantum, hybrid and DA approaches - RQ3}
%	\label{fig:RQ2_hasShots_summary_pie}
%\end{figure}

%
%quantum-inspired and quantum
In Table~\ref{tab:rq2_shots}, we summarize the settings of the 15 primary studies on the number of shots. As shown in the table, there is only one quantum-inspired approach that is based on DA, for which the number of shots was set as 100. There was considerable variation in the configuration of experiments conducted to evaluate quantum and hybrid approaches; the number of shots ranged three orders of magnitude (from 100 to 120,000), and the number of repetitions also differed, with values of 1, 5, 10, 100, and 150 being set. This extreme variation suggests the field lacks standard guidance. It seems that the chosen values appear more driven by practical constraints, such as the financial cost and queue time associated with cloud-based computation resources, etc., than requirements on achieving statistical significance of the problems being solved. 
We would also like to point it out that only six (out of 15) primary studies reported both the number of repetitions and shots. 
The observation once again indicates that there is a lack of standardized reporting guidelines in the field and underscores a critical need for more rigorous reporting standards in conducting empirical studies in quantum optimization. 

%Four QA approaches \cite{kumar2025optimizing, jhaveri2023cloning, trummer2015multiple, niu2024performance} configure the number of shots as 100, 650, 1000 and 120000, respectively.  \tao{Yuechen, I checked paper ID-90, I cannot find the setting of number of shots as 12000. Can you double check?}\yc{The paper claims ``QA was used to take 120,000 repeated samples.''. I am aware that such many samples might not correspond to the shots per run.} The primary study \cite{miranskyy2022using}, which experiments with Grover's algorithm simulator and hardware, sets the number of shots as 1024  (for simulator) and 4000 (for hardware). \tao{Yuechen: can you check the paper to see whether they justify why two settings?}

As shown in Figure~\ref{fig:RQ2_hasShots_summary_bar}, the most common shot counts were 100, 1000, and 1024 (default setting of Qiskit), each of which was used in three primary studies. The lack of a dominant value and the wide range of other settings prevent any meaningful conclusion about a standard or optimal configuration.

\begin{table}
	\small
	\centering
	\caption{Summary of settings of the number of shots - RQ3 
		%\tao{Man and Yuechen: why we report repetitions here? Why do we have "quantum, hybrid" row?}\man{1. "Quantum, Hybrid" indicates that the paper proposed multiple approaches which cover quantum and hybrid types. Now we combine quantum, hybrid and `Quantum, Hybrid'. We could also remove this `Quantum Approach' column. 2. Regarding repetition, since shot settings may also relate to the concept of ``repeat'', it might be worth discussing them together. btw, there are few papers which report both shot and repetition settings.}
		}
	\label{tab:rq2_shots}
	%\resizebox{.9\textwidth}{!}{
		\input{generated_files/RQ2_shots.tex}
		%}
\end{table}

%\begin{figure}
%	\centering
%	\includegraphics[width=0.6\textwidth]{generated_files/RQ2_hasShots_summary_bar.pdf}
%	\caption{Descriptive statistics of settings of the number of shots - RQ3}
%	\label{fig:RQ2_hasShots_summary_bar}
%\end{figure}

\begin{results}[Findings of Number of Shots]
	Of the 28 relevant primary studies, only 15 report the number of shots, which ranged widely (from 100 to 120,000), with no single setting being dominant. More critically, only six primary studies specify both shots and repetitions. This indicates a significant lack of standardized reporting guidelines in the field. 
\end{results}

\subsubsection{Number of Qubits} 
\label{subsec:numberQubits}

\begin{figure}
	\centering
	\includegraphics[width=0.5\textwidth]{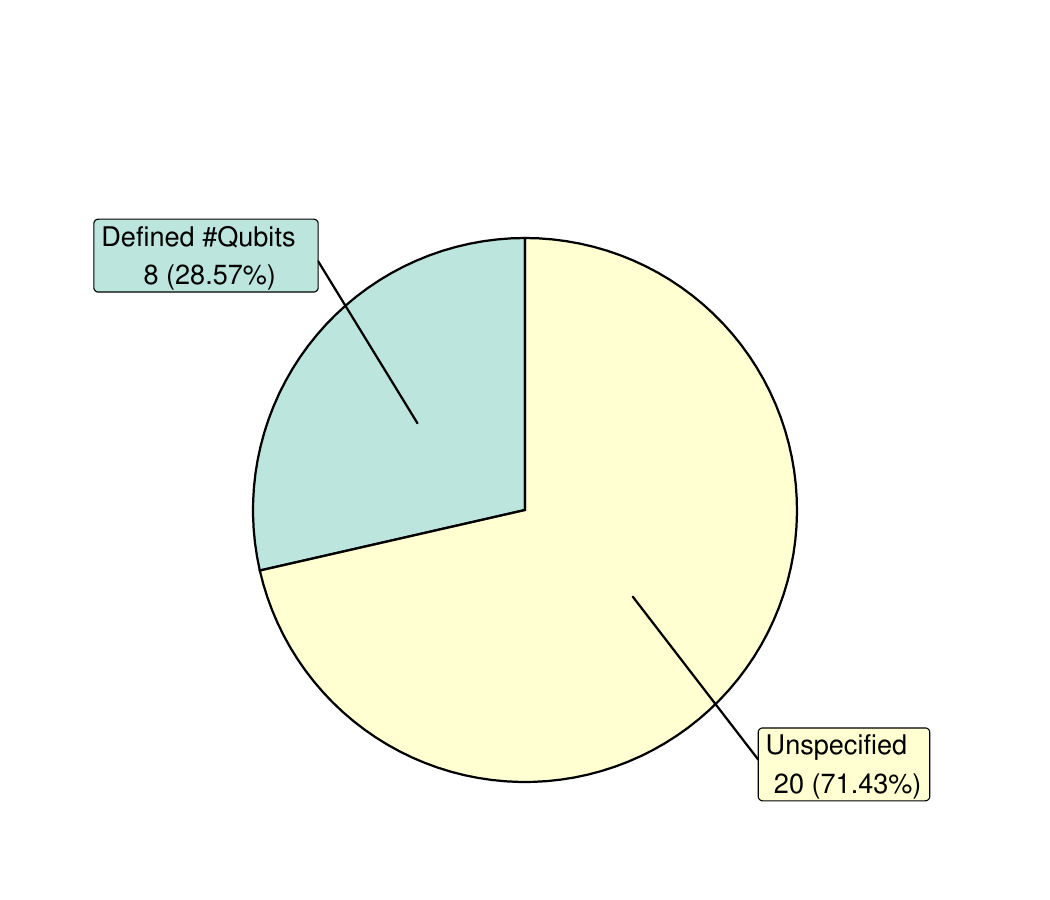}
	\caption{Overview of the primary studies that specify the number of qubits - RQ3 }
	\label{fig:RQ2_hasQubits_summary_pie}	
\end{figure}

In terms of experimental settings, only 8 out of the 29 primary studies employing quantum or hybrid approaches explicitly report the number of qubits used in their experiments, as illustrated in Figure~\ref{fig:RQ2_hasQubits_summary_pie}. %\yc{Missing a figure?} \tao{Man: this figure is commented out. }. 
This low reporting rate is surprising, given the fundamental role that this parameter plays in assessing algorithm scalability and hardware requirements. 
As further shown in Table~\ref{tab:rq2_Qubits}, the number of logical qubits employed ranges from 3 to 247, while the number of physical qubits ranges from a handful to thousands as only reported in \cite{wang2024test}. Notice that, in our context, the number of qubits usually refers to logical qubits, which can be considered as problem-encoding units abstracted from multiple (noisy) physical qubits. Logical qubits provide the meaningful resource metric for scalability, while the actual number of physical qubits required depends on hardware constraints. 

As shown in Table~\ref{tab:rq2_Qubits}, when conducting experiments on (either ideal or noisy) classical simulators\footnote{An ideal classical simulator models a completely fault-tolerant quantum computer without any noise, while a noisy classical computer models a quantum computer in the NISQ era by including realistic noise effects.}, the number of qubits ranges from 4 to 32, which is reasonable because simulating quantum states requires memory and computation that grow exponentially with qubit count (e.g., 20 qubits $\approx$ 1 million amplitudes, 30 qubits $\approx$ 1 billion amplitudes, already demanding for many classical computers).  
When looking the experiments conducted on quantum hardware, from Table~\ref{tab:rq2_Qubits}, one can notice that except for \cite{wang2024test} (reporting both logical and quantum qubit counts) and \cite{bettonte2022quantum, fankhauser2023multiple} (where it remains unclear whether the reported qubit counts refer to logical or physical qubits), the number of logical qubits in the empirical studies ranges from 4 to 247, with the largest number achievable nowadays only on D-Wave quantum annealers. 
In particular, since paper \cite{wang2024test} presents a hybrid approach which employs bootstrap sampling to decompose big problems to small ones to be solved with QA, aiming to optimize the use of qubits. Hence the use of the number of qubits is determined by the proposed approach for a given test case minimization problem.

\begin{table}
	\small
	\centering
	\caption{The number of qubits reported in the empirical studies - RQ3 }
	\label{tab:rq2_Qubits}
	%\resizebox{.9\textwidth}{!}{
		\input{generated_files/RQ2_Qubits.tex}
		%}
		
	\begin{spacing}{0.8}
		\raggedright \footnotesize 
		* and \textsuperscript{\textdagger} indicate that a primary study does not provide sufficient information to determine whether the qubits used in the empirical study are logical or physical, or whether the employed classical simulator is ideal or noisy, respectively.
	\end{spacing}
\end{table}

\begin{results}[Findings of Number of Qubits]
Only 8 out of 29 primary studies
%\yc{The two concrete values are derived from the pie figure that is missing currently.} 
report the number of qubits employed in their empirical studies. The reported logical qubit counts range from 4 to 247 when running experiments on quantum hardware, and from 4 to 32 when running on classical simulators. Notably, only one study explicitly reports both logical and physical qubit counts. Overall, the scale of the experiments in terms of qubit numbers remains limited by computational resources.
\end{results}

\subsubsection{Termination Criteria and Noise}
\label{subsec: TerminationCriteriaNoise}
%Termination criteria
Termination criteria of an optimization algorithm specify when it should stop running. In both classical and quantum optimization, commonly used termination criteria include the number of evaluations, convergence reached, no further improvement, etc. In Table~\ref{tab:rq2_termination}, we present all termination criteria employed by the primary studies and classify them into four categories: convergence, iteration, objective and time. 

%convergence
As shown in Table~\ref{tab:rq2_termination}, various \textit{convergence} indicators were used to in optimization algorithms to solve various problems. For instance, when observing 10 consecutive cost differences less than a given threshold, the employed Quantum Particle Swarm Optimization (QPSO) is considered converged in \cite{jin2015prediction}. We can observe that convergence indicators are domain-dependent, and there is currently no universally accepted standard for defining them. 
%
%Iteration
In addition to convergence, termination criteria in quantum, quantum-inspired, and hybrid optimization approaches often include limits on \textit{the number of generations, iterations, or objective function evaluations}. The maximum number of generations is typically used in population-based or evolutionary algorithms (e.g., quantum-inspired evolutionary algorithm for covering array generation \cite{wagner2019quantum}), which controls how many generations the population evolves before termination; the maximum number of iterations essentially sets an upper bound on how long an algorithm (e.g., DA for join ordering in \cite{schonberger2023digitalannealing} ) continues to update or refine a solution; and the number of objective function evaluations directly bounds the computational effort spent on evaluating solutions, which is only adopted in \cite{kumari2016comparing} for using quantum-inspired evolutionary algorithms for solving the next release problem. 
%
%Objective
Another common termination criterion is the \textit{optimization objective}, which indicates that a satisfactory solution has been found. Examples include discovering a covering array \cite{wagner2019quantum}, identifying a feasible solution \cite{hussein2020quantum}, or achieving an objective function value above a specified threshold \cite{bhatia2019quantum}. 
%Time budget
A \textit{time budget} is also commonly used as a termination criterion, stopping the optimization once a predefined computational time has elapsed, regardless of the solution’s status. Note that the specific time budget often varies depending on the computational hardware employed. 
%For instance, 24 hours were set as the time budget by running a quantum-inspired algorithm in \cite{guo2023effective} and 
%
These criteria are often combined with convergence indicators to define a practical stopping point. For instance, Wagner et al.~\cite{wagner2019quantum} employed both the maximum number of generations and achieving the objective as the termination criteria. 

%noise
Regarding \textit{noise}, only two primary studies considered noise: the primary study \cite{wang2024quantum} considered noise on both a noisy simulator and real quantum hardware in its empirical study while the primary study~\cite{\papersixtyfifth} only considered noise on a noisy simulator. This implies that noise is generally under-explored in current empirical research. None of the primary studies reports noise-related aspects of the simulator or hardware, such as the error rate of gates.

%\begin{figure}[htbp]
%	\centering
%	\includegraphics[width=0.8\textwidth]{generated_files/RQ2_quantum_approach_settings_upset.pdf}
%	\caption{Overview of settings defined in experiments for quantum computing optimization  - RQ2}
%	\label{fig:RQ2_quantum_approaches_settings}
%\end{figure}

\begin{table}
	\small
	\centering
	\caption{Summary of termination criteria - RQ3 }
	\label{tab:rq2_termination}
	\resizebox{.9\textwidth}{!}{
		\input{generated_files/RQ2_terminiation.tex}
	}
	
	\begin{spacing}{0.8}
		\raggedright \footnotesize 
	
	\end{spacing}
\end{table}

\begin{results}[Findings of Termination Criteria and Noise]
	Termination criteria include convergence indicators, which are commonly used. The number of generations, iterations, or objective function evaluations are also reported by some primary studies, while some studies additionally consider time budget. Noise, however, is addressed in only two primary studies, highlighting a significant gap in accounting for quantum hardware imperfections in NISQ era.
\end{results}

%%-------------------------------------------------------------------------%%
\subsection{\rqMetrics}
We classify collected evaluation metrics into four categories: metrics measuring SE problem complexity, approach complexity, effectiveness metrics, and cost and efficiency metrics, which are discussed in the following subsections. 

\subsubsection{Problem complexity}
\label{subsec:ProblemComplexity}
The problem complexity is about how big, complicated, and difficult an SE problem is, and often measured by the specific scale, constraints and structure of a given problem instance. Measuring the problem complexity is important because it provides insight into the scale and computational demands of a solution, which enables meaningful comparisons and informed decisions about feasibility and resource requirements. To this end, we collect all metrics measuring the SE problem complexity reported in the primary studies and summarize them in Table~\ref{tab:rq3_problem_metrics}, along with the SE problems themselves and corresponding SE activities. 

As shown in Table~\ref{tab:rq3_problem_metrics}, in general, we classify problem-specific metrics (e.g., the number of test cases in a test suite, resource constraints, and network topology) into two categories: size and dependency/constraint. All problems were measured with \textit{size}, which is understandable as size is easy to quantify, strongly correlated to the computational difficulty (as the input size grows, the solution space often grows exponentially), etc. If we say that size measures how big the search space it, then dependencies/constraints measure how tangled the search space is. As shown in the table, one can also observe that more than half of the problems' complexity arises from dependencies and constraints, which restrict feasible solutions but also significantly increase the computational hardness of the problem.
In our opinion, SE optimization problems are especially challenging because their complexity is largely attributed to dependencies and constraints, which restrict and fragment the feasible solution space. In contrast, numerical benchmarks such as Max-Cut~\cite{goemans1995improved} gain their difficulty mainly from combinatorial explosion with simpler interaction structures. Thus, SE problems present a different challenge by combining combinatorial complexity with domain-specific constraints (e.g., network topology in node localization problem \cite{tong2023can} and test requirements in test suite minimization problems \cite{bajaj2022Sensorstest}). 

\begin{table}
	\small
	\centering
	\caption{Summary of SE problem complexity metrics}
	\label{tab:rq3_problem_metrics}
	\resizebox{.99\textwidth}{!}{
		\input{generated_files/RQ3_problem_metrics.tex}
		}
\end{table}

\begin{results}[Findings of Problem Complexity]
	Size is the most commonly used measure of problem complexity, while more than half of the primary studies also consider dependencies and constraints, which restrict feasible solutions and increase computational hardness.
\end{results}

%---------------------------------------

\subsubsection{Effectiveness Metrics}
\label{subsec:EffectivenessMetrics}
\begin{figure}
	\centering
	\includegraphics[width=0.6\textwidth]{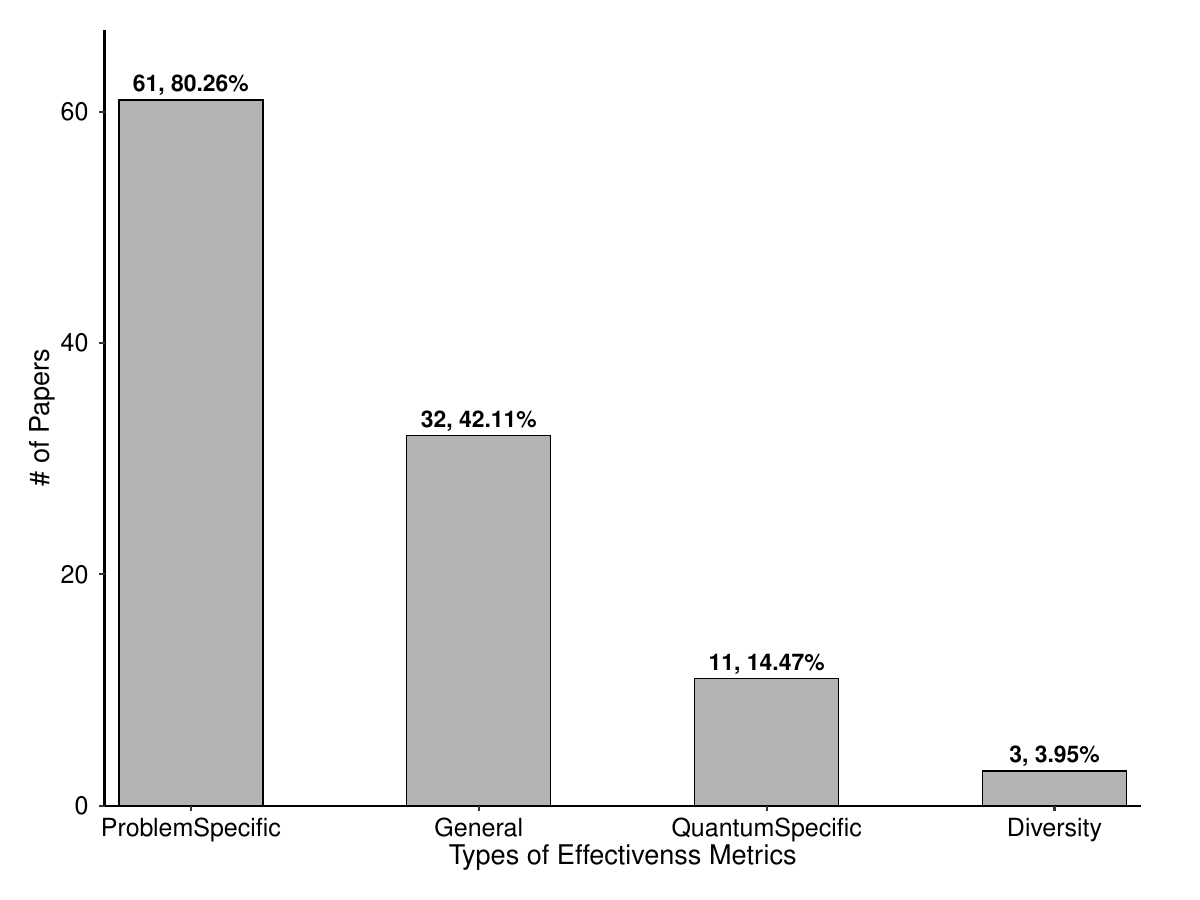}
	\caption{Descriptive statistics of effectiveness metrics - RQ4}
	\label{fig:RQ3_effectiveness_metrics}
\end{figure}

We classify all effectiveness metrics adopted in the primary studies into four categories: problem-specific, general, quantum-specific and diversity. Their descriptive statistics are provided in Figure~\ref{fig:RQ3_effectiveness_metrics}, which shows that problem-specific metrics have been applied by 80.26\% of primary studies, followed by general metrics (42.11\%), quantum-specific metrics (14.47\%), and diversity metrics (3.95\%). 

%\tao{Man: if it is very easy, we can round the data in the figure directly. Otherwise, let's keep like this for now.} \man{It is easy, but it would be better to keep consistent to represent percentages with two digits of precision.}
%

%problem-specific metrics
For readers’ reference, we provide all \textit{problem-specific metrics} in Appendix Table~\ref{tab:rq3_problem_effectivenes_metrics}. These metrics are very diverse (e.g., CPU usage rate, energy consumption, fault detection loss percentage) given their problem-dependent nature; the only metric appearing across multiple primary studies is fitness value, which is particularly designed for optimization problems and commonly applied in SBSE to measure the solution quality.  

%General effectiveness metrics
\textit{General effectiveness} metrics are summarized in Table~\ref{tab:rq3_general_effectiveness_metrics}. As shown in the table, \textit{recall}, \textit{F1 score}, \textit{precision}, \textit{accuracy} and \textit{specificity} are the most frequently used effectiveness metrics, as they directly capture performance in prediction, classification, and optimization tasks. \textit{Root Mean Square Error (RMSE)} is a common regression metric, which measures the average magnitude of error between predicted and actual values. As shown in the table, RMSE is adopted in five primary studies for evaluating the effectiveness of software failure prediction, routing, scheduling, etc. \textit{Approximation Ratio (AR)} has been adopted by three primary studies as a performance indicator to evaluate four optimization approaches. AR measures the quality of an approximate solution relative to the best possible solution and is commonly used to evaluate quantum algorithms. This is mainly because quantum optimization algorithms are inherently heuristic and probabilistic and hence they do not guarantee exact solutions. Hence, AR suits well for measuring the effectiveness of these algorithms and is often used to check whether quantum solutions can match or beat the best classical counterparts. Many other approach effectiveness metrics (e.g., success probability, cost quality, inclusivity) were adopted in only one or two primary studies, indicating that while they are general enough to apply across different problems, their usage is currently sparse and context-dependent. 
\begin{table}
	\small
	\centering
	\caption{Summary of general effectiveness metrics - RQ4. The second column is the number of primary studies.}
	\label{tab:rq3_general_effectiveness_metrics}
	\resizebox{.99\textwidth}{!}{
		\input{generated_files/RQ3_general_effectiveness_metrics.tex}
	}
\end{table}

%quantum-specific effectiveness metrics
\textit{Quantum-specific metrics} are summarized in Table~\ref{tab:rq3_quantum_effectiveness_metrics}. Energy value is the dominant metric, as it directly measures solution quality in quantum optimization with QA and QAOA. 
Energy landscape is considered in \cite{emu2024warm}, which structures configurations regarding the energy of a configuration and the arithmetic mean of the difference of energy between a current configuration and their nearest neighbours. The primary study \cite{schonberger2023ready} adopts three metrics to assess the feasibility of using gate-based quantum computers: circuit depth (number of sequential quantum gate layers), coherence time (duration a qubit retains its quantum state before decoherence), and the number of logical qubits (error-corrected qubits available for computation), which are important to consider in the NISQ era. These metrics are key indicators of quantum hardware capability and algorithm feasibility: circuit depth reflects computational complexity, coherence time limits how long reliable operations can run, and the number of logical qubits indicates the scalability of the proposed QUBO encoding approaches.
The primary study \cite{ammermann2024quantum} introduces two metrics: depth and width of quantum operators. 
The depth of quantum operators corresponds to circuit depth, as the operators in this context represent subroutines.
The width refers to the number of qubits used simultaneously in a subroutine, and hence measures the circuit’s spatial complexity.

\begin{table}
	\small
	\centering
	\caption{Quantum-specific effectiveness metrics - RQ4}
	\label{tab:rq3_quantum_effectiveness_metrics}
	\resizebox{.99\textwidth}{!}{
		\input{generated_files/RQ3_quantum_effectiveness_metrics.tex}
	}
\end{table}

%diversity
Regarding \textit{diversity}, we observe that only three primary studies considered diversity metrics, all of which address multi-objective optimization problems. Specifically, hypervolume (HV), spread, and number of non-dominated solutions are considered in \cite{kumari2016comparing, liu2024quantum}, both of which are quantum-inspired approaches. HV measures the volume of the objective space dominated by a set of solutions; spread quantifies how evenly solutions are distributed along the Pareto front; and the number of non-dominated solutions measures how many distinct, trade-off solutions are available in a solution for a multi-objective optimization problem. These three metrics, especially HV, are commonly applied in SBSE~\cite{ali2020quality,zhang2019uncertaintyTesting}. The primary study \cite{trovato2025reformulating} uses a QUBO formulation with a weighted-sum approach to handle multiple objectives, which inherently transforms the multi-objective optimization problem into a single-objective one. However, that paper still evaluates performance in terms of the Pareto frontier size, which is the number of non-dominated solutions found across runs.

\begin{comment}
\begin{table}
	\small
	\centering
	\caption{TODO\man{remove later.  just put here for easy reference to problem and papers}}
	\label{tab:rq3}
	\resizebox{.99\textwidth}{!}{
		\input{generated_files/RQ3_diversity_effectiveness_metrics.tex}
	}
\end{table}
\end{comment}

\begin{results}[Findings of Effectiveness Metrics]
	Problem-specific effectiveness metrics are highly diverse. Among general metrics, the most frequently used are recall, precision, F1 score, accuracy, and specificity, while RMSE and AR are also common. Time complexity is the most widely considered complexity metric, followed by space complexity. For quantum-specific evaluation, energy value is the dominant quantum-specific metric. Finally, only three primary studies address diversity, using HV, spread, and the number of non-dominated solutions as metrics.  
\end{results}

\subsubsection{Cost and Efficiency Metrics}
\label{subsubsec:cost}
As summarized in Table~\ref{tab:rq3_cost_metrics}, cost metrics are mostly about time (e.g., execution, optimization, training, prediction, embedding, sampling, and pre-processing) cost, but also include economic cost and iteration-based measures such as the mean or total number of iterations. Together, they reflect the practicality of optimization approaches. Execution time is the dominant metric, as it reflects the overall duration of running an algorithm. Optimization time indicates the time overhead for the iterative optimization phase, which is the subprocess of executing an entire optimization algorithm to return a solution, such as the QPU access time for QA. Training and prediction times are only relevant for learning-based tasks, such as security attack identification \cite{barletta2024quantum}. Other types of times include pre-processing time required for solving the test suite minimization problem in \cite{wang2024test} and the QPU-access overhead time via API for addressing the join order problem \cite{nayak2023constructing}, etc. Economic cost is only concerned in \cite{serrano2024minimizing} where quantum computing was used to minimize incidence response time in security management. The mean number of iterations for successful results and the total number of iterations are also used in \cite{wu2020hybrid} and \cite{ren2024dynamic} as cost metrics, reflecting the computational effort required to reach a solution.

\begin{table}
	\small
	\centering
	\caption{Summary of approach cost metrics - RQ4}
	\label{tab:rq3_cost_metrics}
	\resizebox{.99\textwidth}{!}{
		\input{generated_files/RQ3_cost_metrics.tex}
	}
\end{table}

\begin{figure}
	\centering
	\includegraphics[width=0.6\textwidth]{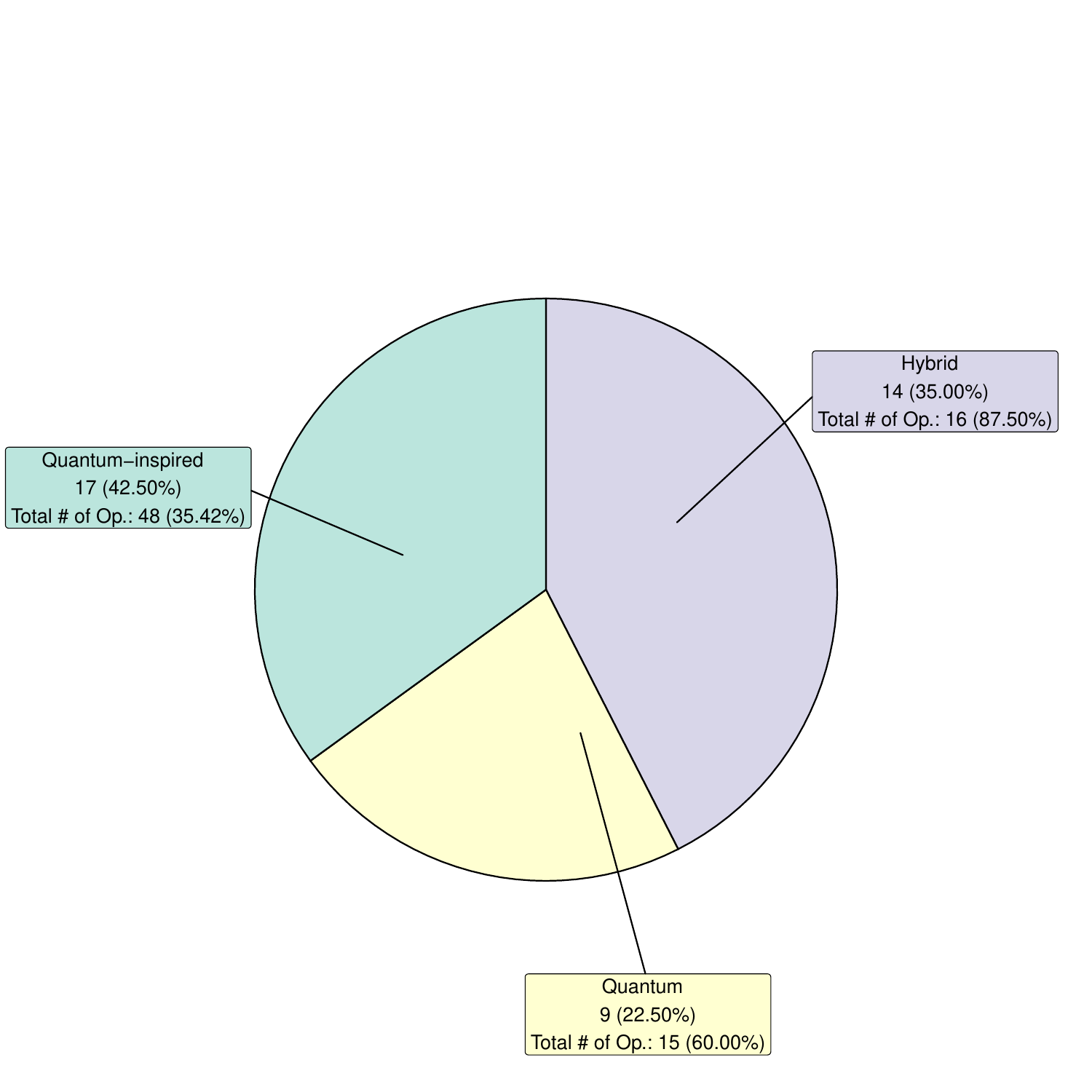}
	\caption{Overview of the optimization approaches that report cost in their evaluations - RQ4}
	\label{fig:RQ3_cost_metrics}
\end{figure}

Among the \finalselected primary studies, as shown in Figure~\ref{fig:RQ3_cost_metrics}, 38 reported resource cost as part of their evaluation: 17 for quantum-inspired, 9 for quantum, and 14 for hybrid optimization approaches.
In terms of types of optimization approaches, resource cost have been reported more frequently in quantum (9 out of 15, 60.00\%) and hybrid (14 out of 16, 87.5\%) optimization than in quantum-inspired ones (17 out of 48, 35.42\%). 

Efficiency represents cost-effectiveness. 
Among the reviewed studies, only one~\cite{\paperseventh} explicitly defines this metric, i.e., Speedup, in the context of addressing scheduling problems.

\begin{comment}
	\begin{table}
		\small
		\centering
		\caption{TODO\man{remove later. just put here for easy reference to problem and papers}}
		\label{tab:rq3}
		\resizebox{.99\textwidth}{!}{
			\input{generated_files/RQ3_efficiency_metrics.tex}
		}
	\end{table}
\end{comment}

\begin{comment}
	%Scalability
	As shown in Figure~\ref{fig:rq3_scalability}, only 52.5\% (i.e., 38) primary studies consider scalability; quantum-inspired approaches dominate in both groups (whether scalability is considered or not), while hybrid and pure quantum approaches are less frequently studied when scalability is considered. 
	
	Scalability metrics include the number of qubits, the number of gates, the number of layers of QAOA, etc. 
	\tao{Man: we miss a table for summarizing scalability metrics}
	\begin{figure}
		\centering
		\includegraphics[width=0.5\textwidth]{generated_files/RQ3_qoptype_scalability_pieDonut_crop.pdf}
		\caption{Descriptive statistics of scalability metrics - RQ3}
		\label{fig:rq3_scalability}
	\end{figure}
	
	\begin{results}[Findings of Approach Metrics]
		XXX
	\end{results}
	
	content...
\end{comment}

\begin{results}[Findings of Cost and Efficiency Metrics]
	Cost metrics are mostly concerned with time, with algorithm execution time dominating other types of time-related measures. 
	Resource cost has been evaluated more frequently in quantum (9 out of 15) and hybrid optimization approaches (14 out of 16) compared to quantum-inspired approaches (17 out of 48).
	Efficiency is primarily reflected in the speedup factor, which, however, is reported in only one primary study employing Grover’s algorithm.
\end{results}

%%-------------------------------------------------------------------------%%
\subsubsection{Approach complexity}
%\man{complexity might not be part of effectiveness}\tao{I moved it here.}
%Complexity metrics
%\man{we may put this at the end of this section as this kind of evaluation does not commonly exist in SE research}\tao{agree, you  move it.}

In contrast to the common practice of evaluating SE approaches through empirical studies, theoretical complexity analysis is often employed to assess the performance of the proposed approach in quantum optimization. 
We summarize the \textit{approach complexity} metrics in Table~\ref{tab:rq3_complexity_effectiveness_metrics}, which can be broadly grouped into three categories: 1) time complexity for algorithm execution, pre-processing, search/optimization, fitness calculation, parameter combination search, and value combination search such as the time complexity of Grover's search adopted in~\cite{\paperfourth} being bounded by $O( {N^2}{\sqrt{N}\text{polylog}(N)})$; 
%(e.g., pre-processing time, which measures preparation overhead; execution time, which measures the duration of running the algorithm on given inputs; optimization time which captures the iterative effort to improve solutions), 
2) space complexity (e.g., number of binary variables \cite{groppe2021optimizing}) like $O((n+m)\log_2 n)$ for that of implementing approaches on a quantum computer, where the number of machines $n$ and the number of transactions $m$ co-determine this complexity \cite{groppe2021optimizing}, 
%other complexity measures such as round complexity (i.e., the number of iterations in the optimization process \cite{hussein2021quantum})\yc{This complexity had already been merged into [Time complexity for search/optimization phases] in the shared table, since it is unnecessary to introduce such a complexity with an unfamiliar name.}  
and 3) code size complexity (i.e., the size of the formula or program to optimize~\cite{groppe2021optimizing}), where this kind of complexity is mentioned in~\cite{groppe2021optimizing} merely, and the code size for the quantum computer is $O(n^2c)$, associating with the number of conflicts $c$ apart from the above-mentioned $n$.

The majority of the primary studies that reported on approach complexity focused on the time complexity of algorithm execution (19 studies). This focus is understandable, as execution time is the most widely used indicator of cost and is easier to measure and compare with baselines than other complexity types such as space or preprocessing time. 
Only four studies focused on space complexity, three on time complexity for pre-processing, and three on time complexity for search/optimization, while all other types were addressed by only one study each. Especially, the primary study~\cite{guo2023effective} breaks down the complexity into three parts: fitness calculation, parameter and value combination search to analyze the performance of the multi-learning-based QPSO algorithm in the context of combinatorial testing.
Nevertheless, all these metrics are important, as they jointly reflect the scalability of optimization approaches and their practical feasibility, given that time and resource requirements directly influence the deployability of quantum, quantum-inspired, and hybrid optimization methods. Therefore, whenever possible, new algorithmic proposals should include a comprehensive complexity analysis covering not only execution time but also other relevant aspects such as space and pre-processing costs.

\begin{table}
	\small
	\centering
	\caption{Summary of the approach complexity metrics - RQ4}
	\label{tab:rq3_complexity_effectiveness_metrics}
	\resizebox{.99\textwidth}{!}{
		\input{generated_files/RQ3_complexity_effectiveness_metrics.tex}
	}
\end{table}
\begin{results}[Findings of Approach Complexity]
	Existing studies on approach complexity primarily focus on time complexity, especially that of algorithm execution. Future empirical studies would benefit from also considering space complexity and other time-related aspects, such as pre-processing time.
\end{results}

%%-------------------------------------------------------------------------%%
\subsection{\rqBenchmarks}
\label{sec:benchmarks}
\begin{figure}
	\centering
	\begin{subfigure}{0.55\textwidth}
		\centering
		\includegraphics[width=\textwidth]{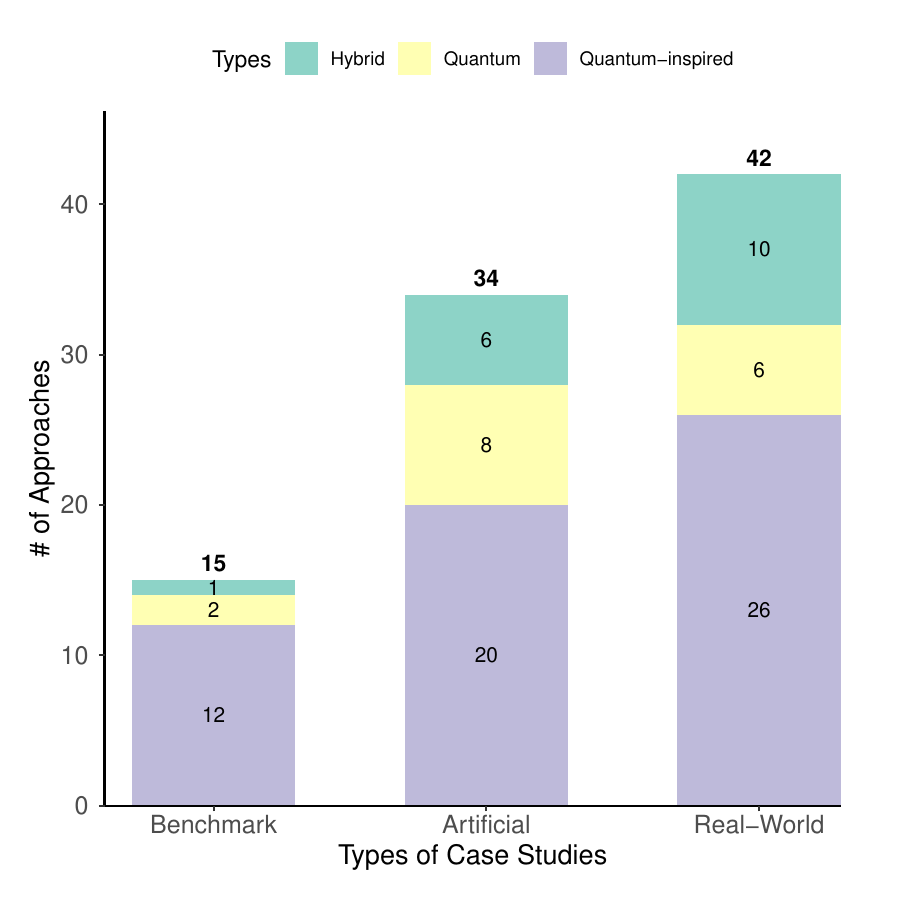}
		\caption{Types of case studies}
		\label{fig:RQ4_types_caseStudies}
	\end{subfigure}
	\hfill
	\begin{subfigure}{0.4\textwidth}
		\centering
			\includegraphics[width=\textwidth]{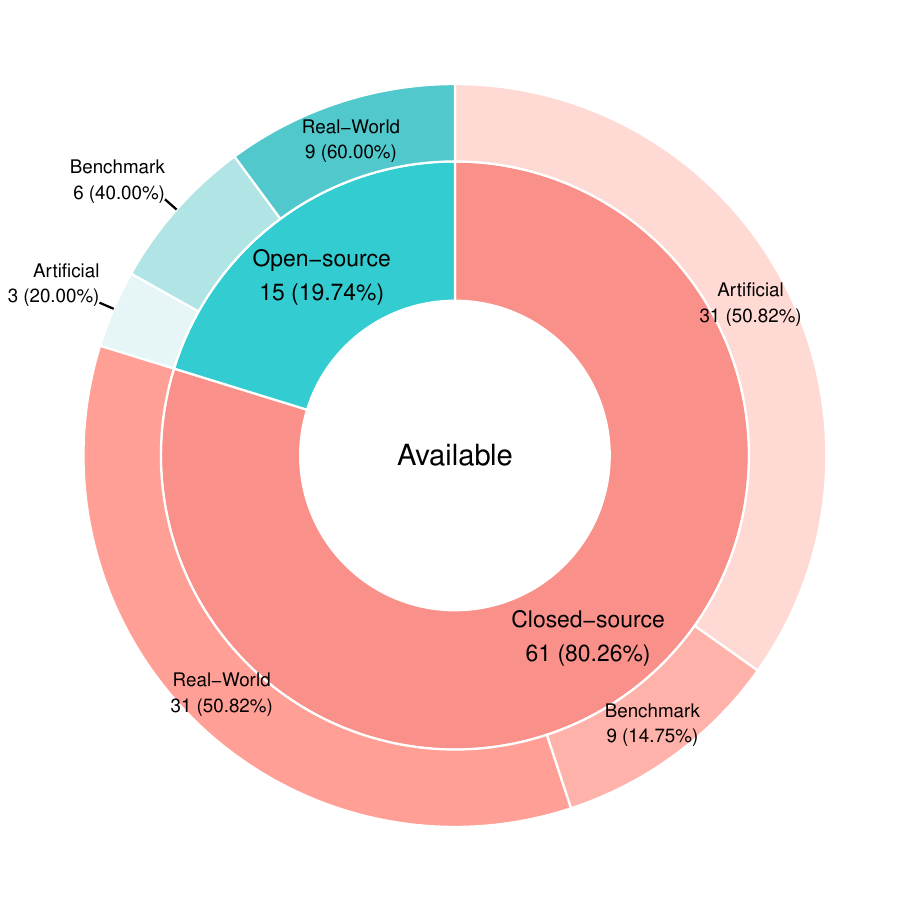}
		\caption{Descriptive statistics of primary studies with open and closed case studies. Note that a primary study may include multiple types of case studies, so the total percentage can exceed 100\%. 
			%\tao{The counts here are the number of case studies, not primary studies?\man{updated. tao, please double check}}
			}
		\label{fig:RQ4_open_closed_case studies}
	\end{subfigure}
	\hfill
	\begin{subfigure}{0.6\textwidth}
		\centering
		\includegraphics[width=\textwidth]{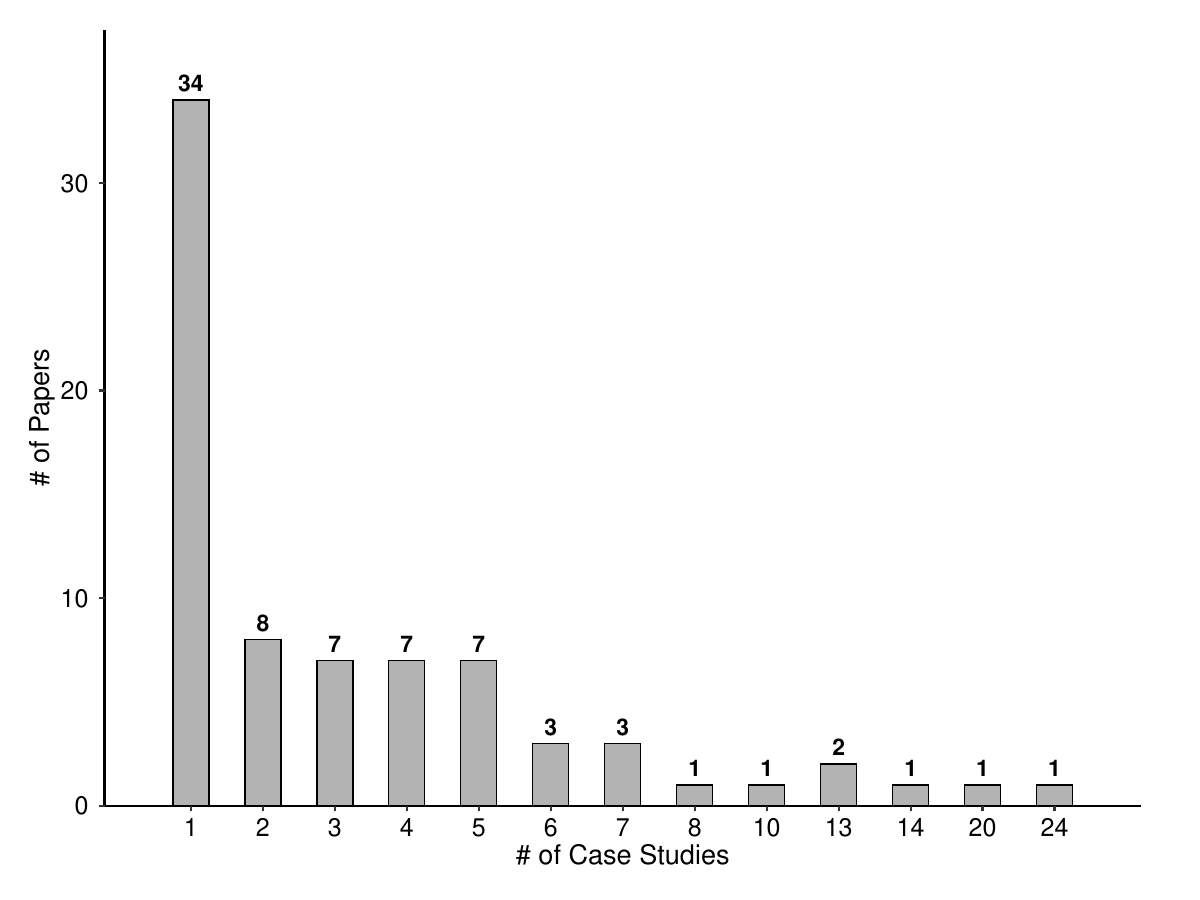}
	\caption{Number of case studies}
	\label{fig:RQ4_number_caseStudies}
	\end{subfigure}
	
	\caption{Overview of settings of case studies - RQ3}
	\label{fig:overall_casestudies_figure}
\end{figure}

When evaluating quantum, quantum-inspired, and hybrid optimization approaches, case studies used for experiments can usually be grouped into three main types: benchmarks, artificial, and real-world. Benchmarks are standardized sets for fair comparison; artificial problems are synthetically generated to test scalability and robustness, and real-world case studies are typically from practical domains to show applicability and impact. 
As shown in Figure~\ref{fig:RQ4_types_caseStudies}, real-world case studies were the most commonly adopted, followed by artificial case studies, and then benchmarks. This trend can be attributed to the nature of SE research, where problems are often grounded in practical contexts and therefore require real-world case studies. Moreover, many of these case studies have already been used to evaluate classical optimization approaches, making them natural candidates for assessing quantum-inspired and hybrid counterparts. We also notice that artificial case studies were adopted the most in quantum approaches, which is mainly because current quantum hardware put limits on solving real-world problems, so researchers opted for small and controlled artificial case studies.

%\begin{figure}
%	\centering
%	\includegraphics[width=0.5\textwidth]{generated_files/RQ4_qoptype_benchmark_groupedbar.pdf}
%	\caption{Types of Case Studies}
%	\label{fig:RQ4_types_caseStudies}
%\end{figure}

In Figure~\ref{fig:RQ4_number_caseStudies}, we show the counts of case studies, which clearly indicate that most empirical studies use a single case study (34), while 8, 7, 7, and 7 primary studies use 2, 3, 4, and 5 case studies, respectively. This shows that the effort and complexity involved in conducting empirical studies in this field: using a single case study is simpler and more feasible, while including multiple case studies requires additional resources. 

%\begin{figure}
%	\centering
%	\includegraphics[width=0.8\textwidth]{generated_files/RQ4_num_of_casestudies_summary_bar.pdf}
%	\caption{Number of Case Studies\man{need to discuss with yuechen about the paper with 0 case study}\yc{ID-61 does not include the empirical study, even a complete toy example. This paper only focuses on the theoretical approach for QUBO encoding.}\tao{So, we remove this primary study? }}
%	\label{fig:RQ4_number_caseStudies}
%\end{figure}

As presented in Figure~\ref{fig:RQ4_open_closed_case studies},
%\yc{What do the percentages in the pie figure refer to? For example, why the following holds: 31(50.82\%) for Real-world, 9(14.74\%) for Benchmark, Artificial 31(50.82\%), but 50.82+14.74+50.82=116.38>100.00?}, 
among 
%90 reported case studies
\finalselected primary studies, only 
%22.22\% is
15 studies (19.74\%) conducted experiments with open-source case studies, among which 
%11, 6 and 3 
9, 6, and 3 are real-world, benchmark and artificial case studies, respectively. 
Out of the 61 primary studies with closed case studies, 31 used real-world, 31 used artificial, and 9 used benchmark case studies.
Over half (50.82\%) is artificial which is somewhat surprising given that artificial problems are generally easier to share and reproduce. This indicates that, even for synthetic problems, researchers frequently keep the data or experimental setups proprietary. We further present all open source case studies in Table~\ref{tab:rq4_open_benchmark} for future references. 

%\begin{figure}
%	\centering
%	\includegraphics[width=0.5\textwidth]{generated_files/RQ4_open_close_type_pieDonut_crop.pdf}
%	\caption{Descriptive statistics of open vs. closed case studies \tao{The counts here are the number of case studies, not primary studies?}}
%	\label{fig:RQ4_open_closed_case studies}
%\end{figure}

\begin{table}
	\small
	\centering
	\caption{Summary of open-source case studies}
		 %\tao{Yuechen: I see VLDB'97 and SIGMOD'22. Are they actually references? The table needs to informative but also need to be self-contained. We perhaps can keep: the counts of each type of case studies, the short names? I assume we have 20 open source case studies as reported in Figure 12.} \yc{The mentioned two are references, as I copied the texts originally from the primary papers. I agree that descriptions can be retained in our benchmark for readers of interest, whereas it seems too detailed to display all of them in the main body.} \tao{Yuechen and Man: we update the table? }
	\label{tab:rq4_open_benchmark}
	\resizebox{.99\textwidth}{!}{
		\input{generated_files/RQ4_open_benchmark.tex}
	}
\end{table}

\begin{results}[Findings of Employed Case Studies]
 Real-world case studies are the most commonly selected in evaluating quantum-inspired and hybrid approaches, while artificial case studies are selected the most for evaluating pure quantum approaches, with benchmarks being the least used across all three; Most primary studies choose to use a single case study; Among the 
 %90 case studies, only 20 are open-source, of which 11 are real-world cases.
 \finalselected primary studies,
only 15 employed open-source case studies, of which 9 are real-world cases.
\end{results}

\subsection{\rqBaselines}
\label{sec:baselines}
\begin{figure}
	\centering
	\includegraphics[width=0.6\textwidth]{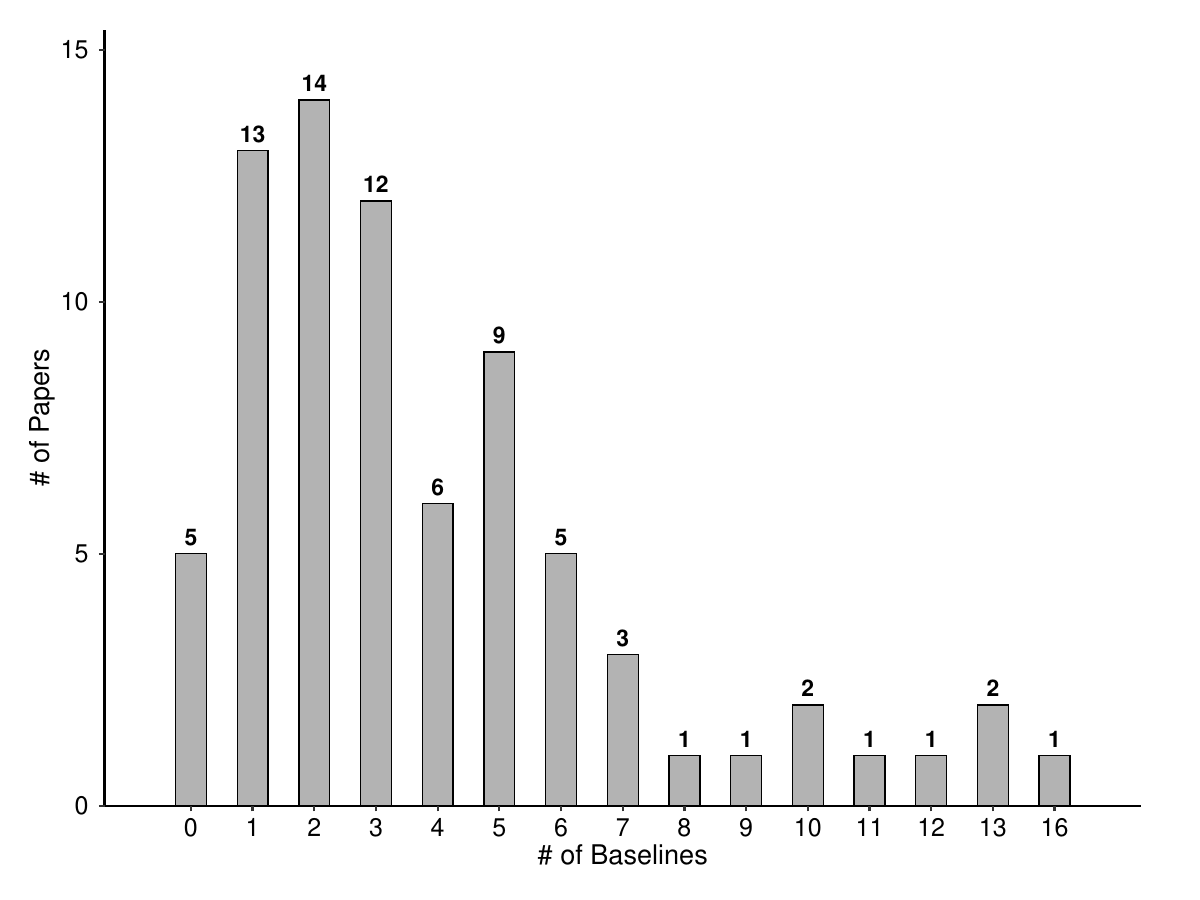}
	\caption{Descriptive statistics of baselines employed in primary studies}
	\label{fig:RQ5_number_baselines}
\end{figure}

Baselines are critical in evaluating optimization solutions, as they provide reference points against which the effectiveness, efficiency, and potential advantages of quantum, quantum-inspired, and hybrid approaches can be assessed.
From Figure~\ref{fig:RQ5_number_baselines}, we observe that most primary studies employ 1–3 baselines, while five studies do not use any baseline for comparison. Only a few studies consider more than seven baselines. In classical SE optimization, using baselines is standard practice, and the same expectation should apply for evaluating quantum, quantum-inspired, and hybrid solutions. There is no justification for omitting baselines, since classical optimization methods are always available as points of reference. 

\begin{figure}
	\centering
	\includegraphics[width=0.5\textwidth]{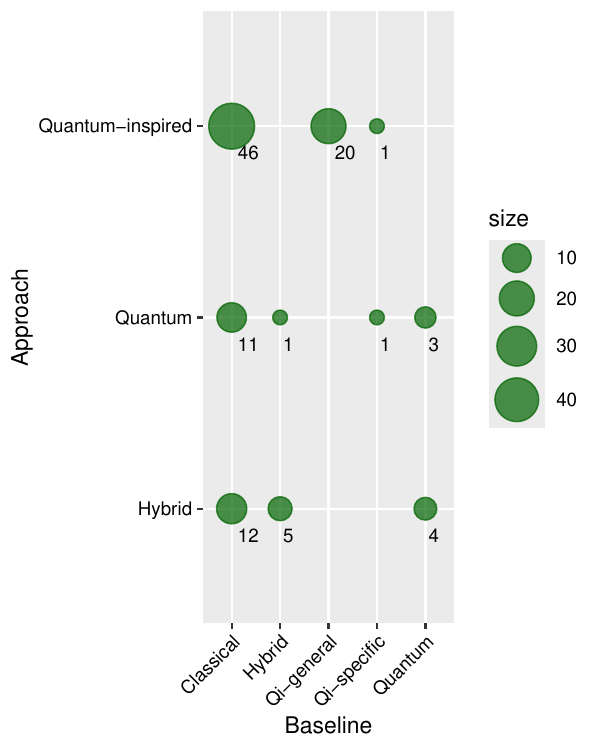}
	\caption{Alignment of optimization approaches with baseline types}
	\label{fig:RQ5_approach_vs_baselines}
\end{figure}

As shown in Figure~\ref{fig:RQ5_approach_vs_baselines}, we classify all baselines into five categories: classical baselines, hybrid, quantum-inspired general baselines (shorten as \textit{qi-general} in the figure) that can run on a general CPU (e.g., using QPSO as a baseline in~\cite{gan2024research}), quantum-inspired specific baselines (shorten as \textit{qi-specific} in the figure) (e.g., using DA in~\cite{schonberger2023digitalannealing,saxena2024constrained}, which must be executed on specific classical hardware), and quantum baselines. 
From the figure, we observe that 69 approaches (46+11+12) were evaluated against classical baselines, 
and 20 against quantum-inspired general baselines,
%29 (21+4+4) 
%and 7 (3+4) against quantum baselines, 
with only a few studies opting for quantum, hybrid or quantum-inspired specific baselines. This is reasonable, as quantum-inspired, quantum and hybrid solutions must demonstrate their superiority by comparing with classical optimization approaches. 
In particular, quantum-inspired approaches were most often evaluated against classical baselines, and quantum-inspired general baselines.
%, and quantum baselines. \tao{Yuechen: can you check why quantum-inspired approaches were evaluated by comparing with 21 quantum baselines? It does not sound rational to me. }\yc{I did not find any [Q] baseline corresponding to [classical] approaches in shared table. Is this confused by [Qi-general] and [Qi-specific]?\tao{In the figure, the Quantum-inspired row has 21 Quantum baseline.}}  
Quantum approaches largely opt for classical baselines with few compared with quantum baselines. For instance Simulated Annealing (SA) is a widely used classical baseline, as four primary studies~\cite{mandal2024evaluating,bettonte2022quantum,nayak2023constructing,niu2024performance} adopt SA to compare with QA. In comparison, the applicable and available quantum algorithms are relatively scarce, and the paper~\cite{niu2024performance}, for example, adopts four quantum baselines based on different QUBO encoding approaches for QA. Similarly, most hybrid approaches are evaluated against classical baselines, with only a few using hybrid or quantum baselines. Taking an example of~\cite{emu2024warm}, there are three classical baselines (i.e., Deep Q-Networks, Integer Linear Programming, and SA) and only one quantum baseline (i.e., QA) in comparison with the proposed hybrid approach that adopts warm start or cold start for QA. However, we notice a case where a quantum approach was evaluated with one quantum-inspired specific baseline: in ~\cite{saxena2024constrained}, DA with the original QUBO encoding was established as a baseline, against the QA with the proposed constrained quadratic model. 

%From Prof. Cai: Figure 9 might serve as a core of the paper and could be upgraded to show a three-dimensional view: 
%	Subject approaches under evaluation (assessment)
%	Subject baselines (programs) adopted
%	Evaluation (assessment) criteria (metrics), related to RQs

\begin{results}[Findings of Employed Baselines]
	 Most primary studies employ 1–3 baselines. Five types of baselines are used (i.e., classical, hybrid, quantum-inspired general, quantum-inspired specific, and quantum) with classical baselines being the most commonly adopted. Classical optimization approaches have been the most widely adopted baselines by all optimization approaches. 
\end{results}

\subsection{\rqTooling}
\label{sec:tooling}
Tooling plays a crucial role in empirical studies conducted to address software engineering optimization problems, as it enables reproducibility and fair comparison of approaches. Hence, in Table~\ref{tab:rq4_open_tool}, we summarize all tools made available by the primary studies. 
Interestingly, all of these tools are implementations of quantum or hybrid approaches.
%, with the exception of the one proposed in \cite{schonberger2023quantum} for solving the join order problem. 
This is likely because quantum and hybrid approaches often require custom implementations.
%~\yc{(1) I have excluded this paper [81]. (2) What is meant by ``the exception''? Should we discuss why only one study on quantum-inspired algorithms made its artifact available?} \tao{Yuechen: if this is the case, then there is no need to mention this exception. }
In particular, the encoding of SE problems into QUBO or Ising requires special handling to respect hardware constraints and hybrid integration, which motivates researchers to develop and release new tools. 
%
%\tao{Yuechen: this is a bit weird to me. Is it also because there exist platforms that already implement quantum-inspired algorithms like QPSO?}\yc{I do not really think so. As we can see, many of the studies on quantum-inspired optimization algorithms did not make their artifacts available. At least, this holds true for our collected primary papers.}

%
\begin{table}
	\small
	\centering
	\caption{Summary of the tools made publicly available in the primary studies}
	\label{tab:rq4_open_tool}
	\resizebox{.99\textwidth}{!}{
		\input{generated_files/RQ5_open_tool.tex}
	}
\end{table}

\begin{results}[Findings of Available Tools]
	Among the final selected primary studies, only 12 (15.79\%) made their tools publicly available. 
	However, in most SE publication venues, it is typically required to share tools, raw experimental data, etc. in order to validate, reproduce, reuse, or build upon existing research. This is especially critical in quantum optimization for SE, as various backends (e.g., specialized hardware, simulators, or hybrid classical–quantum frameworks) are involved. Transparency enables the SE community to verify findings, evaluate approaches, and assess implementations across different backends, facilitating the evaluation and comparison of various quantum optimization approaches on these platforms. 
\end{results}

%% file: generated_files/RQ2_hyperparameters.tex
\begin{tabular} {p{0.2\textwidth} p{0.4\textwidth} p{0.1\textwidth} p{0.38\textwidth}} \\ 
\toprule 
 Quantum Opt. Type & Parameters (\# of Papers) & Category & Papers\\ 
\midrule  
Quantum-inspired  & \emph{Population size} (31) & Classical & \cite{ \papertenth,\papersixteenth,\papertwentieth,\papertwentysecond,\papertwentythird,\papertwentysixth,\papertwentyeighth,\papertwentyninth,\paperthirtieth,\paperthirtyfirst,\paperthirtysecond,\paperthirtythird,\paperthirtyfourth,\paperthirtysixth,\paperthirtyeighth,\paperthirtyninth,\paperfortieth,\paperfortyfirst,\paperfortyfourth,\papersixtyseventh,\paperseventyfirst,\paperseventysecond,\paperseventythird,\paperseventyfourth,\paperseventyfifth,\paperseventysixth,\paperseventyseventh,\paperseventyninth,\papereightieth,\papereightysecond,\papereightyfifth }\\ 
 & \emph{Initial qubit states} (12) & Quantum & \cite{ \papersixteenth,\papertwentieth,\papertwentysecond,\papertwentyeighth,\paperthirtyninth,\paperfortieth,\paperfortysecond,\papersixtysixth,\paperseventyseventh,\papereightieth,\papereightyfifth,\paperninetysecond }\\ 
 & \emph{Parameter for updating quantum rotation angles} (12) & Quantum & \cite{ \papersixteenth,\papertwentieth,\papertwentyeighth,\papertwentyninth,\papersixtysixth,\papersixtyseventh,\paperseventysecond,\paperseventythird,\paperseventyseventh,\papereightieth,\papereightyfifth,\paperninetysecond }\\ 
 & \emph{Mutation setting} (10) & Classical & \cite{ \papertwentysecond,\papertwentyeighth,\papersixtysixth,\papersixtyseventh,\paperseventieth,\paperseventysecond,\paperseventyseventh,\paperseventyninth,\papereightieth,\papereightyfifth }\\ 
 & \emph{Contraction-expansion coefficient} (9) & Quantum & \cite{ \paperthirtysecond,\paperthirtythird,\paperthirtyfourth,\paperthirtyeighth,\paperfortyfourth,\paperseventyfirst,\paperseventyfifth,\paperseventysixth,\papereightythird }\\ 
 & \emph{Objective weights} (6) & Both & \cite{ \paperfortieth,\paperfortyfirst,\paperseventieth,\paperseventyninth,\papereightysecond,\paperninetysecond }\\ 
 & \emph{Crossover setting} (5) & Classical & \cite{ \papertwentysecond,\papertwentyeighth,\papersixtyseventh,\paperseventysecond,\paperseventyseventh }\\ 
 & \emph{Dataset split ratio} (2) & Both & \cite{ \paperseventyfirst,\paperseventythird }\\ 
\midrule  
Quantum, Hybrid  & \emph{Penalty coefficient} (13) & Quantum & \cite{ \paperfirst,\papersecond,\papertwentyfifth,\papertwentyseventh,\paperfiftysecond,\paperfiftythird,\paperfiftyfourth,\paperfiftyeighth,\papersixtieth,\papersixtythird,\papersixtyeighth,\papereightyfourth,\paperninetieth }\\ 
 & \emph{QAOA layers} (7) & Quantum & \cite{ \papersecond,\paperthirteenth,\papertwentyseventh,\paperfiftyfourth,\paperfiftyeighth,\papereightyfourth,\paperninetythird }\\ 
 & \emph{Selection of classical optimizers for variational algorithms} (7) & Quantum & \cite{ \papersecond,\paperthirteenth,\papertwentyseventh,\paperfiftyfourth,\paperfiftyeighth,\papersixtyninth,\papereightyfourth }\\ 
 & \emph{Objective weights} (5) & Both & \cite{ \papereleventh,\paperthirteenth,\papersixtythird,\papersixtyfourth,\papersixtyeighth }\\ 
 & \emph{Dataset split ratio} (4) & Both & \cite{ \papersixth,\papereleventh,\paperfiftysixth,\papersixtyninth }\\ 
\bottomrule 
\end{tabular} 

%% file: psedo_code/repeats_vs_shots.tex
\KwIn{
    Repetitions of experiments $r$; 
    Shots per run $s$; 
    Target quantum system $Q$
} 
\KwOut{Experimental metrics $M$} 

$t \leftarrow 0$\;
$M \leftarrow \varnothing$\;
\While{$t < r$}{
    $t \leftarrow t + 1$\;
    \tcc{Submit the circuit to the backend once; it is executed $s$ times to collect measurement statistics.} 
    $o \leftarrow \textbf{running\_backend}(Q, s)$\;  
    \tcc{For each group of $s$ measurement outcomes, we can calculate a corresponding metric value.} 
    $m \leftarrow \textbf{output\_analysis}(o)$\;
    $M \leftarrow M \cup \{m\}$\;
}
\Return $M$\;

%% file: generated_files/RQ2_shots.tex
\begin{tabular} {l l l } \\ 
\toprule 
Approach Type & Shots & Backends/Algorithms: Repetition \\ 
\midrule  
Quantum-inspired & 100 & DA hardware: \emph{Repetition}(10) \cite{ \paperfifth }\\ 
Quantum, Hybrid & 100 & QA hardware: \emph{Repetition}(10) \cite{ \papersixtyfourth }, \cite{ \papereleventh }\\ 
 & 256 & QAOA simulator: \cite{ \papersecond }\\ 
 & 500 & QAOA simulator: \cite{ \papereightyfourth }\\ 
 & 650 & QA hardware: \emph{Repetition}(10) \cite{ \paperfirst }\\ 
 & 900 & QA hardware: \cite{ \paperninetythird }\\ 
 & 1000 & QA hardware: \cite{ \papertwentyseventh,\paperfiftythird,\paperfiftyfourth }\\ 
 & 1024 & QAOA simulator: \cite{ \papertwentyseventh }; Grover's algorithm simulator: \cite{ \papersixtyfifth }; QAOA hardware: \cite{ \paperninetythird }\\ 
 & 3000 & QA hardware: \cite{ \papersixth,\papersixtyninth }\\ 
 & 4000 & Grover's algorithm hardware: \cite{ \papersixtyfifth }\\ 
 & 5000 & QA hardware: \emph{Repetition}(10) \cite{ \papersixtieth }\\ 
 & 10000 & QA simulator: \cite{ \paperfiftyfourth }; QA hardware: \cite{ \paperninetythird }\\ 
 & 120000 & QA hardware: \emph{Repetition}(5; 150) \cite{ \paperninetieth }\\ 
\bottomrule 
\end{tabular} 

%% file: generated_files/RQ2_qubits.tex
\begin{tabular} {l p{0.4\textwidth} l }\\ 
\toprule 
Backend & Qubit Setting & Papers\\ 
\midrule  
Ideal classical simulator & 6 to 32 logical qubits & \cite{ \papersecond }\\ 
 & 3 to 8 logical qubits & \cite{ \paperfourth }\\ 
 & 7 to 16 logical qubits & \cite{ \paperthirteenth }\\ 
 & 4 logical qubits & \cite{ \papersixtyfifth }\\ 
Noisy classical simulator & 7 logical qubits & \cite{ \paperthirteenth }\\ 
 & 4 logical qubits & \cite{ \papersixtyfifth }\\ 
Quantum hardware & 7 logical qubits & \cite{ \paperthirteenth }\\ 
 & 12 and 15 qubits\textsuperscript{*} & \cite{ \papertwentyseventh }\\ 
 & 4 to 247 logical qubits & \cite{ \paperfiftyfourth }\\ 
 & 4 to 14 qubits\textsuperscript{*} & \cite{ \paperfiftyeighth }\\ 
 & 10 to 160 logical qubits & \cite{ \papersixtyfourth }\\ 
 & 4 logical qubits & \cite{ \papersixtyfifth }\\ 
Classical simulator\textsuperscript{\textdagger} & 4 to 247 logical qubits & \cite{ \paperfiftyfourth }\\ 
\bottomrule 
\end{tabular} 

%% file: generated_files/RQ2_terminiation.tex
\begin{tabular} {l p{0.4\textwidth} p{0.5\textwidth} } \\ 
\toprule 
Category & Termination Criteria & Value \\ 
\midrule  
Convergence & Indicator for convergence & 10 consecutive cost difference less than a threshold~\cite{ \paperseventyfirst }; No fitness improvement for 3 consecutive iterations~\cite{ \paperthirteenth }; No improvement of solution over a given number of iterations~\cite{ \paperseventysecond }; Stop with a tolerance of $10^{-12}$ for the globally optimum values~\cite{ \paperseventyfourth }; The absolute difference between two adjacent steps less than 0.001 for 10 times~\cite{ \paperthirtieth }; The absolute difference between two adjacent steps less than an un-specified threshold for 10 times~\cite{ \paperthirtysecond }\\ 
Iteration & Number of maximum generations & $5\times 10^5$ for parameter tuning and $3 \times 10^6$ for approach evaluation~\cite{ \papersixtysixth }; 10~\cite{ \paperseventyninth }; 100 or 500~\cite{ \papereightieth }; 400 or 500~\cite{ \paperninetysecond }; 500~\cite{ \paperseventyseventh,\papereightysecond }; 60~\cite{ \papereightyfifth }\\ 
 & Number of maximum iterations & $10^8$~\cite{ \paperfifth }; 100~\cite{ \paperthird,\papertwentysecond,\papertwentythird,\papertwentyninth,\paperthirtysixth,\paperseventyfirst }; 100 or 200~\cite{ \paperthirtyninth }; 1000~\cite{ \papersecond,\papertenth,\papertwentysixth,\paperthirtythird,\paperthirtyeighth,\paperfortyfirst,\paperfortysecond,\paperfortyfourth }; 1000 (QAOA classical optimizer only)~\cite{ \paperfiftyeighth }; 200~\cite{ \paperthirtysecond,\papersixtyseventh }; 30~\cite{ \paperthirteenth,\paperthirtyfourth }; 30 for VQC~\cite{ \papersixtyninth }; 300~\cite{ \paperthirtieth,\paperseventythird }; 40~\cite{ \paperseventysixth }; 50, 60 or 200~\cite{ \paperthirtyfirst }; 50000~\cite{ \paperseventyfourth }; 75~\cite{ \papertwentieth }; 90~\cite{ \paperseventyfifth }; Adaptive~\cite{ \papersixteenth,\papersixtyfifth }\\ 
 & Number of objective function evaluations & $10^4$~\cite{ \papertwentyeighth }\\ 
Objective & Optimized objective/solution achieved & A covering array found~\cite{ \papersixtysixth }; a solution being found~\cite{ \paperseventyseventh }; All the t-way schemes covered~\cite{ \paperseventysixth }; All the t-way schemes covered and the complete test suite meeting the constraints~\cite{ \paperseventyfifth }; The objective function value greater than a threshold~\cite{ \paperninetyfirst }\\ 
Time & Time budget & 24h~\cite{ \papertwentieth }; 300s~\cite{ \paperthirtyfifth }; 30min~\cite{ \paperseventysecond }; 60s~\cite{ \paperfifth,\paperfiftieth }\\ 
\bottomrule 
\end{tabular} 

%% file: generated_files/RQ3_problem_metrics.tex
\begin{tabular} {p{0.3\textwidth} p{1.1\textwidth}} \\ 
\toprule 
Problem & Metrics\\ 
\midrule  
\multicolumn{2}{l}{\cellcolor{black!10!white}\textbf{Software Maintainance}} \\ 
Code Clone Detection & \textbf{Dependency/Constraint}: Abstract Syntax Tree~\cite{ \paperfirst }\\ 
 & \textbf{Size}: Nodes~\cite{ \paperfirst }\\ 
\multicolumn{2}{l}{\cellcolor{black!10!white}\textbf{Software Configuration Management}} \\ 
Configuration Selection & \textbf{Dependency/Constraint}: Clauses~\cite{ \papersecond }, Literals~\cite{ \papersecond }\\ 
 & \textbf{Size}: Features~\cite{ \papersecond }\\ 
Configuration Prioritization & \textbf{Dependency/Constraint}: Clauses~\cite{ \papersecond }, Literals~\cite{ \papersecond }\\ 
 & \textbf{Size}: Features~\cite{ \papersecond }\\ 
\multicolumn{2}{l}{\cellcolor{black!10!white}\textbf{Software Engineering Operations}} \\ 
Scheduling Problem & \textbf{Dependency/Constraint}: Conflicts~\cite{ \paperseventh }, Environment Constraints~\cite{ \papertwentyfourth }\\ 
 & \textbf{Size}: Virtual Machines~\cite{ \paperthird,\papertwentyfourth,\paperfortyfirst,\paperseventysecond }, Cpu Cores~\cite{ \paperseventh }, Transactions~\cite{ \paperseventh }, Length Of Transactions~\cite{ \paperseventh }, Jobs~\cite{ \papertwentyfourth }, Tasks~\cite{ \paperfortyfirst,\papereightysecond,\paperninetyfirst }, Nodes~\cite{ \paperseventysecond,\papereightysecond,\paperninetyfirst }, Server Types~\cite{ \paperseventysecond }\\ 
Join Order Problem & \textbf{Dependency/Constraint}: Joins~\cite{ \paperfifth,\paperfiftysecond,\paperninetythird }, Graph Type~\cite{ \paperfifth,\paperfiftysecond,\paperninetythird }, Predicates~\cite{ \paperfifth }, Join Tree Structure~\cite{ \paperfiftieth }, Join Predicates~\cite{ \paperfiftieth,\paperninetythird }\\ 
 & \textbf{Size}: Relations~\cite{ \paperfifth,\paperfiftieth,\paperfiftysecond,\paperfiftyfourth,\paperninetythird }, Joins~\cite{ \paperfiftieth }\\ 
Security Attack Identification & \textbf{Size}: Instances~\cite{ \papertwentysecond }, Classes~\cite{ \papertwentysecond }, Attributes~\cite{ \papertwentysecond }, Features~\cite{ \paperfiftyfifth }, Records~\cite{ \papersixtyninth }\\ 
Node Localization & \textbf{Dependency/Constraint}: Communication Radius~\cite{ \papertwentythird }\\ 
 & \textbf{Size}: Network Size~\cite{ \papertwentythird }, Nodes~\cite{ \papertwentythird }\\ 
Security Management & \textbf{Dependency/Constraint}: Interdependencies Between Incidents And Controls~\cite{ \papertwentyfifth }\\ 
 & \textbf{Size}: Incidents~\cite{ \papertwentyfifth }, Instances~\cite{ \papersixtyeighth }\\ 
Controller Placement & \textbf{Dependency/Constraint}: Network Links~\cite{ \paperthirtyfirst }\\ 
 & \textbf{Size}: Controllers~\cite{ \paperthirtyfirst,\paperfortyfourth,\papereightythird }, Nodes~\cite{ \paperthirtyfirst,\paperfortyfourth,\papereightythird }, Possible Placements~\cite{ \paperfortyfourth }, Dimensions Of Benchmark Functions~\cite{ \paperfortyfourth }\\ 
Communication Optimization & \textbf{Dependency/Constraint}: Network Topology~\cite{ \paperthirtyfifth }\\ 
 & \textbf{Size}: Network Size~\cite{ \paperthirtyfifth }, Nodes~\cite{ \paperthirtyfifth }, Package Size~\cite{ \paperthirtyfifth }\\ 
Service Deployment & \textbf{Dependency/Constraint}: Resource Constraint~\cite{ \paperthirtysixth }, Network Topology~\cite{ \paperthirtyninth }\\ 
 & \textbf{Size}: Nodes~\cite{ \paperthirtysixth,\paperthirtyninth }, Service Platforms~\cite{ \paperthirtysixth }, Service Types~\cite{ \paperthirtysixth }\\ 
Service Configuration & \textbf{Size}: Nodes~\cite{ \paperfortieth }, Service Requests~\cite{ \paperfortieth }\\ 
Container Management & \textbf{Size}: Servers~\cite{ \paperfortythird }, Requests~\cite{ \paperfortythird }, Services~\cite{ \paperfortythird }\\ 
Resource Optimization & \textbf{Size}: Sensors~\cite{ \paperfortyfifth }, Sensor Parameters~\cite{ \paperfortyfifth }, Objects~\cite{ \paperfortyfifth }\\ 
Multiple Query Optimization & \textbf{Size}: Plans~\cite{ \paperfiftythird,\paperfiftyeighth }, Queries~\cite{ \paperfiftythird,\paperfiftyeighth }\\ 
Database Index Selection & \textbf{Size}: Index Candidates~\cite{ \papersixtieth }\\ 
Software Failure Prediction & \textbf{Size}: Failure Data~\cite{ \papersixtyseventh }\\ 
Allocation Problem & \textbf{Size}: Tasks~\cite{ \paperseventieth,\papereightysixth }, Agents~\cite{ \paperseventieth }, Instances~\cite{ \papereightysixth }, Nodes~\cite{ \papereightysixth }, Service Function Chain Length~\cite{ \papereightyseventh }, Iot Hosts~\cite{ \papereightyseventh }, Metaslice Length~\cite{ \papereightyeighth }, Hosts~\cite{ \papereightyeighth }\\ 
Queuing Delay Optimization & \textbf{Size}: Sources~\cite{ \paperseventyninth }\\ 
Network Delay Optimization & \textbf{Size}: Nodes~\cite{ \papereightieth }, Links~\cite{ \papereightieth }\\ 
Routing Optimization & \textbf{Size}: Nodes~\cite{ \papereightyfourth,\paperninetysecond }, Links~\cite{ \papereightyfourth }, Sensors~\cite{ \papereightyninth }, Objects~\cite{ \papereightyninth }\\ 
Clustering Protocol Optimization & \textbf{Size}: Nodes~\cite{ \papereightyfifth }\\ 
\multicolumn{2}{l}{\cellcolor{black!10!white}\textbf{Software Quality}} \\ 
DTC Problem & \textbf{Dependency/Constraint}: Constraints~\cite{ \paperfourth }\\ 
 & \textbf{Size}: Nodes~\cite{ \paperfourth }\\ 
Software Failure Prediction & \textbf{Dependency/Constraint}: Attributes~\cite{ \paperthirtieth }, Percentage Of Defect-Prone Instances~\cite{ \paperthirtieth }\\ 
 & \textbf{Size}: Features~\cite{ \papereleventh,\papertwelfth }, Dimensions Of Benchmark Functions~\cite{ \papertwentysixth }, The Datasets~\cite{ \papertwentysixth }, Instances~\cite{ \paperthirtieth }, Failure Data~\cite{ \papersixtyseventh }, Modules~\cite{ \paperseventyfirst }, Dimensions Of The Input Nodes~\cite{ \paperseventythird }\\ 
WCET Evaluation & \textbf{Dependency/Constraint}: If Conditions~\cite{ \papertwentyseventh }, Switch Choices~\cite{ \papertwentyseventh }\\ 
 & \textbf{Size}: Basic Blocks~\cite{ \papertwentyseventh }\\ 
Software Reliability Growth Model & \textbf{Size}: Problem Parameters~\cite{ \paperthirtysecond }\\ 
\multicolumn{2}{l}{\cellcolor{black!10!white}\textbf{Software Security}} \\ 
Security Attack Identification & \textbf{Size}: Instances~\cite{ \papertwentysecond }, Classes~\cite{ \papertwentysecond }, Attributes~\cite{ \papertwentysecond }, Dataset Size~\cite{ \paperfortysecond }, Features~\cite{ \paperfiftyfifth,\paperfiftysixth }, Records~\cite{ \papersixtyninth }\\ 
Security Management & \textbf{Dependency/Constraint}: Interdependencies Between Incidents And Controls~\cite{ \papertwentyfifth }\\ 
 & \textbf{Size}: Incidents~\cite{ \papertwentyfifth }, Instances~\cite{ \papersixtyeighth }\\ 
\multicolumn{2}{l}{\cellcolor{black!10!white}\textbf{Software Testing}} \\ 
Test Suite Minimization & \textbf{Dependency/Constraint}: Test Requirements~\cite{ \papersixteenth,\paperthirtyseventh,\paperthirtyeighth,\paperseventyseventh }\\ 
 & \textbf{Size}: Test Cases~\cite{ \paperthirteenth,\papersixteenth,\paperthirtythird,\paperthirtyseventh,\paperthirtyeighth,\papersixtyfourth,\paperseventyseventh }, Components In A Test Case~\cite{ \paperseventyfourth }\\ 
Test Case Prioritization & \textbf{Dependency/Constraint}: Test Requirements~\cite{ \paperthirtyeighth }\\ 
 & \textbf{Size}: Test Cases~\cite{ \paperthirtythird,\paperthirtyeighth }\\ 
Test Case Selection & \textbf{Dependency/Constraint}: Test Requirements~\cite{ \paperthirtyeighth }\\ 
 & \textbf{Size}: Test Cases~\cite{ \paperthirteenth,\paperthirtyeighth,\papersixtythird }, Statements To Be Covered~\cite{ \papersixtythird }\\ 
Covering Array Generation & \textbf{Dependency/Constraint}: Forbidden Tuples~\cite{ \paperthirtyfourth }, Constraints~\cite{ \paperseventyfifth }\\ 
 & \textbf{Size}: Parameters For Software Under Test~\cite{ \papertwentieth,\paperthirtyfourth,\paperseventyfifth,\paperseventysixth }, Discrete Values Per Parameter~\cite{ \papertwentieth,\paperthirtyfourth,\paperseventyfifth,\paperseventysixth }, Covering Strength~\cite{ \papertwentieth,\paperthirtyfourth,\papersixtysixth,\paperseventyfifth,\paperseventysixth }, Size Of The Binary Covering Array~\cite{ \papersixtysixth }\\ 
Dynamic Testing & \textbf{Size}: Database Size~\cite{ \papersixtyfifth }\\ 
\multicolumn{2}{l}{\cellcolor{black!10!white}\textbf{Software Requirements}} \\ 
Software Requirement Selection & \textbf{Dependency/Constraint}: Density Of The Customer-Requirements Matrix~\cite{ \papertwentyeighth }, Interactions Constraints~\cite{ \papertwentyninth }\\ 
 & \textbf{Size}: Customers~\cite{ \papertwentyeighth,\papertwentyninth }, Requirements~\cite{ \papertwentyeighth,\papertwentyninth }\\ 
\multicolumn{2}{l}{\cellcolor{black!10!white}\textbf{Software Engineering Management}} \\ 
Scheduling Problem & \textbf{Size}: Employees~\cite{ \paperfortyeighth }, Tasks~\cite{ \paperfortyeighth }\\ 
\multicolumn{2}{l}{\cellcolor{black!10!white}\textbf{Software Design}} \\ 
Feature Selection & \textbf{Size}: Features~\cite{ \paperninetieth }\\ 
\bottomrule 
\end{tabular} 

%% file: generated_files/RQ3_general_effectiveness_metrics.tex
\begin{tabular} {p{0.3\textwidth} p{0.02\textwidth} p{0.88\textwidth} } \\ 
\toprule 
General  Metric & \# & Problems \\ 
\midrule  
Recall & 12 & DTC Problem~\cite{ \paperfourth }, Security Attack Identification~\cite{ \papersixth,\paperfiftyfifth,\paperfiftysixth,\papersixtyninth }, Software Failure Prediction~\cite{ \papertwentysixth,\paperseventyfirst }, Container Management~\cite{ \paperfortythird }, Resource Optimization~\cite{ \paperfortyfifth }, Allocation Problem~\cite{ \papereightysixth }, Routing Optimization~\cite{ \papereightyninth }, Scheduling Problem~\cite{ \paperninetyfirst }\\ 
F1 Score & 11 & Security Attack Identification~\cite{ \papersixth,\papertwentysecond,\paperfiftyfifth,\paperfiftysixth,\papersixtyninth }, Software Failure Prediction~\cite{ \papereleventh,\papertwentysixth,\paperthirtieth }, Container Management~\cite{ \paperfortythird }, Resource Optimization~\cite{ \paperfortyfifth }, Routing Optimization~\cite{ \papereightyninth }\\ 
Precision & 11 & DTC Problem~\cite{ \paperfourth }, Security Attack Identification~\cite{ \papersixth,\paperfiftyfifth,\paperfiftysixth,\papersixtyninth }, Software Failure Prediction~\cite{ \papertwentysixth }, Container Management~\cite{ \paperfortythird }, Resource Optimization~\cite{ \paperfortyfifth }, Allocation Problem~\cite{ \papereightysixth }, Routing Optimization~\cite{ \papereightyninth }, Scheduling Problem~\cite{ \paperninetyfirst }\\ 
Accuracy & 10 & Security Attack Identification~\cite{ \papersixth,\papertwentysecond,\paperfortysecond,\paperfiftyfifth,\paperfiftysixth,\papersixtyninth }, Software Failure Prediction~\cite{ \papereleventh,\papertwentysixth }, Container Management~\cite{ \paperfortythird }, Routing Optimization~\cite{ \papereightyninth }\\ 
Specificity & 6 & Container Management~\cite{ \paperfortythird }, Resource Optimization~\cite{ \paperfortyfifth }, Software Failure Prediction~\cite{ \paperseventyfirst }, Allocation Problem~\cite{ \papereightysixth }, Routing Optimization~\cite{ \papereightyninth }, Scheduling Problem~\cite{ \paperninetyfirst }\\ 
Root Mean Square Error & 5 & Software Failure Prediction~\cite{ \papertwentysixth,\paperseventythird }, Allocation Problem~\cite{ \papereightysixth }, Routing Optimization~\cite{ \papereightyninth }, Scheduling Problem~\cite{ \paperninetyfirst }\\ 
Approximation Ratio & 3 & Test Case Selection~\cite{ \paperthirteenth }, Test Suite Minimization~\cite{ \paperthirteenth }, WCET Evaluation~\cite{ \papertwentyseventh }, Multiple Query Optimization~\cite{ \paperfiftyeighth }\\ 
Area Under The Roc Curve & 2 & Software Failure Prediction~\cite{ \paperthirtieth,\paperseventyfirst }\\ 
Average Relative Prediction Error & 2 & Software Failure Prediction~\cite{ \papersixtyseventh,\paperseventythird }\\ 
Coverage & 2 & Allocation Problem~\cite{ \papereightysixth }, Scheduling Problem~\cite{ \paperninetyfirst }\\ 
Mean Absolute Error & 2 & Allocation Problem~\cite{ \papereightysixth }, Scheduling Problem~\cite{ \paperninetyfirst }\\ 
Optimal Solution Probability & 2 & WCET Evaluation~\cite{ \papertwentyseventh }, Multiple Query Optimization~\cite{ \paperfiftyeighth }\\ 
Relative Absolute Error & 2 & Allocation Problem~\cite{ \papereightysixth }, Scheduling Problem~\cite{ \paperninetyfirst }\\ 
Reliability & 2 & Resource Optimization~\cite{ \paperfortyfifth }, Routing Optimization~\cite{ \papereightyninth }\\ 
Root Relative Squared Error & 2 & Allocation Problem~\cite{ \papereightysixth }, Scheduling Problem~\cite{ \paperninetyfirst }\\ 
Success Probability & 2 & DTC Problem~\cite{ \paperfourth }, Test Suite Minimization~\cite{ \paperthirtyseventh }\\ 
Average Error & 1 & Software Reliability Growth Model~\cite{ \paperthirtysecond }\\ 
Cost Quality & 1 & Configuration Selection~\cite{ \papersecond }, Configuration Prioritization~\cite{ \papersecond }\\ 
Fairness Index & 1 & Service Deployment~\cite{ \paperthirtysixth }\\ 
Generational Distance & 1 & Software Requirement Selection~\cite{ \papertwentyeighth }\\ 
Inclusivity & 1 & Test Case Prioritization~\cite{ \paperthirtyeighth }, Test Case Selection~\cite{ \paperthirtyeighth }, Test Suite Minimization~\cite{ \paperthirtyeighth }\\ 
Kappa Coefficient & 1 & Routing Optimization~\cite{ \papereightyninth }\\ 
Mean Absolute Percentage Error & 1 & Software Failure Prediction~\cite{ \papertwentysixth }\\ 
Mean Squared Errors & 1 & Software Reliability Growth Model~\cite{ \paperthirtysecond }\\ 
Number Of Non-Dominated Solutions & 1 & Test Case Selection~\cite{ \papersixtythird }\\ 
Number Of Optimal Solutions & 1 & Scheduling Problem~\cite{ \paperseventh }\\ 
Number Of Suboptimal Solutions & 1 & Scheduling Problem~\cite{ \paperseventh }\\ 
Percentage Of Optimal Solutions & 1 & Join Order Problem~\cite{ \paperfiftyfourth }\\ 
Percentage Of Valid Solutions & 1 & Join Order Problem~\cite{ \paperfiftyfourth }\\ 
Proportion Of Optimal Solutions & 1 & Join Order Problem~\cite{ \paperninetythird }\\ 
Proportion Of Valid Solutions & 1 & Join Order Problem~\cite{ \paperninetythird }\\ 
Rank-Biased Overlap & 1 & Configuration Selection~\cite{ \papersecond }, Configuration Prioritization~\cite{ \papersecond }\\ 
Relative Error & 1 & Software Reliability Growth Model~\cite{ \paperthirtysecond }\\ 
Sum Of Intra-Cluster Distance & 1 & Security Attack Identification~\cite{ \papertwentysecond }\\ 
Top-10 Normalized Discounted Cumulative Gain & 1 & Feature Selection~\cite{ \paperninetieth }\\ 
Validity Quality & 1 & Configuration Selection~\cite{ \papersecond }, Configuration Prioritization~\cite{ \papersecond }\\ 
\bottomrule 
\end{tabular} 

%% file: generated_files/RQ3_quantum_effectiveness_metrics.tex
\begin{tabular} {p{0.3\textwidth} p{0.02\textwidth} p{0.88\textwidth} } \\ 
\toprule 
Quantum-specific  Metric & \# & Problems \\ 
\midrule  
Energy Value & 9 & Code Clone Detection~\cite{ \paperfirst }, Configuration Selection~\cite{ \papersecond }, Configuration Prioritization~\cite{ \papersecond }, Scheduling Problem~\cite{ \paperseventh }, Software Failure Prediction~\cite{ \papereleventh }, Security Management~\cite{ \papertwentyfifth,\papersixtyeighth }, WCET Evaluation~\cite{ \papertwentyseventh }, Join Order Problem~\cite{ \paperfiftysecond }, Feature Selection~\cite{ \paperninetieth }\\ 
Circuit Depth & 1 & Join Order Problem~\cite{ \paperninetythird }\\ 
Coherence Time & 1 & Join Order Problem~\cite{ \paperninetythird }\\ 
Depth Of A Quantum Operator & 1 & Configuration Selection~\cite{ \papersecond }, Configuration Prioritization~\cite{ \papersecond }\\ 
Energy Landscape & 1 & Allocation Problem~\cite{ \papereightyeighth }\\ 
Number Of Logical Qubits & 1 & Join Order Problem~\cite{ \paperninetythird }\\ 
Width Of A Quantum Operator & 1 & Configuration Selection~\cite{ \papersecond }, Configuration Prioritization~\cite{ \papersecond }\\ 
\bottomrule 
\end{tabular} 

%% file: generated_files/RQ3_cost_metrics.tex
\begin{tabular} {p{0.3\textwidth} p{0.02\textwidth} p{0.88\textwidth} } \\ 
\toprule 
Cost  Metric & \# & Problems \\ 
\midrule  
Execution Time & 26 & Join Order Problem~\cite{ \paperfifth,\paperfiftyfourth }, Test Case Selection~\cite{ \paperthirteenth,\papersixtythird }, Test Suite Minimization~\cite{ \paperthirteenth,\papersixtyfourth }, Covering Array Generation~\cite{ \papertwentieth,\paperthirtyfourth,\papersixtysixth,\paperseventyfifth,\paperseventysixth }, Node Localization~\cite{ \papertwentythird }, Security Management~\cite{ \papertwentyfifth,\papersixtyeighth }, Software Requirement Selection~\cite{ \papertwentyeighth }, Service Deployment~\cite{ \paperthirtyninth }, Service Configuration~\cite{ \paperfortieth }, Container Management~\cite{ \paperfortythird }, Scheduling Problem~\cite{ \paperfortyeighth }, Security Attack Identification~\cite{ \paperfiftyfifth }, Software Failure Prediction~\cite{ \paperseventyfirst }, Network Delay Optimization~\cite{ \papereightieth }, Routing Optimization~\cite{ \papereightyfourth,\papereightyninth }, Allocation Problem~\cite{ \papereightyseventh,\papereightyeighth }, Feature Selection~\cite{ \paperninetieth }\\ 
Optimization Time & 8 & Code Clone Detection~\cite{ \paperfirst }, Join Order Problem~\cite{ \paperfifth,\paperninetythird }, Software Failure Prediction~\cite{ \papereleventh }, Security Management~\cite{ \papertwentyfifth }, Multiple Query Optimization~\cite{ \paperfiftythird,\paperfiftyeighth }, Test Suite Minimization~\cite{ \papersixtyfourth }\\ 
Training Time & 4 & Security Attack Identification~\cite{ \papersixth,\paperfiftysixth,\papersixtyninth }, Software Failure Prediction~\cite{ \paperseventythird }\\ 
Prediction Time & 3 & Security Attack Identification~\cite{ \papersixth,\paperfiftysixth,\papersixtyninth }\\ 
Embedding Time & 2 & Database Index Selection~\cite{ \papersixtieth }, Test Suite Minimization~\cite{ \papersixtyfourth }\\ 
Sampling Time & 2 & Join Order Problem~\cite{ \paperfiftyfourth,\paperninetythird }\\ 
Economic Cost & 1 & Security Management~\cite{ \papertwentyfifth }\\ 
Mean Iterations For Successful Results & 1 & Software Failure Prediction~\cite{ \papertwentysixth }\\ 
Number Of Iterations & 1 & DTC Problem~\cite{ \paperfourth }\\ 
Pre-Processing Time & 1 & Test Suite Minimization~\cite{ \papersixtyfourth }\\ 
Programming Time & 1 & Join Order Problem~\cite{ \paperfiftyfourth }\\ 
QPU-Access Overhead Time By The Solver API & 1 & Join Order Problem~\cite{ \paperfiftyfourth }\\ 
Single-Sample Readout Time & 1 & Join Order Problem~\cite{ \paperfiftyfourth }\\ 
\bottomrule 
\end{tabular} 

%% file: generated_files/RQ3_complexity_effectiveness_metrics.tex
\begin{tabular} {p{0.3\textwidth} p{0.02\textwidth} p{0.88\textwidth} } \\ 
\toprule 
Complexity  Metric & \# & Problems \\ 
\midrule  
Time Complexity For Algorithm Execution & 19 & DTC Problem~\cite{ \paperfourth }, Scheduling Problem~\cite{ \paperseventh,\papereightysecond }, Test Case Selection~\cite{ \paperthirteenth,\paperthirtyeighth }, Test Suite Minimization~\cite{ \paperthirteenth,\paperthirtyeighth }, Security Attack Identification~\cite{ \papertwentysecond }, Software Requirement Selection~\cite{ \papertwentyninth }, Software Reliability Growth Model~\cite{ \paperthirtysecond }, Service Deployment~\cite{ \paperthirtysixth }, Test Case Prioritization~\cite{ \paperthirtyeighth }, Controller Placement~\cite{ \paperfortyfourth,\papereightythird }, Multiple Query Optimization~\cite{ \paperfiftyeighth }, Dynamic Testing~\cite{ \papersixtyfifth }, Software Failure Prediction~\cite{ \paperseventyfirst }, Covering Array Generation~\cite{ \paperseventyfifth }, Network Delay Optimization~\cite{ \papereightieth }, Clustering Protocol Optimization~\cite{ \papereightyfifth }, Allocation Problem~\cite{ \papereightyeighth }, Routing Optimization~\cite{ \papereightyninth }\\ 
Space Complexity & 4 & Scheduling Problem~\cite{ \paperseventh }, Covering Array Generation~\cite{ \papertwentieth,\paperseventyfifth }, Software Requirement Selection~\cite{ \papertwentyninth }\\ 
Time Complexity For Preprocessing & 3 & Scheduling Problem~\cite{ \paperseventh }, Test Case Selection~\cite{ \paperthirteenth }, Test Suite Minimization~\cite{ \paperthirteenth }, Multiple Query Optimization~\cite{ \paperfiftythird }\\ 
Time Complexity For Search/Optimization Phases & 3 & DTC Problem~\cite{ \paperfourth }, Test Case Selection~\cite{ \paperthirteenth }, Test Suite Minimization~\cite{ \paperthirteenth,\paperthirtyseventh }\\ 
Code Size Complexity & 1 & Scheduling Problem~\cite{ \paperseventh }\\ 
Time Complexity Of Fitness Calculation & 1 & Covering Array Generation~\cite{ \papertwentieth }\\ 
Time Complexity Of Parameter Combination Search & 1 & Covering Array Generation~\cite{ \papertwentieth }\\ 
Time Complexity Of Value Combination Search & 1 & Covering Array Generation~\cite{ \papertwentieth }\\ 
\bottomrule 
\end{tabular} 

%% file: generated_files/RQ4_open_benchmark.tex
\begin{tabular} {p{0.3\textwidth} p{0.05\textwidth} p{0.5\textwidth} p{0.5\textwidth}} \\ 
\toprule 
Problem & Paper & Description & URLs \\ 
\midrule  
Join Order Problem & \cite{ \paperfifth } & \textbf{Benchmark}: The involved cases are TPC-H, TPC-DS, LDBC BI, SQLite and the Join Order Benchmark. & \url{https://github.com/lfd/vldb24/tree/main/base/Experiments/Problems/benchmarks}\\ 
 & \cite{ \paperfiftieth } & \textbf{Benchmark}: There are three dataset, i.e., JOB: The Join Order Benchmark (JOB) contains queries on the IMDB dataset; Chain and Cycle: Two synthetic join benchmarks including join queries of chain and cycle graph structures with different numbers of relations. & \url{https://github.com/lfd/vldb24/tree/main}\\ 
 & \cite{ \paperfiftyfourth } & \textbf{Real-World}: ErgastF14 dataset contains information about the Formula One series. & \url{http://ergast.com/mrd/}\\ 
 & \cite{ \paperninetythird } & \textbf{Benchmark, Artificial}: The Chain, Star, Cycle graphs are included, which are generated based on the seminal method in VLDB'97.The Clique graphs are included, whose generation is related to SIGMOD'22. & \url{https://github.com/lfd/sigmod23-reproduction}\\ 
Scheduling Problem & \cite{ \paperfortyfirst } & \textbf{Benchmark}: Planet Lab is a real system provide workload traces & \url{https://github.com/beloglazov/planetlab-workload-traces}\\ 
 &  & \textbf{Real-World}: NASA Ames iPSC/860 is a real parallel workload. & \url{https://www.cs.huji.ac.il/labs/parallel/workload/l_nasa_ipsc/}\\ 
 & \cite{ \paperseventysecond } & \textbf{Benchmark}: Instance: Benchmark 855 & \url{https://www.researchgate.net/publication/341480343_Benchmark_855}\\ 
 &  & \textbf{Benchmark}: Numerical experiments: Benchmark90,
Benchmark306, Benchmark855 and Benchmark1020. & \url{https://www.researchgate.net/profile/Piotr-Dryja}\\ 
Security Attack Identification & \cite{ \papersixth } & \textbf{Real-World}: A SQL database (Automotive Attack Dataset (AAD)) contains 361 security attacks (621 attack steps) from public sources. & \url{https://github.com/IEEM-HsKA/AAD}\\ 
 & \cite{ \papertwentysecond } & \textbf{Benchmark}: UCI is a repository for machine learning, which include Iris, Glass, Wine, Cancer, Vowel, CMC and Vehicle & \url{https://archive.ics.uci.edu/}\\ 
 &  & \textbf{Real-World}: Four samples from KDD Cup 99 dataset are used in this paer, where KDD Cup 99 is the most widely used dataset for the evaluation of anomaly detection methods.  & \url{https://kdd.ics.uci.edu/databases/kddcup99/kddcup99.html}\\ 
 & \cite{ \paperfiftyfifth } & \textbf{Real-World}: Telemetry datasets are used for IoT and IIoT sensors. & \url{https://research.unsw.edu.au/projects/toniot-datasets}\\ 
Test Case Selection/Test Suite Minimization & \cite{ \paperthirteenth } & \textbf{Real-World}: There are four case studies for software testing, i.e., Paint Control (90 test cases), IOF/ROL (1941 test cases), GSDTSR (5555 test cases), and ELEVATOR (1925 test cases). & \url{https://doi.org/10.5281/zenodo.13911651}\\ 
Test Case Selection & \cite{ \papersixtythird } & \textbf{Real-World}: (1) 4 programs (i.e., flex, grep, gzip, and sed) from SIR;
(2) PaintControl and GSDTSR case studies. & \url{https://github.com/AntonioTrovato/SelectQA}\\ 
Test Suite Minimization & \cite{ \papersixtyfourth } & \textbf{Real-World}: Three industrial case studies, i.e., PaintControl; GSDTSR; IOF/ROL. & \url{https://zenodo.org/records/13273345}; \url{https://github.com/AsmarMuqeet/BootQA}\\ 
Configuration Selection/Configuration Prioritization & \cite{ \papersecond } & \textbf{Artificial}: There are 20 problem instances generated by the tool FeatureIDE. The feature costs are random integers from 10 to 100. & \url{https://github.com/KIT-TVA/qc-configuration-problem}\\ 
Dynamic Testing & \cite{ \papersixtyfifth } & \textbf{Artificial}: The artifact merely includes a toy example used in this paper & \url{https://zenodo.org/records/7075159}; \url{https://github.com/miranska/qc-dynamic-testing/blob/v0.1.1/example.ipynb}\\ 
Security Management & \cite{ \papersixtyeighth } & \textbf{Real-World}: The dataset used presents 50 incidents, together with the possible courses of action to
respond, the time needed and the associated sustainability label. & \url{https://github.com/GSYAtools/QISS}\\ 
\bottomrule 
\end{tabular} 

%% file: generated_files/RQ5_open_tool.tex
\begin{tabular} {p{0.3\textwidth} p{0.05\textwidth} p{0.18\textwidth} p{0.6\textwidth}} \\ 
\toprule 
Problem & Paper & Approach & URLs \\ 
\midrule  
Join Order Problem & \cite{ \paperfifth } & Quantum-Inspired & \url{https://github.com/lfd/vldb24}\\ 
 & \cite{ \paperfiftysecond } & Quantum & \url{https://anonymous.4open.science/r/Q-Join-PODS25}\\ 
 & \cite{ \paperninetythird } & Quantum/Hybrid & \url{https://github.com/lfd/sigmod23-reproduction}\\ 
Dynamic Testing & \cite{ \papersixtyfifth } & Quantum & \url{https://github.com/miranska/qc-dynamic-testing/blob/v0.1.1/example.ipynb}\\ 
 &  & Quantum & \url{https://zenodo.org/records/7075159}\\ 
Test Suite Minimization & \cite{ \papersixtyfourth } & Hybrid & \url{https://github.com/AsmarMuqeet/BootQA}\\ 
 &  & Hybrid & \url{https://zenodo.org/records/13273345}\\ 
Test Case Selection & \cite{ \papersixtythird } & Hybrid & \url{https://github.com/AntonioTrovato/SelectQA}\\ 
Test Case Selection/Test Suite Minimization & \cite{ \paperthirteenth } & Hybrid & \url{https://doi.org/10.5281/zenodo.13911651}\\ 
Configuration Selection/Configuration Prioritization & \cite{ \papersecond } & Hybrid & \url{https://github.com/KIT-TVA/qc-configuration-problem}\\ 
DTC Problem & \cite{ \paperfourth } & Quantum & \url{https://github.com/jiawei-95/tosem-QDTCSA-artifact}\\ 
Multiple Query Optimization & \cite{ \paperfiftyeighth } & Hybrid & \url{https://drive.google.com/file/d/1MiZTBzbrm8_SVnruGcPa9hX1PC_W8wO3/view}\\ 
Scheduling Problem & \cite{ \paperseventh } & Quantum & \url{https://github.com/luposdate/OptimizingTransactionSchedulesWithSilq}\\ 
Security Management & \cite{ \papersixtyeighth } & Quantum & \url{https://github.com/GSYAtools/QISS}\\ 
\midrule  
Total (Percentage) & \multicolumn{3}{l}{12 (15.79\%)  } \\ 
\bottomrule 
\end{tabular} 

%% file: related.tex
%%%%%%%%%%%%%%%%%%%%%%%%%%%%%%%%%%%%%%%%%%%%%%%%%%%%%%%%%%%%%%%%%%%%%%%%%%%%

%SBSE
In SE, optimization problems are often framed as search problems, hence, the Search-Based Software Engineering (SBSE) community was formed and successfully addressed many SE optimization problems such test generation and optimization~\cite{zhang2021adaptive}, requirements prioritization~\cite{zhang2020uncertainty}, next release problems~\cite{durillo2011study} and code refactoring~\cite{mariani2017systematic}. 
Empirical studies are a common practice in SBSE, providing the foundation for understanding the cost-effectiveness of different search algorithms and evaluating their performance across diverse SE tasks. Such empirical studies offer systematic insights that guide the design, comparison, and improvement of optimization approaches in practice. 
After more than a decade of development, the SBSE community has accumulated a substantial body of knowledge on how empirical studies should be conducted. Guidelines have been provided on choosing appropriate quality indicators \cite{ali2020quality}, determining the number of experiment repetitions~\cite{arcuri2014hitchhiker}, tuning parameters~\cite{arcuri2013parameter}, and other aspects of experimental design. 

%Quantum-inspired in SE
Empirical studies have been conducted to evaluate the application of quantum-inspired algorithms in SE, as evidenced by our study. For instance, the primary study~\cite{kumari2016comparing} reports an empirical evaluation comparing three quantum-inspired multi-objective evolutionary algorithms, demonstrating their effectiveness in addressing the software requirements selection problem. However, we are unable to find any studies that survey, review, or investigate how empirical studies should be conducted to evaluate quantum-inspired algorithms for solving SE problems.

%Quantum in SE
Though Quantum Software Engineering (QSE), as an emerging interdisciplinary field that focuses on developing, designing, testing and maintaining software solutions tailored for quantum computing systems, has received increasing attention over the last five years, with contributions such as the survey by Murillo et al.~\cite{murillo2025quantum} and the roadmap proposed by Pezzè et al.~\cite{pezze20252030} providing comprehensive overviews of the field, only some empirical practices have been accumulated. While these works discuss methodologies, challenges, and potential applications of QSE, there remains a lack of a systematic body of knowledge on how empirical studies should be designed and conducted to rigorously evaluate quantum algorithms in solving SE problems. 

%beyond SE

%----------------------------------------------------------

%% file: arxiv_quantumsurvey.bbl

%% file: appendix.tex
\section*{Appendix}

\begin{table}
	\small
	\centering
	\caption{Problem-specific output metrics}
	\label{tab:rq3_problem_effectivenes_metrics}
	\resizebox{\textwidth}{!}{
		\input{generated_files/RQ3_problem_effectiveness_metrics.tex}
	}
\end{table}

%% file: generated_files/RQ3_problem_effectiveness_metrics.tex
\begin{tabular} {p{0.3\textwidth} p{0.9\textwidth} } \\ 
\toprule 
Problem &  Problem-specific  Metrics \\ 
\midrule  
Scheduling Problem & CPU Usage Rate~\cite{ \paperthird }, Energy Consumption~\cite{ \paperthird,\papereightysecond }, Task Completion Time~\cite{ \paperthird }, Response Time~\cite{ \paperfourteenth,\paperfortyfirst }, Throughput~\cite{ \paperfourteenth,\paperfortyfirst }, Resource Utilization~\cite{ \paperfourteenth,\papertwentyfourth,\paperfortyfirst,\papereightysecond }, Request Error Rate~\cite{ \paperfourteenth }, Energe Consumption~\cite{ \papertwentyfourth,\paperninetyfirst }, Makespan~\cite{ \papertwentyfourth,\paperfortyfirst,\papereightysecond }, Degree Of Imbalance~\cite{ \papertwentyfourth }, Cost Of Task Execution~\cite{ \papertwentyfourth }, Fitness Value~\cite{ \paperfortyfirst,\paperfortyeighth }, Profit~\cite{ \paperfortyfirst }, Service Level Agreement Violation Rate~\cite{ \paperfortyfirst }, Task Rejection Rate~\cite{ \paperfortyfirst }, Output Probabilities~\cite{ \paperseventysecond }, Free Capacity Of The Ram In The Bottleneck Computer~\cite{ \paperseventysecond }, Free Capacity Of The Disc Storage In The Bottleneck Computer~\cite{ \paperseventysecond }, Reliability Of The Cloud~\cite{ \paperseventysecond }, Electric Power Of The Cloud~\cite{ \paperseventysecond }, Economic Cost Of Hosts~\cite{ \paperseventysecond }, Communication Capability Of The Bottleneck Node~\cite{ \paperseventysecond }, Processing Workload Of The Bottleneck CPU~\cite{ \paperseventysecond }, P-Norm Between The Solution And The Ideal Point~\cite{ \paperseventysecond }, Temporal Delay~\cite{ \paperninetyfirst }\\ 
Join Order Problem & Solution Cost~\cite{ \paperfifth,\paperfiftieth,\paperfiftysecond }, Percentage Of Queries Retrieving Optimal Join Orders~\cite{ \paperfiftyfourth }\\ 
Security Attack Identification & Detection Rate~\cite{ \papertwentysecond,\paperfortysecond }, Silhouette Value~\cite{ \paperfortysecond }, False Alarm Rate~\cite{ \paperfortysecond }\\ 
Test Suite Minimization & Test Reduction Percentage~\cite{ \papertenth,\paperthirtythird }, Coverage Loss Percentage~\cite{ \papertenth,\paperthirtythird }, Fault Detection Loss Percentage~\cite{ \papertenth,\paperthirtyeighth }, Cost Reduction Percentage~\cite{ \papertenth,\paperthirtythird,\paperthirtyeighth }, Average Percentage Of Fault Detection~\cite{ \papertenth,\paperthirtythird,\paperthirtyeighth }, Average Percentage Of Fault Detection With Cost~\cite{ \papertenth,\paperthirtythird }, Average Percentage Of Statement Coverage~\cite{ \papertenth,\paperthirtythird,\paperthirtyeighth }, Average Percentage Of Statement Coverage With Cost~\cite{ \papertenth,\paperthirtythird }, Fitness Value~\cite{ \paperthirteenth,\paperthirtythird,\papersixtyfourth,\paperseventyfourth,\paperseventyseventh }, Solution Cost~\cite{ \papersixteenth }, Fault Detection Capability Loss Percentage~\cite{ \paperthirtythird }, Size Of The Reduced Test Suite~\cite{ \paperthirtyseventh,\paperseventyseventh }, Test Selection Percentage~\cite{ \paperthirtyeighth }, Reduced Number Of Components In A Test Case~\cite{ \paperseventyfourth }, Number Of Detected Faults~\cite{ \paperseventyseventh }\\ 
Test Case Prioritization & Test Reduction Percentage~\cite{ \papertenth,\paperthirtythird }, Coverage Loss Percentage~\cite{ \papertenth,\paperthirtythird }, Fault Detection Loss Percentage~\cite{ \papertenth,\paperthirtyeighth }, Cost Reduction Percentage~\cite{ \papertenth,\paperthirtythird,\paperthirtyeighth }, Average Percentage Of Fault Detection~\cite{ \papertenth,\paperthirtythird,\paperthirtyeighth }, Average Percentage Of Fault Detection With Cost~\cite{ \papertenth,\paperthirtythird }, Average Percentage Of Statement Coverage~\cite{ \papertenth,\paperthirtythird,\paperthirtyeighth }, Average Percentage Of Statement Coverage With Cost~\cite{ \papertenth,\paperthirtythird }, Fault Detection Capability Loss Percentage~\cite{ \paperthirtythird }, Fitness Value~\cite{ \paperthirtythird }, Test Selection Percentage~\cite{ \paperthirtyeighth }\\ 
Software Failure Prediction & Number Of Reduced Features~\cite{ \papereleventh }, Number Of Correct Clusters~\cite{ \papertwelfth }, Number Of Incorrect Clusters~\cite{ \papertwelfth }, Error Rate~\cite{ \papertwelfth }, Optimal Value~\cite{ \papertwentysixth }, Mean Optimal Value~\cite{ \papertwentysixth }, Number Of Hidden Nodes~\cite{ \papertwentysixth }, Fitness Value~\cite{ \papersixtyseventh }\\ 
Test Case Selection & Fitness Value~\cite{ \paperthirteenth }, Test Selection Percentage~\cite{ \paperthirtyeighth }, Fault Detection Loss Percentage~\cite{ \paperthirtyeighth }, Cost Reduction Percentage~\cite{ \paperthirtyeighth }, Average Percentage Of Fault Detection~\cite{ \paperthirtyeighth }, Average Percentage Of Statement Coverage~\cite{ \paperthirtyeighth }, Execution Cost Of The Test Suite~\cite{ \papersixtythird }, Failure Rate Of The Test Suite~\cite{ \papersixtythird }\\ 
Covering Array Generation & Size Of The Generated Test Suite~\cite{ \papertwentieth,\paperthirtyfourth,\paperseventyfifth,\paperseventysixth }, Coverage Of Covering Arrays~\cite{ \papersixtysixth }, Number Of Successful Covering Array Generations~\cite{ \papersixtysixth }\\ 
Node Localization & Localization Error~\cite{ \papertwentythird }, Number Of Localized Nodes~\cite{ \papertwentythird }\\ 
Security Management & Number Of Selected Incidents~\cite{ \papertwentyfifth }, Solution Cost~\cite{ \papersixtyeighth }, Number Of Selected Courses Of Action~\cite{ \papersixtyeighth }\\ 
Software Requirement Selection & Development Cost~\cite{ \papertwentyeighth,\papertwentyninth }, Customer Satisfaction~\cite{ \papertwentyeighth,\papertwentyninth }\\ 
Controller Placement & Total Latency In Software Defined Networking~\cite{ \paperthirtyfirst }, Degree Of Imbalance~\cite{ \paperfortyfourth,\papereightythird }, Switch-To-Controller Latency~\cite{ \paperfortyfourth,\papereightythird }, Controller-To-Controller Latency~\cite{ \paperfortyfourth,\papereightythird }, Fitness Value~\cite{ \paperfortyfourth,\papereightythird }, Placement Nodes~\cite{ \paperfortyfourth }\\ 
Software Reliability Growth Model & Number Of Faults~\cite{ \paperthirtysecond }\\ 
Communication Optimization & End-To-End Delay~\cite{ \paperthirtyfifth }, Energy Consumption~\cite{ \paperthirtyfifth }, Package Delivery Rate~\cite{ \paperthirtyfifth }, Expected Transmission Rate~\cite{ \paperthirtyfifth }\\ 
Service Deployment & Network Overall Delay~\cite{ \paperthirtysixth }, Delay Of Traffic~\cite{ \paperthirtysixth }, Satisfaction Degree~\cite{ \paperthirtysixth }, Ratio Of Service Nodes~\cite{ \paperthirtysixth,\paperthirtyninth }, Transport Delay~\cite{ \paperthirtyninth }\\ 
Service Configuration & End-To-End Delay~\cite{ \paperfortieth }, Success Rate Of Service Reconfiguration~\cite{ \paperfortieth }\\ 
Container Management & Energy Efficiency~\cite{ \paperfortythird }, Number Of Active Servers~\cite{ \paperfortythird }, Communication Cost~\cite{ \paperfortythird }\\ 
Resource Optimization & Temporal Delay~\cite{ \paperfortyfifth }\\ 
Multiple Query Optimization & Solution Cost~\cite{ \paperfiftythird }\\ 
Database Index Selection & Solution Quality~\cite{ \papersixtieth }\\ 
Dynamic Testing & Output Probabilities~\cite{ \papersixtyfifth }\\ 
Allocation Problem & Task Completion Time~\cite{ \paperseventieth }, Agent Load~\cite{ \paperseventieth }, Energe Consumption~\cite{ \papereightysixth }, Temporal Delay~\cite{ \papereightysixth }, Resource Utilization~\cite{ \papereightyseventh }, Output Probabilities~\cite{ \papereightyseventh }, Resource Consumption Cost~\cite{ \papereightyeighth }, Proportion Of Deployed Metaslice~\cite{ \papereightyeighth }, Service Acceptance Rate~\cite{ \papereightyeighth }\\ 
Queuing Delay Optimization & Fitness Value~\cite{ \paperseventyninth }, Arrival Rate~\cite{ \paperseventyninth }\\ 
Network Delay Optimization & End-To-End Delay~\cite{ \papereightieth }\\ 
Routing Optimization & Output Probabilities~\cite{ \papereightyfourth }, Battery Cost~\cite{ \papereightyfourth }, Throughput~\cite{ \papereightyfourth }, Number Of Hops~\cite{ \papereightyfourth }, Solution Cost~\cite{ \papereightyfourth,\papereightyninth }, Temporal Delay~\cite{ \papereightyninth }, Fitness Value~\cite{ \paperninetysecond }, Network Lifetime~\cite{ \paperninetysecond }\\ 
Clustering Protocol Optimization & Number Of Living Nodes~\cite{ \papereightyfifth }, Time When The First Node Dies~\cite{ \papereightyfifth }, Time When A Specific Number Of Nodes Die~\cite{ \papereightyfifth }, Network Lifetime~\cite{ \papereightyfifth }, Node Energy~\cite{ \papereightyfifth }, Packet Transmission Delay~\cite{ \papereightyfifth }\\ 
\bottomrule 
\end{tabular} 

%% file: arxiv_quantumsurvey.bbl
\begin{thebibliography}{107}

%%% the Web address.

\ifx \showCODEN    \undefined \def \showCODEN     #1{\unskip}     \fi
\ifx \showDOI      \undefined \def \showDOI       #1{#1}\fi
\ifx \showISBNx    \undefined \def \showISBNx     #1{\unskip}     \fi
\ifx \showISBNxiii \undefined \def \showISBNxiii  #1{\unskip}     \fi
\ifx \showISSN     \undefined \def \showISSN      #1{\unskip}     \fi
\ifx \showLCCN     \undefined \def \showLCCN      #1{\unskip}     \fi
\ifx \shownote     \undefined \def \shownote      #1{#1}          \fi
\ifx \showarticletitle \undefined \def \showarticletitle #1{#1}   \fi
\ifx \showURL      \undefined \def \showURL       {\relax}        \fi
% invisible to TeX
\providecommand\bibfield[2]{#2}
\providecommand\bibinfo[2]{#2}
\providecommand\natexlab[1]{#1}
\providecommand\showeprint[2][]{arXiv:#2}

\bibitem[\protect\citeauthoryear{Abbas, Ambainis, Augustino, B{\"a}rtschi,
  Buhrman, Coffrin, Cortiana, Dunjko, Egger, Elmegreen, et~al\mbox{.}}{Abbas
  et~al\mbox{.}}{2024}]%
        {abbas2024challenges}
\bibfield{author}{\bibinfo{person}{Amira Abbas}, \bibinfo{person}{Andris
  Ambainis}, \bibinfo{person}{Brandon Augustino}, \bibinfo{person}{Andreas
  B{\"a}rtschi}, \bibinfo{person}{Harry Buhrman}, \bibinfo{person}{Carleton
  Coffrin}, \bibinfo{person}{Giorgio Cortiana}, \bibinfo{person}{Vedran
  Dunjko}, \bibinfo{person}{Daniel~J Egger}, \bibinfo{person}{Bruce~G
  Elmegreen}, {et~al\mbox{.}}} \bibinfo{year}{2024}\natexlab{}.
\newblock \showarticletitle{Challenges and opportunities in quantum
  optimization}.
\newblock \bibinfo{journal}{\emph{Nature Reviews Physics}}
  (\bibinfo{year}{2024}), \bibinfo{pages}{1--18}.
\newblock


\bibitem[\protect\citeauthoryear{Ahanger, Dahan, Tariq, and Ullah}{Ahanger
  et~al\mbox{.}}{2022}]%
        {ahanger2022quantum}
\bibfield{author}{\bibinfo{person}{Tariq~Ahamed Ahanger}, \bibinfo{person}{Fadl
  Dahan}, \bibinfo{person}{Usman Tariq}, {and} \bibinfo{person}{Imdad Ullah}.}
  \bibinfo{year}{2022}\natexlab{}.
\newblock \showarticletitle{Quantum inspired task optimization for IoT edge fog
  computing environment}.
\newblock \bibinfo{journal}{\emph{Mathematics}} \bibinfo{volume}{11},
  \bibinfo{number}{1} (\bibinfo{year}{2022}), \bibinfo{pages}{156}.
\newblock


\bibitem[\protect\citeauthoryear{Ali, Arcaini, Pradhan, Safdar, and Yue}{Ali
  et~al\mbox{.}}{2020}]%
        {ali2020quality}
\bibfield{author}{\bibinfo{person}{Shaukat Ali}, \bibinfo{person}{Paolo
  Arcaini}, \bibinfo{person}{Dipesh Pradhan}, \bibinfo{person}{Safdar~Aqeel
  Safdar}, {and} \bibinfo{person}{Tao Yue}.} \bibinfo{year}{2020}\natexlab{}.
\newblock \showarticletitle{Quality indicators in search-based software
  engineering: An empirical evaluation}.
\newblock \bibinfo{journal}{\emph{ACM Transactions on Software Engineering and
  Methodology (TOSEM)}} \bibinfo{volume}{29}, \bibinfo{number}{2}
  (\bibinfo{year}{2020}), \bibinfo{pages}{1--29}.
\newblock


\bibitem[\protect\citeauthoryear{Ammermann, Brenneisen, Bittner, and
  Schaefer}{Ammermann et~al\mbox{.}}{2024}]%
        {ammermann2024quantum}
\bibfield{author}{\bibinfo{person}{Joshua Ammermann},
  \bibinfo{person}{Fabian~Jakob Brenneisen}, \bibinfo{person}{Tim Bittner},
  {and} \bibinfo{person}{Ina Schaefer}.} \bibinfo{year}{2024}\natexlab{}.
\newblock \showarticletitle{Quantum Solution for Configuration Selection and
  Prioritization}. In \bibinfo{booktitle}{\emph{Proceedings of the 5th ACM/IEEE
  International Workshop on Quantum Software Engineering}}.
  \bibinfo{pages}{21--28}.
\newblock


\bibitem[\protect\citeauthoryear{Arcuri and Briand}{Arcuri and Briand}{2011}]%
        {arcuri2011practical}
\bibfield{author}{\bibinfo{person}{Andrea Arcuri} {and} \bibinfo{person}{Lionel
  Briand}.} \bibinfo{year}{2011}\natexlab{}.
\newblock \showarticletitle{A practical guide for using statistical tests to
  assess randomized algorithms in software engineering}. In
  \bibinfo{booktitle}{\emph{Proceedings of the 33rd international conference on
  software engineering}}. \bibinfo{pages}{1--10}.
\newblock


\bibitem[\protect\citeauthoryear{Arcuri and Briand}{Arcuri and Briand}{2014}]%
        {arcuri2014hitchhiker}
\bibfield{author}{\bibinfo{person}{Andrea Arcuri} {and} \bibinfo{person}{Lionel
  Briand}.} \bibinfo{year}{2014}\natexlab{}.
\newblock \showarticletitle{A hitchhiker's guide to statistical tests for
  assessing randomized algorithms in software engineering}.
\newblock \bibinfo{journal}{\emph{Software Testing, Verification and
  Reliability}} \bibinfo{volume}{24}, \bibinfo{number}{3}
  (\bibinfo{year}{2014}), \bibinfo{pages}{219--250}.
\newblock


\bibitem[\protect\citeauthoryear{Arcuri and Fraser}{Arcuri and Fraser}{2013}]%
        {arcuri2013parameter}
\bibfield{author}{\bibinfo{person}{Andrea Arcuri} {and} \bibinfo{person}{Gordon
  Fraser}.} \bibinfo{year}{2013}\natexlab{}.
\newblock \showarticletitle{Parameter tuning or default values? An empirical
  investigation in search-based software engineering}.
\newblock \bibinfo{journal}{\emph{Empirical Software Engineering}}
  \bibinfo{volume}{18}, \bibinfo{number}{3} (\bibinfo{year}{2013}),
  \bibinfo{pages}{594--623}.
\newblock


\bibitem[\protect\citeauthoryear{Bajaj and Abraham}{Bajaj and Abraham}{2022}]%
        {bajaj2022test}
\bibfield{author}{\bibinfo{person}{Anu Bajaj} {and} \bibinfo{person}{Ajith
  Abraham}.} \bibinfo{year}{2022}\natexlab{}.
\newblock \showarticletitle{Test case prioritization and reduction using hybrid
  quantum-behaved particle swarm optimization}. In
  \bibinfo{booktitle}{\emph{2022 IEEE Congress on Evolutionary Computation
  (CEC)}}. IEEE, \bibinfo{pages}{1--8}.
\newblock


\bibitem[\protect\citeauthoryear{Bajaj, Abraham, Ratnoo, and Gabralla}{Bajaj
  et~al\mbox{.}}{2022a}]%
        {bajaj2022Sensorstest}
\bibfield{author}{\bibinfo{person}{Anu Bajaj}, \bibinfo{person}{Ajith Abraham},
  \bibinfo{person}{Saroj Ratnoo}, {and} \bibinfo{person}{Lubna~Abdelkareim
  Gabralla}.} \bibinfo{year}{2022}\natexlab{a}.
\newblock \showarticletitle{Test case prioritization, selection, and reduction
  using improved quantum-behaved particle swarm optimization}.
\newblock \bibinfo{journal}{\emph{Sensors}} \bibinfo{volume}{22},
  \bibinfo{number}{12} (\bibinfo{year}{2022}), \bibinfo{pages}{4374}.
\newblock


\bibitem[\protect\citeauthoryear{Bajaj, Sangwan, and Abraham}{Bajaj
  et~al\mbox{.}}{2022b}]%
        {bajaj2022improved}
\bibfield{author}{\bibinfo{person}{Anu Bajaj}, \bibinfo{person}{Om~Prakash
  Sangwan}, {and} \bibinfo{person}{Ajith Abraham}.}
  \bibinfo{year}{2022}\natexlab{b}.
\newblock \showarticletitle{Improved novel bat algorithm for test case
  prioritization and minimization}.
\newblock \bibinfo{journal}{\emph{Soft Computing}} \bibinfo{volume}{26},
  \bibinfo{number}{22} (\bibinfo{year}{2022}), \bibinfo{pages}{12393--12419}.
\newblock


\bibitem[\protect\citeauthoryear{Balicki}{Balicki}{2021}]%
        {balicki2021many}
\bibfield{author}{\bibinfo{person}{Jerzy Balicki}.}
  \bibinfo{year}{2021}\natexlab{}.
\newblock \showarticletitle{Many-objective quantum-inspired particle swarm
  optimization algorithm for placement of virtual machines in smart computing
  cloud}.
\newblock \bibinfo{journal}{\emph{Entropy}} \bibinfo{volume}{24},
  \bibinfo{number}{1} (\bibinfo{year}{2021}), \bibinfo{pages}{58}.
\newblock


\bibitem[\protect\citeauthoryear{Barletta, Caivano, Catalano, and
  De~Vincentiis}{Barletta et~al\mbox{.}}{2024}]%
        {barletta2024quantum}
\bibfield{author}{\bibinfo{person}{Vita~Santa Barletta},
  \bibinfo{person}{Danilo Caivano}, \bibinfo{person}{Christian Catalano}, {and}
  \bibinfo{person}{Mirko De~Vincentiis}.} \bibinfo{year}{2024}\natexlab{}.
\newblock \showarticletitle{Quantum-based automotive threat intelligence and
  countermeasures}. In \bibinfo{booktitle}{\emph{Proceedings of the 28th
  International Conference on Evaluation and Assessment in Software
  Engineering}}. \bibinfo{pages}{548--554}.
\newblock


\bibitem[\protect\citeauthoryear{Barletta, Caivano, De~Vincentiis, Magr{\`\i},
  and Piccinno}{Barletta et~al\mbox{.}}{2022}]%
        {barletta2022quantum}
\bibfield{author}{\bibinfo{person}{Vita~Santa Barletta},
  \bibinfo{person}{Danilo Caivano}, \bibinfo{person}{Mirko De~Vincentiis},
  \bibinfo{person}{Alessio Magr{\`\i}}, {and} \bibinfo{person}{Antonio
  Piccinno}.} \bibinfo{year}{2022}\natexlab{}.
\newblock \showarticletitle{Quantum optimization for IoT security detection}.
  In \bibinfo{booktitle}{\emph{International Symposium on Ambient
  Intelligence}}. Springer, \bibinfo{pages}{187--196}.
\newblock


\bibitem[\protect\citeauthoryear{Barletta, Caivano, Lako, and Pal}{Barletta
  et~al\mbox{.}}{2023}]%
        {barletta2023quantum}
\bibfield{author}{\bibinfo{person}{Vita~Santa Barletta},
  \bibinfo{person}{Danilo Caivano}, \bibinfo{person}{Alfred Lako}, {and}
  \bibinfo{person}{Anibrata Pal}.} \bibinfo{year}{2023}\natexlab{}.
\newblock \showarticletitle{Quantum as a service architecture for security in a
  smart city}. In \bibinfo{booktitle}{\emph{International Conference on the
  Quality of Information and Communications Technology}}. Springer,
  \bibinfo{pages}{76--89}.
\newblock


\bibitem[\protect\citeauthoryear{Bettonte, Gilbert, Vert, Louise, and
  Sirdey}{Bettonte et~al\mbox{.}}{2022}]%
        {bettonte2022quantum}
\bibfield{author}{\bibinfo{person}{Gabriella Bettonte},
  \bibinfo{person}{Valentin Gilbert}, \bibinfo{person}{Daniel Vert},
  \bibinfo{person}{St{\'e}phane Louise}, {and} \bibinfo{person}{Renaud
  Sirdey}.} \bibinfo{year}{2022}\natexlab{}.
\newblock \showarticletitle{Quantum approaches for WCET-related optimization
  problems}. In \bibinfo{booktitle}{\emph{International Conference on
  Computational Science}}. Springer, \bibinfo{pages}{202--217}.
\newblock


\bibitem[\protect\citeauthoryear{Bhatia and Sood}{Bhatia and Sood}{2020}]%
        {bhatia2020quantum}
\bibfield{author}{\bibinfo{person}{Munish Bhatia} {and}
  \bibinfo{person}{Sandeep~K Sood}.} \bibinfo{year}{2020}\natexlab{}.
\newblock \showarticletitle{Quantum computing-inspired network optimization for
  IoT applications}.
\newblock \bibinfo{journal}{\emph{IEEE Internet of Things Journal}}
  \bibinfo{volume}{7}, \bibinfo{number}{6} (\bibinfo{year}{2020}),
  \bibinfo{pages}{5590--5598}.
\newblock


\bibitem[\protect\citeauthoryear{Bhatia, Sood, and Kaur}{Bhatia
  et~al\mbox{.}}{2019}]%
        {bhatia2019quantum}
\bibfield{author}{\bibinfo{person}{Munish Bhatia}, \bibinfo{person}{Sandeep~K
  Sood}, {and} \bibinfo{person}{Simranpreet Kaur}.}
  \bibinfo{year}{2019}\natexlab{}.
\newblock \showarticletitle{Quantum-based predictive fog scheduler for IoT
  applications}.
\newblock \bibinfo{journal}{\emph{Computers in Industry}}
  \bibinfo{volume}{111} (\bibinfo{year}{2019}), \bibinfo{pages}{51--67}.
\newblock


\bibitem[\protect\citeauthoryear{Blanco, Santos-Olmo, and S{\'a}nchez}{Blanco
  et~al\mbox{.}}{2024}]%
        {blanco2024qiss}
\bibfield{author}{\bibinfo{person}{Carlos Blanco}, \bibinfo{person}{Antonio
  Santos-Olmo}, {and} \bibinfo{person}{Luis~Enrique S{\'a}nchez}.}
  \bibinfo{year}{2024}\natexlab{}.
\newblock \showarticletitle{QISS: Quantum-enhanced sustainable security
  incident handling in the IoT}.
\newblock \bibinfo{journal}{\emph{Information}} \bibinfo{volume}{15},
  \bibinfo{number}{4} (\bibinfo{year}{2024}), \bibinfo{pages}{181}.
\newblock


\bibitem[\protect\citeauthoryear{Caivano, De~Vincentiis, Nitti, and
  Pal}{Caivano et~al\mbox{.}}{2022}]%
        {caivano2022quantum}
\bibfield{author}{\bibinfo{person}{Danilo Caivano}, \bibinfo{person}{Mirko
  De~Vincentiis}, \bibinfo{person}{Federica Nitti}, {and}
  \bibinfo{person}{Anibrata Pal}.} \bibinfo{year}{2022}\natexlab{}.
\newblock \showarticletitle{Quantum optimization for fast CAN bus intrusion
  detection}. In \bibinfo{booktitle}{\emph{Proceedings of the 1st international
  workshop on quantum programming for software engineering}}.
  \bibinfo{pages}{15--18}.
\newblock


\bibitem[\protect\citeauthoryear{Chen, Qi, Chen, Chen, and Cheng}{Chen
  et~al\mbox{.}}{2020}]%
        {chen2020quantum}
\bibfield{author}{\bibinfo{person}{Junwen Chen}, \bibinfo{person}{Xuemei Qi},
  \bibinfo{person}{Linfeng Chen}, \bibinfo{person}{Fulong Chen}, {and}
  \bibinfo{person}{Guihua Cheng}.} \bibinfo{year}{2020}\natexlab{}.
\newblock \showarticletitle{Quantum-inspired ant lion optimized hybrid k-means
  for cluster analysis and intrusion detection}.
\newblock \bibinfo{journal}{\emph{Knowledge-Based Systems}}
  \bibinfo{volume}{203} (\bibinfo{year}{2020}), \bibinfo{pages}{106167}.
\newblock


\bibitem[\protect\citeauthoryear{Deepalakshmi and Chandran}{Deepalakshmi and
  Chandran}{2022}]%
        {deepalakshmi2022optimized}
\bibfield{author}{\bibinfo{person}{J Deepalakshmi} {and} \bibinfo{person}{M
  Chandran}.} \bibinfo{year}{2022}\natexlab{}.
\newblock \showarticletitle{An optimized clustering model for heterogeneous
  cross-project defect prediction using Quantum Crow search}. In
  \bibinfo{booktitle}{\emph{2022 1st International Conference on Software
  Engineering and Information Technology (ICoSEIT)}}. IEEE,
  \bibinfo{pages}{30--35}.
\newblock


\bibitem[\protect\citeauthoryear{Deng, Wan, and Guo}{Deng
  et~al\mbox{.}}{2022}]%
        {deng2022research}
\bibfield{author}{\bibinfo{person}{Lijuan Deng}, \bibinfo{person}{Long Wan},
  {and} \bibinfo{person}{Jian Guo}.} \bibinfo{year}{2022}\natexlab{}.
\newblock \showarticletitle{Research on security anomaly detection for big data
  platforms based on quantum optimization clustering}.
\newblock \bibinfo{journal}{\emph{Mathematical Problems in Engineering}}
  \bibinfo{volume}{2022}, \bibinfo{number}{1} (\bibinfo{year}{2022}),
  \bibinfo{pages}{4805035}.
\newblock


\bibitem[\protect\citeauthoryear{Dornala, Ponnapalli, Sai, Koteru, Koteru, and
  Koteru}{Dornala et~al\mbox{.}}{2023}]%
        {dornala2023quantum}
\bibfield{author}{\bibinfo{person}{Raghunadha~Reddi Dornala},
  \bibinfo{person}{Sudhir Ponnapalli}, \bibinfo{person}{Kalakoti~Thriveni Sai},
  \bibinfo{person}{Siva Rama Krishna~Reddi Koteru}, \bibinfo{person}{Rami~Reddy
  Koteru}, {and} \bibinfo{person}{Bhavani Koteru}.}
  \bibinfo{year}{2023}\natexlab{}.
\newblock \showarticletitle{Quantum based Fault-Tolerant Load Balancing in
  Cloud Computing with Quantum Computing}. In \bibinfo{booktitle}{\emph{2023
  3rd International Conference on Innovative Mechanisms for Industry
  Applications (ICIMIA)}}. IEEE, \bibinfo{pages}{1153--1160}.
\newblock


\bibitem[\protect\citeauthoryear{Durillo, Zhang, Alba, Harman, and
  Nebro}{Durillo et~al\mbox{.}}{2011}]%
        {durillo2011study}
\bibfield{author}{\bibinfo{person}{Juan~J Durillo}, \bibinfo{person}{Yuanyuan
  Zhang}, \bibinfo{person}{Enrique Alba}, \bibinfo{person}{Mark Harman}, {and}
  \bibinfo{person}{Antonio~J Nebro}.} \bibinfo{year}{2011}\natexlab{}.
\newblock \showarticletitle{A study of the bi-objective next release problem}.
\newblock \bibinfo{journal}{\emph{Empirical Software Engineering}}
  \bibinfo{volume}{16}, \bibinfo{number}{1} (\bibinfo{year}{2011}),
  \bibinfo{pages}{29--60}.
\newblock


\bibitem[\protect\citeauthoryear{Emu, Choudhury, and Salomaa}{Emu
  et~al\mbox{.}}{2022}]%
        {emu2022resource}
\bibfield{author}{\bibinfo{person}{Mahzabeen Emu}, \bibinfo{person}{Salimur
  Choudhury}, {and} \bibinfo{person}{Kai Salomaa}.}
  \bibinfo{year}{2022}\natexlab{}.
\newblock \showarticletitle{Resource optimization of sfc embedding for iot
  networks using quantum computing}. In \bibinfo{booktitle}{\emph{2022 IEEE
  27th International Workshop on Computer Aided Modeling and Design of
  Communication Links and Networks (CAMAD)}}. IEEE, \bibinfo{pages}{83--88}.
\newblock


\bibitem[\protect\citeauthoryear{Emu, Choudhury, and Salomaa}{Emu
  et~al\mbox{.}}{2024}]%
        {emu2024warm}
\bibfield{author}{\bibinfo{person}{Mahzabeen Emu}, \bibinfo{person}{Salimur
  Choudhury}, {and} \bibinfo{person}{Kai Salomaa}.}
  \bibinfo{year}{2024}\natexlab{}.
\newblock \showarticletitle{Warm and cold start quantum annealing for metaverse
  resource optimization}.
\newblock \bibinfo{journal}{\emph{IEEE Open Journal of the Communications
  Society}}  \bibinfo{volume}{5} (\bibinfo{year}{2024}),
  \bibinfo{pages}{1057--1071}.
\newblock


\bibitem[\protect\citeauthoryear{Fankhauser, Sol{\`e}r, F{\"u}chslin, and
  Stockinger}{Fankhauser et~al\mbox{.}}{2023}]%
        {fankhauser2023multiple}
\bibfield{author}{\bibinfo{person}{Tobias Fankhauser}, \bibinfo{person}{Marc~E
  Sol{\`e}r}, \bibinfo{person}{Rudolf~Marcel F{\"u}chslin}, {and}
  \bibinfo{person}{Kurt Stockinger}.} \bibinfo{year}{2023}\natexlab{}.
\newblock \showarticletitle{Multiple query optimization using a gate-based
  quantum computer}.
\newblock \bibinfo{journal}{\emph{IEEE Access}}  \bibinfo{volume}{11}
  (\bibinfo{year}{2023}), \bibinfo{pages}{114031--114043}.
\newblock


\bibitem[\protect\citeauthoryear{Farhi, Goldstone, and Gutmann}{Farhi
  et~al\mbox{.}}{2014}]%
        {farhi2014quantum}
\bibfield{author}{\bibinfo{person}{Edward Farhi}, \bibinfo{person}{Jeffrey
  Goldstone}, {and} \bibinfo{person}{Sam Gutmann}.}
  \bibinfo{year}{2014}\natexlab{}.
\newblock \showarticletitle{A quantum approximate optimization algorithm}.
\newblock \bibinfo{journal}{\emph{arXiv preprint arXiv:1411.4028}}
  (\bibinfo{year}{2014}).
\newblock


\bibitem[\protect\citeauthoryear{Gan and Li}{Gan and Li}{2024}]%
        {gan2024research}
\bibfield{author}{\bibinfo{person}{Weimin Gan} {and} \bibinfo{person}{Jianhui
  Li}.} \bibinfo{year}{2024}\natexlab{}.
\newblock \showarticletitle{Research on Genetic Quantum Particle Swarm
  Optimization Fusion Algorithm in Virtual Machine Scheduling}. In
  \bibinfo{booktitle}{\emph{Proceedings of the 2nd International Conference on
  Educational Knowledge and Informatization}}. \bibinfo{pages}{136--142}.
\newblock


\bibitem[\protect\citeauthoryear{Goemans and Williamson}{Goemans and
  Williamson}{1995}]%
        {goemans1995improved}
\bibfield{author}{\bibinfo{person}{Michel~X Goemans} {and}
  \bibinfo{person}{David~P Williamson}.} \bibinfo{year}{1995}\natexlab{}.
\newblock \showarticletitle{Improved approximation algorithms for maximum cut
  and satisfiability problems using semidefinite programming}.
\newblock \bibinfo{journal}{\emph{Journal of the ACM (JACM)}}
  \bibinfo{volume}{42}, \bibinfo{number}{6} (\bibinfo{year}{1995}),
  \bibinfo{pages}{1115--1145}.
\newblock


\bibitem[\protect\citeauthoryear{Griffiths and Schroeter}{Griffiths and
  Schroeter}{2018}]%
        {griffiths2018introduction}
\bibfield{author}{\bibinfo{person}{David~J Griffiths} {and}
  \bibinfo{person}{Darrell~F Schroeter}.} \bibinfo{year}{2018}\natexlab{}.
\newblock \bibinfo{booktitle}{\emph{Introduction to quantum mechanics}}.
\newblock \bibinfo{publisher}{Cambridge university press}.
\newblock


\bibitem[\protect\citeauthoryear{Groppe and Groppe}{Groppe and Groppe}{2021}]%
        {groppe2021optimizing}
\bibfield{author}{\bibinfo{person}{Sven Groppe} {and} \bibinfo{person}{Jinghua
  Groppe}.} \bibinfo{year}{2021}\natexlab{}.
\newblock \showarticletitle{Optimizing transaction schedules on universal
  quantum computers via code generation for grover’s search algorithm}. In
  \bibinfo{booktitle}{\emph{Proceedings of the 25th International Database
  Engineering \& Applications Symposium}}. \bibinfo{pages}{149--156}.
\newblock


\bibitem[\protect\citeauthoryear{Grover}{Grover}{1996}]%
        {grover1996fast}
\bibfield{author}{\bibinfo{person}{Lov~K Grover}.}
  \bibinfo{year}{1996}\natexlab{}.
\newblock \showarticletitle{A fast quantum mechanical algorithm for database
  search}. In \bibinfo{booktitle}{\emph{Proceedings of the twenty-eighth annual
  ACM symposium on Theory of computing}}. \bibinfo{pages}{212--219}.
\newblock


\bibitem[\protect\citeauthoryear{Guo, Song, and Zhou}{Guo
  et~al\mbox{.}}{2022}]%
        {guo2022synergic}
\bibfield{author}{\bibinfo{person}{Xu Guo}, \bibinfo{person}{Xiaoyu Song},
  {and} \bibinfo{person}{Jian-tao Zhou}.} \bibinfo{year}{2022}\natexlab{}.
\newblock \showarticletitle{A synergic quantum particle swarm optimisation for
  constrained combinatorial test generation}.
\newblock \bibinfo{journal}{\emph{IET software}} \bibinfo{volume}{16},
  \bibinfo{number}{3} (\bibinfo{year}{2022}), \bibinfo{pages}{279--300}.
\newblock


\bibitem[\protect\citeauthoryear{Guo, Song, Zhou, and Wang}{Guo
  et~al\mbox{.}}{2023a}]%
        {guo2023tolerance}
\bibfield{author}{\bibinfo{person}{Xu Guo}, \bibinfo{person}{Xiaoyu Song},
  \bibinfo{person}{Jian-tao Zhou}, {and} \bibinfo{person}{Feiyu Wang}.}
  \bibinfo{year}{2023}\natexlab{a}.
\newblock \showarticletitle{A tolerance-based memetic algorithm for constrained
  covering array generation}.
\newblock \bibinfo{journal}{\emph{Memetic Computing}} \bibinfo{volume}{15},
  \bibinfo{number}{3} (\bibinfo{year}{2023}), \bibinfo{pages}{319--340}.
\newblock


\bibitem[\protect\citeauthoryear{Guo, Song, Zhou, Wang, and Tang}{Guo
  et~al\mbox{.}}{2023b}]%
        {guo2023effective}
\bibfield{author}{\bibinfo{person}{Xu Guo}, \bibinfo{person}{Xiaoyu Song},
  \bibinfo{person}{Jian-tao Zhou}, \bibinfo{person}{Feiyu Wang}, {and}
  \bibinfo{person}{Kecheng Tang}.} \bibinfo{year}{2023}\natexlab{b}.
\newblock \showarticletitle{An Effective Approach to High Strength Covering
  Array Generation in Combinatorial Testing}.
\newblock \bibinfo{journal}{\emph{IEEE Transactions on Software Engineering}}
  \bibinfo{volume}{49}, \bibinfo{number}{10} (\bibinfo{year}{2023}),
  \bibinfo{pages}{4566--4593}.
\newblock


\bibitem[\protect\citeauthoryear{Guo, Song, Zhou, Wang, Tang, and Wang}{Guo
  et~al\mbox{.}}{2023c}]%
        {guo2023memetic}
\bibfield{author}{\bibinfo{person}{Xu Guo}, \bibinfo{person}{Xiaoyu Song},
  \bibinfo{person}{Jian-tao Zhou}, \bibinfo{person}{Feiyu Wang},
  \bibinfo{person}{Kecheng Tang}, {and} \bibinfo{person}{Zhuowei Wang}.}
  \bibinfo{year}{2023}\natexlab{c}.
\newblock \showarticletitle{A memetic algorithm for high-strength covering
  array generation}.
\newblock \bibinfo{journal}{\emph{IET Software}} \bibinfo{volume}{17},
  \bibinfo{number}{4} (\bibinfo{year}{2023}), \bibinfo{pages}{538--553}.
\newblock


\bibitem[\protect\citeauthoryear{Han and Kim}{Han and Kim}{2002}]%
        {han2002quantum}
\bibfield{author}{\bibinfo{person}{Kuk-Hyun Han} {and}
  \bibinfo{person}{Jong-Hwan Kim}.} \bibinfo{year}{2002}\natexlab{}.
\newblock \showarticletitle{Quantum-inspired evolutionary algorithm for a class
  of combinatorial optimization}.
\newblock \bibinfo{journal}{\emph{IEEE transactions on evolutionary
  computation}} \bibinfo{volume}{6}, \bibinfo{number}{6}
  (\bibinfo{year}{2002}), \bibinfo{pages}{580--593}.
\newblock


\bibitem[\protect\citeauthoryear{Harman, Mansouri, and Zhang}{Harman
  et~al\mbox{.}}{2012}]%
        {harman2012search}
\bibfield{author}{\bibinfo{person}{Mark Harman}, \bibinfo{person}{S~Afshin
  Mansouri}, {and} \bibinfo{person}{Yuanyuan Zhang}.}
  \bibinfo{year}{2012}\natexlab{}.
\newblock \showarticletitle{Search-based software engineering: Trends,
  techniques and applications}.
\newblock \bibinfo{journal}{\emph{ACM Computing Surveys (CSUR)}}
  \bibinfo{volume}{45}, \bibinfo{number}{1} (\bibinfo{year}{2012}),
  \bibinfo{pages}{1--61}.
\newblock


\bibitem[\protect\citeauthoryear{Hidary}{Hidary}{2021}]%
        {hidary2021quantum}
\bibfield{author}{\bibinfo{person}{Jack~D Hidary}.}
  \bibinfo{year}{2021}\natexlab{}.
\newblock \bibinfo{booktitle}{\emph{Quantum computing: an applied approach}}.
  Vol.~\bibinfo{volume}{1}.
\newblock \bibinfo{publisher}{Springer}.
\newblock


\bibitem[\protect\citeauthoryear{Hussein, Younes, and Abdelmoez}{Hussein
  et~al\mbox{.}}{2020}]%
        {hussein2020quantum}
\bibfield{author}{\bibinfo{person}{Hager Hussein}, \bibinfo{person}{Ahmed
  Younes}, {and} \bibinfo{person}{Walid Abdelmoez}.}
  \bibinfo{year}{2020}\natexlab{}.
\newblock \showarticletitle{Quantum-inspired genetic algorithm for solving the
  test suite minimization problem}.
\newblock \bibinfo{journal}{\emph{WSEAS Transactions on Computers}}
  \bibinfo{volume}{19} (\bibinfo{year}{2020}).
\newblock


\bibitem[\protect\citeauthoryear{Hussein, Younes, and Abdelmoez}{Hussein
  et~al\mbox{.}}{2021}]%
        {hussein2021quantum}
\bibfield{author}{\bibinfo{person}{Hager Hussein}, \bibinfo{person}{Ahmed
  Younes}, {and} \bibinfo{person}{Walid Abdelmoez}.}
  \bibinfo{year}{2021}\natexlab{}.
\newblock \showarticletitle{Quantum algorithm for solving the test suite
  minimization problem}.
\newblock \bibinfo{journal}{\emph{Cogent Engineering}} \bibinfo{volume}{8},
  \bibinfo{number}{1} (\bibinfo{year}{2021}), \bibinfo{pages}{1882116}.
\newblock


\bibitem[\protect\citeauthoryear{J., Adedoyin, Ambrosiano, Anisimov, Casper,
  Chennupati, Coffrin, Djidjev, Gunter, Karra, Lemons, Lin, Malyzhenkov,
  Mascarenas, Mniszewski, Nadiga, O’malley, Oyen, Pakin, Prasad, Roberts,
  Romero, Santhi, Sinitsyn, Swart, Wendelberger, Yoon, Zamora, Zhu, Eidenbenz,
  B\"{a}rtschi, Coles, Vuffray, and Lokhov}{J. et~al\mbox{.}}{2022}]%
        {Abhijith2022Quantum}
\bibfield{author}{\bibinfo{person}{Abhijith J.}, \bibinfo{person}{Adetokunbo
  Adedoyin}, \bibinfo{person}{John Ambrosiano}, \bibinfo{person}{Petr
  Anisimov}, \bibinfo{person}{William Casper}, \bibinfo{person}{Gopinath
  Chennupati}, \bibinfo{person}{Carleton Coffrin}, \bibinfo{person}{Hristo
  Djidjev}, \bibinfo{person}{David Gunter}, \bibinfo{person}{Satish Karra},
  \bibinfo{person}{Nathan Lemons}, \bibinfo{person}{Shizeng Lin},
  \bibinfo{person}{Alexander Malyzhenkov}, \bibinfo{person}{David Mascarenas},
  \bibinfo{person}{Susan Mniszewski}, \bibinfo{person}{Balu Nadiga},
  \bibinfo{person}{Daniel O’malley}, \bibinfo{person}{Diane Oyen},
  \bibinfo{person}{Scott Pakin}, \bibinfo{person}{Lakshman Prasad},
  \bibinfo{person}{Randy Roberts}, \bibinfo{person}{Phillip Romero},
  \bibinfo{person}{Nandakishore Santhi}, \bibinfo{person}{Nikolai Sinitsyn},
  \bibinfo{person}{Pieter~J. Swart}, \bibinfo{person}{James~G. Wendelberger},
  \bibinfo{person}{Boram Yoon}, \bibinfo{person}{Richard Zamora},
  \bibinfo{person}{Wei Zhu}, \bibinfo{person}{Stephan Eidenbenz},
  \bibinfo{person}{Andreas B\"{a}rtschi}, \bibinfo{person}{Patrick~J. Coles},
  \bibinfo{person}{Marc Vuffray}, {and} \bibinfo{person}{Andrey~Y. Lokhov}.}
  \bibinfo{year}{2022}\natexlab{}.
\newblock \showarticletitle{Quantum Algorithm and Implementations for
  Beginners}.
\newblock \bibinfo{journal}{\emph{ACM Transactions on Quantum Computing}}
  \bibinfo{volume}{3}, \bibinfo{number}{4}, Article \bibinfo{articleno}{18}
  (\bibinfo{date}{July} \bibinfo{year}{2022}), \bibinfo{numpages}{92}~pages.
\newblock
\urldef\tempurl%
\url{https://doi.org/10.1145/3517340}
\showDOI{\tempurl}


\bibitem[\protect\citeauthoryear{Jain and Sharma}{Jain and Sharma}{2023}]%
        {jain2023quantum}
\bibfield{author}{\bibinfo{person}{Richa Jain} {and} \bibinfo{person}{Neelam
  Sharma}.} \bibinfo{year}{2023}\natexlab{}.
\newblock \showarticletitle{A quantum inspired hybrid SSA--GWO algorithm for
  SLA based task scheduling to improve QoS parameter in cloud computing}.
\newblock \bibinfo{journal}{\emph{Cluster Computing}} \bibinfo{volume}{26},
  \bibinfo{number}{6} (\bibinfo{year}{2023}), \bibinfo{pages}{3587--3610}.
\newblock


\bibitem[\protect\citeauthoryear{Jhaveri, Krone-Martins, and Lopes}{Jhaveri
  et~al\mbox{.}}{2023}]%
        {jhaveri2023cloning}
\bibfield{author}{\bibinfo{person}{Samyak Jhaveri}, \bibinfo{person}{Alberto
  Krone-Martins}, {and} \bibinfo{person}{Cristina~V Lopes}.}
  \bibinfo{year}{2023}\natexlab{}.
\newblock \showarticletitle{Cloning and beyond: A quantum solution to duplicate
  code}. In \bibinfo{booktitle}{\emph{Proceedings of the 2023 ACM SIGPLAN
  International Symposium on New Ideas, New Paradigms, and Reflections on
  Programming and Software}}. \bibinfo{pages}{32--49}.
\newblock


\bibitem[\protect\citeauthoryear{Jin}{Jin}{2021}]%
        {jin2021cross}
\bibfield{author}{\bibinfo{person}{Cong Jin}.} \bibinfo{year}{2021}\natexlab{}.
\newblock \showarticletitle{Cross-project software defect prediction based on
  domain adaptation learning and optimization}.
\newblock \bibinfo{journal}{\emph{Expert Systems with Applications}}
  \bibinfo{volume}{171} (\bibinfo{year}{2021}), \bibinfo{pages}{114637}.
\newblock


\bibitem[\protect\citeauthoryear{Jin and Jin}{Jin and Jin}{2015}]%
        {jin2015prediction}
\bibfield{author}{\bibinfo{person}{Cong Jin} {and} \bibinfo{person}{Shu-Wei
  Jin}.} \bibinfo{year}{2015}\natexlab{}.
\newblock \showarticletitle{Prediction approach of software fault-proneness
  based on hybrid artificial neural network and quantum particle swarm
  optimization}.
\newblock \bibinfo{journal}{\emph{Applied Soft Computing}}
  \bibinfo{volume}{35} (\bibinfo{year}{2015}), \bibinfo{pages}{717--725}.
\newblock


\bibitem[\protect\citeauthoryear{Jin and Jin}{Jin and Jin}{2016}]%
        {jin2016parameter}
\bibfield{author}{\bibinfo{person}{Cong Jin} {and} \bibinfo{person}{Shu-Wei
  Jin}.} \bibinfo{year}{2016}\natexlab{}.
\newblock \showarticletitle{Parameter optimization of software reliability
  growth model with S-shaped testing-effort function using improved swarm
  intelligent optimization}.
\newblock \bibinfo{journal}{\emph{Applied Soft Computing}}
  \bibinfo{volume}{40} (\bibinfo{year}{2016}), \bibinfo{pages}{283--291}.
\newblock


\bibitem[\protect\citeauthoryear{Kiruthiga and Vennila}{Kiruthiga and
  Vennila}{2019}]%
        {kiruthiga2019enriched}
\bibfield{author}{\bibinfo{person}{G Kiruthiga} {and} \bibinfo{person}{S~Mary
  Vennila}.} \bibinfo{year}{2019}\natexlab{}.
\newblock \showarticletitle{An enriched chaotic quantum whale optimization
  algorithm based job scheduling in cloud computing environment}.
\newblock \bibinfo{journal}{\emph{International Journal of Advanced Trends in
  Computer Science and Engineering}} \bibinfo{volume}{6}, \bibinfo{number}{4}
  (\bibinfo{year}{2019}), \bibinfo{pages}{1753--1760}.
\newblock


\bibitem[\protect\citeauthoryear{Krishna and Saha}{Krishna and Saha}{2023}]%
        {krishna2023optimal}
\bibfield{author}{\bibinfo{person}{Gopal Krishna} {and}
  \bibinfo{person}{Anish~Kumar Saha}.} \bibinfo{year}{2023}\natexlab{}.
\newblock \showarticletitle{Optimal sensor spacing in IoT network based on
  quantum computing technology}.
\newblock \bibinfo{journal}{\emph{International Journal of Parallel, Emergent
  and Distributed Systems}} \bibinfo{volume}{38}, \bibinfo{number}{1}
  (\bibinfo{year}{2023}), \bibinfo{pages}{58--84}.
\newblock


\bibitem[\protect\citeauthoryear{Kumar~Jaiswal and Prajapati}{Kumar~Jaiswal and
  Prajapati}{2025}]%
        {kumar2025optimizing}
\bibfield{author}{\bibinfo{person}{Updesh Kumar~Jaiswal} {and}
  \bibinfo{person}{Amarjeet Prajapati}.} \bibinfo{year}{2025}\natexlab{}.
\newblock \showarticletitle{Optimizing the software test case through
  physics-informed particle-based method}.
\newblock \bibinfo{journal}{\emph{International Journal of System Assurance
  Engineering and Management}} (\bibinfo{year}{2025}), \bibinfo{pages}{1--18}.
\newblock


\bibitem[\protect\citeauthoryear{Kumari and Srinivas}{Kumari and
  Srinivas}{2016}]%
        {kumari2016comparing}
\bibfield{author}{\bibinfo{person}{A~Charan Kumari} {and} \bibinfo{person}{K
  Srinivas}.} \bibinfo{year}{2016}\natexlab{}.
\newblock \showarticletitle{Comparing the performance of olutionary algorithms
  for the solution of software requirements selection problem}.
\newblock \bibinfo{journal}{\emph{Information and Software Technology}}
  \bibinfo{volume}{76} (\bibinfo{year}{2016}), \bibinfo{pages}{31--64}.
\newblock


\bibitem[\protect\citeauthoryear{LaRose, Mari, Kaiser, Karalekas, Alves,
  Czarnik, El~Mandouh, Gordon, Hindy, Robertson, et~al\mbox{.}}{LaRose
  et~al\mbox{.}}{2022}]%
        {larose2022mitiq}
\bibfield{author}{\bibinfo{person}{Ryan LaRose}, \bibinfo{person}{Andrea Mari},
  \bibinfo{person}{Sarah Kaiser}, \bibinfo{person}{Peter~J Karalekas},
  \bibinfo{person}{Andre~A Alves}, \bibinfo{person}{Piotr Czarnik},
  \bibinfo{person}{Mohamed El~Mandouh}, \bibinfo{person}{Max~H Gordon},
  \bibinfo{person}{Yousef Hindy}, \bibinfo{person}{Aaron Robertson},
  {et~al\mbox{.}}} \bibinfo{year}{2022}\natexlab{}.
\newblock \showarticletitle{Mitiq: A software package for error mitigation on
  noisy quantum computers}.
\newblock \bibinfo{journal}{\emph{Quantum}}  \bibinfo{volume}{6}
  (\bibinfo{year}{2022}), \bibinfo{pages}{774}.
\newblock


\bibitem[\protect\citeauthoryear{Li, Liu, and Xu}{Li et~al\mbox{.}}{2019}]%
        {li2019quantum}
\bibfield{author}{\bibinfo{person}{Fei Li}, \bibinfo{person}{Min Liu}, {and}
  \bibinfo{person}{Gaowei Xu}.} \bibinfo{year}{2019}\natexlab{}.
\newblock \showarticletitle{A quantum ant colony multi-objective routing
  algorithm in WSN and its application in a manufacturing environment}.
\newblock \bibinfo{journal}{\emph{Sensors}} \bibinfo{volume}{19},
  \bibinfo{number}{15} (\bibinfo{year}{2019}), \bibinfo{pages}{3334}.
\newblock


\bibitem[\protect\citeauthoryear{Liu, Zhou, Zhou, Luo, and Wei}{Liu
  et~al\mbox{.}}{2024}]%
        {liu2024quantum}
\bibfield{author}{\bibinfo{person}{Bo Liu}, \bibinfo{person}{Guo Zhou},
  \bibinfo{person}{Yongquan Zhou}, \bibinfo{person}{Qifang Luo}, {and}
  \bibinfo{person}{Yuanfei Wei}.} \bibinfo{year}{2024}\natexlab{}.
\newblock \showarticletitle{Quantum-inspired multi-objective African vultures
  optimization algorithm with hierarchical structure for software requirement}.
\newblock \bibinfo{journal}{\emph{Cluster Computing}} \bibinfo{volume}{27},
  \bibinfo{number}{8} (\bibinfo{year}{2024}), \bibinfo{pages}{11317--11345}.
\newblock


\bibitem[\protect\citeauthoryear{Liu, Li, Zhang, Xu, Xiao, and Zhou}{Liu
  et~al\mbox{.}}{2022}]%
        {liu2022hpcp}
\bibfield{author}{\bibinfo{person}{Yang Liu}, \bibinfo{person}{Chaoqun Li},
  \bibinfo{person}{Yao Zhang}, \bibinfo{person}{Mengying Xu},
  \bibinfo{person}{Jing Xiao}, {and} \bibinfo{person}{Jie Zhou}.}
  \bibinfo{year}{2022}\natexlab{}.
\newblock \showarticletitle{HPCP-QCWOA: High performance clustering protocol
  based on quantum clone whale optimization algorithm in integrated energy
  system}.
\newblock \bibinfo{journal}{\emph{Future Generation Computer Systems}}
  \bibinfo{volume}{135} (\bibinfo{year}{2022}), \bibinfo{pages}{315--332}.
\newblock


\bibitem[\protect\citeauthoryear{Lou, Jiang, Shen, and Wang}{Lou
  et~al\mbox{.}}{2018}]%
        {lou2018failure}
\bibfield{author}{\bibinfo{person}{Jungang Lou}, \bibinfo{person}{Yunliang
  Jiang}, \bibinfo{person}{Qing Shen}, {and} \bibinfo{person}{Ruiqin Wang}.}
  \bibinfo{year}{2018}\natexlab{}.
\newblock \showarticletitle{Failure prediction by relevance vector regression
  with improved quantum-inspired gravitational search}.
\newblock \bibinfo{journal}{\emph{Journal of Network and Computer
  Applications}}  \bibinfo{volume}{103} (\bibinfo{year}{2018}),
  \bibinfo{pages}{171--177}.
\newblock


\bibitem[\protect\citeauthoryear{Lubinski, Coffrin, McGeoch, Sathe,
  Apanavicius, Bernal~Neira, Consortium, et~al\mbox{.}}{Lubinski
  et~al\mbox{.}}{2024}]%
        {lubinski2024optimization}
\bibfield{author}{\bibinfo{person}{Thomas Lubinski}, \bibinfo{person}{Carleton
  Coffrin}, \bibinfo{person}{Catherine McGeoch}, \bibinfo{person}{Pratik
  Sathe}, \bibinfo{person}{Joshua Apanavicius}, \bibinfo{person}{David
  Bernal~Neira}, \bibinfo{person}{Quantum Economic~Development Consortium},
  {et~al\mbox{.}}} \bibinfo{year}{2024}\natexlab{}.
\newblock \showarticletitle{Optimization applications as quantum performance
  benchmarks}.
\newblock \bibinfo{journal}{\emph{ACM Transactions on Quantum Computing}}
  \bibinfo{volume}{5}, \bibinfo{number}{3} (\bibinfo{year}{2024}),
  \bibinfo{pages}{1--44}.
\newblock


\bibitem[\protect\citeauthoryear{Mandal, Nadim, Roy, Roy, and Schneider}{Mandal
  et~al\mbox{.}}{2024}]%
        {mandal2024evaluating}
\bibfield{author}{\bibinfo{person}{Ashis~Kumar Mandal}, \bibinfo{person}{Md
  Nadim}, \bibinfo{person}{Chanchal~K Roy}, \bibinfo{person}{Banani Roy}, {and}
  \bibinfo{person}{Kevin~A Schneider}.} \bibinfo{year}{2024}\natexlab{}.
\newblock \showarticletitle{Evaluating the Performance of a D-Wave Quantum
  Annealing System for Feature Subset Selection in Software Defect Prediction}.
  In \bibinfo{booktitle}{\emph{2024 IEEE International Conference on Quantum
  Computing and Engineering (QCE)}}, Vol.~\bibinfo{volume}{2}. IEEE,
  \bibinfo{pages}{103--108}.
\newblock


\bibitem[\protect\citeauthoryear{Mariani and Vergilio}{Mariani and
  Vergilio}{2017}]%
        {mariani2017systematic}
\bibfield{author}{\bibinfo{person}{Thain{\'a} Mariani} {and}
  \bibinfo{person}{Silvia~Regina Vergilio}.} \bibinfo{year}{2017}\natexlab{}.
\newblock \showarticletitle{A systematic review on search-based refactoring}.
\newblock \bibinfo{journal}{\emph{Information and Software Technology}}
  \bibinfo{volume}{83} (\bibinfo{year}{2017}), \bibinfo{pages}{14--34}.
\newblock


\bibitem[\protect\citeauthoryear{Miranskyy}{Miranskyy}{2022}]%
        {miranskyy2022using}
\bibfield{author}{\bibinfo{person}{Andriy Miranskyy}.}
  \bibinfo{year}{2022}\natexlab{}.
\newblock \showarticletitle{Using quantum computers to speed up dynamic testing
  of software}. In \bibinfo{booktitle}{\emph{Proceedings of the 1st
  International Workshop on Quantum Programming for Software Engineering}}.
  \bibinfo{pages}{26--31}.
\newblock


\bibitem[\protect\citeauthoryear{Muniswamy and Vignesh}{Muniswamy and
  Vignesh}{2024}]%
        {muniswamy2024joint}
\bibfield{author}{\bibinfo{person}{Saravanan Muniswamy} {and}
  \bibinfo{person}{Radhakrishnan Vignesh}.} \bibinfo{year}{2024}\natexlab{}.
\newblock \showarticletitle{Joint optimization of load balancing and resource
  allocation in cloud environment using optimal container management strategy}.
\newblock \bibinfo{journal}{\emph{Concurrency and Computation: Practice and
  Experience}} \bibinfo{volume}{36}, \bibinfo{number}{12}
  (\bibinfo{year}{2024}), \bibinfo{pages}{e8035}.
\newblock


\bibitem[\protect\citeauthoryear{Muqeet, Yue, Ali, and Arcaini}{Muqeet
  et~al\mbox{.}}{2024}]%
        {muqeet2024mitigating}
\bibfield{author}{\bibinfo{person}{Asmar Muqeet}, \bibinfo{person}{Tao Yue},
  \bibinfo{person}{Shaukat Ali}, {and} \bibinfo{person}{Paolo Arcaini}.}
  \bibinfo{year}{2024}\natexlab{}.
\newblock \showarticletitle{Mitigating noise in quantum software testing using
  machine learning}.
\newblock \bibinfo{journal}{\emph{IEEE Transactions on Software Engineering}}
  (\bibinfo{year}{2024}).
\newblock


\bibitem[\protect\citeauthoryear{Murillo, Garcia-Alonso, Moguel, Barzen,
  Leymann, Ali, Yue, Arcaini, P{\'e}rez-Castillo, Garc{\'\i}a-Rodr{\'\i}guez~de
  Guzm{\'a}n, et~al\mbox{.}}{Murillo et~al\mbox{.}}{2025}]%
        {murillo2025quantum}
\bibfield{author}{\bibinfo{person}{Juan~Manuel Murillo}, \bibinfo{person}{Jose
  Garcia-Alonso}, \bibinfo{person}{Enrique Moguel}, \bibinfo{person}{Johanna
  Barzen}, \bibinfo{person}{Frank Leymann}, \bibinfo{person}{Shaukat Ali},
  \bibinfo{person}{Tao Yue}, \bibinfo{person}{Paolo Arcaini},
  \bibinfo{person}{Ricardo P{\'e}rez-Castillo}, \bibinfo{person}{Ignacio
  Garc{\'\i}a-Rodr{\'\i}guez~de Guzm{\'a}n}, {et~al\mbox{.}}}
  \bibinfo{year}{2025}\natexlab{}.
\newblock \showarticletitle{Quantum software engineering: Roadmap and
  challenges ahead}.
\newblock \bibinfo{journal}{\emph{ACM Transactions on Software Engineering and
  Methodology}} \bibinfo{volume}{34}, \bibinfo{number}{5}
  (\bibinfo{year}{2025}), \bibinfo{pages}{1--48}.
\newblock


\bibitem[\protect\citeauthoryear{Naik, Priyanka, and Ansari}{Naik
  et~al\mbox{.}}{2025}]%
        {naik2025energy}
\bibfield{author}{\bibinfo{person}{Banavath~Balaji Naik},
  \bibinfo{person}{Bollu Priyanka}, {and} \bibinfo{person}{Sarfaraj~Alam
  Ansari}.} \bibinfo{year}{2025}\natexlab{}.
\newblock \showarticletitle{Energy-efficient task offloading and efficient
  resource allocation for edge computing: a quantum inspired particle swarm
  optimization approach}.
\newblock \bibinfo{journal}{\emph{Cluster Computing}} \bibinfo{volume}{28},
  \bibinfo{number}{3} (\bibinfo{year}{2025}), \bibinfo{pages}{155}.
\newblock


\bibitem[\protect\citeauthoryear{Nayak, Rehfeld, Winker, Warnke,
  {\c{C}}alikyilmaz, and Groppe}{Nayak et~al\mbox{.}}{2023}]%
        {nayak2023constructing}
\bibfield{author}{\bibinfo{person}{Nitin Nayak}, \bibinfo{person}{Jan Rehfeld},
  \bibinfo{person}{Tobias Winker}, \bibinfo{person}{Benjamin Warnke},
  \bibinfo{person}{Umut {\c{C}}alikyilmaz}, {and} \bibinfo{person}{Sven
  Groppe}.} \bibinfo{year}{2023}\natexlab{}.
\newblock \showarticletitle{Constructing optimal bushy join trees by solving
  qubo problems on quantum hardware and simulators}. In
  \bibinfo{booktitle}{\emph{Proceedings of the international workshop on big
  data in emergent distributed environments}}. \bibinfo{pages}{1--7}.
\newblock


\bibitem[\protect\citeauthoryear{Nielsen and Chuang}{Nielsen and
  Chuang}{2010}]%
        {nielsen2010quantum}
\bibfield{author}{\bibinfo{person}{Michael~A Nielsen} {and}
  \bibinfo{person}{Isaac~L Chuang}.} \bibinfo{year}{2010}\natexlab{}.
\newblock \bibinfo{booktitle}{\emph{Quantum computation and quantum
  information}}.
\newblock \bibinfo{publisher}{Cambridge university press}.
\newblock


\bibitem[\protect\citeauthoryear{Niu, Li, Deng, Sanderson, and Ren}{Niu
  et~al\mbox{.}}{2024}]%
        {niu2024performance}
\bibfield{author}{\bibinfo{person}{Jiayang Niu}, \bibinfo{person}{Jie Li},
  \bibinfo{person}{Ke Deng}, \bibinfo{person}{Mark Sanderson}, {and}
  \bibinfo{person}{Yongli Ren}.} \bibinfo{year}{2024}\natexlab{}.
\newblock \showarticletitle{Performance-Driven QUBO for Recommender Systems on
  Quantum Annealers}.
\newblock \bibinfo{journal}{\emph{arXiv preprint arXiv:2410.15272}}
  (\bibinfo{year}{2024}).
\newblock


\bibitem[\protect\citeauthoryear{Ou, Li, Chen, Wu, Tsai, Yan, and Chang}{Ou
  et~al\mbox{.}}{2023}]%
        {ou2023quantum}
\bibfield{author}{\bibinfo{person}{Chia-Ho Ou}, \bibinfo{person}{Yu-Hong Li},
  \bibinfo{person}{Chih-Yu Chen}, \bibinfo{person}{Chi-Hsuan Wu},
  \bibinfo{person}{Yu-Chen Tsai}, \bibinfo{person}{Zhi-You Yan}, {and}
  \bibinfo{person}{Ching-Ray Chang}.} \bibinfo{year}{2023}\natexlab{}.
\newblock \showarticletitle{Quantum-inspired optimization for task scheduling
  in software development projects}. In \bibinfo{booktitle}{\emph{2023 IEEE
  International Conference on Quantum Computing and Engineering (QCE)}},
  Vol.~\bibinfo{volume}{2}. IEEE, \bibinfo{pages}{348--349}.
\newblock


\bibitem[\protect\citeauthoryear{Pathak, Mani, and Chatterjee}{Pathak
  et~al\mbox{.}}{2023}]%
        {pathak2023opposition}
\bibfield{author}{\bibinfo{person}{Sanjai Pathak}, \bibinfo{person}{Ashish
  Mani}, {and} \bibinfo{person}{Amlan Chatterjee}.}
  \bibinfo{year}{2023}\natexlab{}.
\newblock \showarticletitle{An Opposition Learning-based Quantum Inspired Salp
  Swarm Optimization for The Multiobjective Controller Placement Problem}. In
  \bibinfo{booktitle}{\emph{2023 10th IEEE Uttar Pradesh Section International
  Conference on Electrical, Electronics and Computer Engineering (UPCON)}},
  Vol.~\bibinfo{volume}{10}. IEEE, \bibinfo{pages}{761--768}.
\newblock


\bibitem[\protect\citeauthoryear{Pathak, Mani, and Chatterjee}{Pathak
  et~al\mbox{.}}{2024}]%
        {pathak2024cooperative}
\bibfield{author}{\bibinfo{person}{Sanjai Pathak}, \bibinfo{person}{Ashish
  Mani}, {and} \bibinfo{person}{Amlan Chatterjee}.}
  \bibinfo{year}{2024}\natexlab{}.
\newblock \showarticletitle{A Cooperative Hybrid Quantum-inspired Salp Swarm
  and Differential Evolution for solving CEC 2022 Benchmark and Controller
  Placement Problems in Software Defined Networks}.
\newblock \bibinfo{journal}{\emph{IEEE Access}} (\bibinfo{year}{2024}).
\newblock


\bibitem[\protect\citeauthoryear{Pezz{\`e}, Abrah{\~a}o, Penzenstadler,
  Poshyvanyk, Roychoudhury, and Yue}{Pezz{\`e} et~al\mbox{.}}{2025}]%
        {pezze20252030}
\bibfield{author}{\bibinfo{person}{Mauro Pezz{\`e}}, \bibinfo{person}{Silvia
  Abrah{\~a}o}, \bibinfo{person}{Birgit Penzenstadler}, \bibinfo{person}{Denys
  Poshyvanyk}, \bibinfo{person}{Abhik Roychoudhury}, {and} \bibinfo{person}{Tao
  Yue}.} \bibinfo{year}{2025}\natexlab{}.
\newblock \showarticletitle{A 2030 Roadmap for Software Engineering}.
\newblock \bibinfo{journal}{\emph{ACM Transactions on Software Engineering and
  Methodology}} \bibinfo{volume}{34}, \bibinfo{number}{5}
  (\bibinfo{year}{2025}), \bibinfo{pages}{1--55}.
\newblock


\bibitem[\protect\citeauthoryear{Qiao, Liu, Lu, Li, Yan, and Yao}{Qiao
  et~al\mbox{.}}{2020}]%
        {qiao2020novel}
\bibfield{author}{\bibinfo{person}{Wenxin Qiao}, \bibinfo{person}{Yicen Liu},
  \bibinfo{person}{Yu Lu}, \bibinfo{person}{Xi Li}, \bibinfo{person}{Jie Yan},
  {and} \bibinfo{person}{Zhigang Yao}.} \bibinfo{year}{2020}\natexlab{}.
\newblock \showarticletitle{A novel approach for service function chain
  embedding in cloud datacenter networks}.
\newblock \bibinfo{journal}{\emph{IEEE Communications Letters}}
  \bibinfo{volume}{25}, \bibinfo{number}{4} (\bibinfo{year}{2020}),
  \bibinfo{pages}{1134--1138}.
\newblock


\bibitem[\protect\citeauthoryear{Qiao, Lu, Lu, Meng, and Liu}{Qiao
  et~al\mbox{.}}{2021}]%
        {fi13100260}
\bibfield{author}{\bibinfo{person}{Wenxin Qiao}, \bibinfo{person}{Hao Lu},
  \bibinfo{person}{Yu Lu}, \bibinfo{person}{Lijie Meng}, {and}
  \bibinfo{person}{Yicen Liu}.} \bibinfo{year}{2021}\natexlab{}.
\newblock \showarticletitle{A Dynamic Service Reconfiguration Method for
  Satellite–Terrestrial Integrated Networks}.
\newblock \bibinfo{journal}{\emph{Future Internet}} \bibinfo{volume}{13},
  \bibinfo{number}{10} (\bibinfo{year}{2021}).
\newblock
\showISSN{1999-5903}
\urldef\tempurl%
\url{https://doi.org/10.3390/fi13100260}
\showDOI{\tempurl}


\bibitem[\protect\citeauthoryear{Rani, Babbar, Kaur, and Ali~Khan}{Rani
  et~al\mbox{.}}{2023}]%
        {rani2023novel}
\bibfield{author}{\bibinfo{person}{Shalli Rani}, \bibinfo{person}{Himanshi
  Babbar}, \bibinfo{person}{Pardeep Kaur}, {and} \bibinfo{person}{Asif
  Ali~Khan}.} \bibinfo{year}{2023}\natexlab{}.
\newblock \showarticletitle{A novel approach of localization with single mobile
  anchor using quantum-based Salp swarm algorithm in wireless sensor networks}.
\newblock \bibinfo{journal}{\emph{Soft Computing}} (\bibinfo{year}{2023}),
  \bibinfo{pages}{1--15}.
\newblock


\bibitem[\protect\citeauthoryear{Ren, Sui, Cheng, Feng, and Zhao}{Ren
  et~al\mbox{.}}{2024}]%
        {ren2024dynamic}
\bibfield{author}{\bibinfo{person}{Jiawei Ren}, \bibinfo{person}{Yulei Sui},
  \bibinfo{person}{Xiao Cheng}, \bibinfo{person}{Yuan Feng}, {and}
  \bibinfo{person}{Jianjun Zhao}.} \bibinfo{year}{2024}\natexlab{}.
\newblock \showarticletitle{Dynamic Transitive Closure-Based Static Analysis
  through the Lens of Quantum Search}.
\newblock \bibinfo{journal}{\emph{ACM Transactions on Software Engineering and
  Methodology}} \bibinfo{volume}{33}, \bibinfo{number}{5}
  (\bibinfo{year}{2024}), \bibinfo{pages}{1--29}.
\newblock


\bibitem[\protect\citeauthoryear{Roy, Pachuau, Singh, and Saha}{Roy
  et~al\mbox{.}}{2024}]%
        {roy2024quantum}
\bibfield{author}{\bibinfo{person}{Arnab Roy}, \bibinfo{person}{Joseph~L
  Pachuau}, \bibinfo{person}{Nongmeikapam~Brajabidhu Singh}, {and}
  \bibinfo{person}{Anish~Kumar Saha}.} \bibinfo{year}{2024}\natexlab{}.
\newblock \showarticletitle{Quantum inspired genetic algorithm and optimization
  of queuing delay}. In \bibinfo{booktitle}{\emph{TENCON 2024-2024 IEEE Region
  10 Conference (TENCON)}}. IEEE, \bibinfo{pages}{1929--1933}.
\newblock


\bibitem[\protect\citeauthoryear{Sato and Katsube}{Sato and Katsube}{2025}]%
        {sato2025bug}
\bibfield{author}{\bibinfo{person}{Naoto Sato} {and} \bibinfo{person}{Ryota
  Katsube}.} \bibinfo{year}{2025}\natexlab{}.
\newblock \showarticletitle{Bug-locating Method based on Statistical Testing
  for Quantum Programs}.
\newblock \bibinfo{journal}{\emph{IEEE Transactions on Software Engineering}}
  (\bibinfo{year}{2025}).
\newblock


\bibitem[\protect\citeauthoryear{Saxena, Sabek, and Spedalieri}{Saxena
  et~al\mbox{.}}{2024}]%
        {saxena2024constrained}
\bibfield{author}{\bibinfo{person}{Pranshi Saxena}, \bibinfo{person}{Ibrahim
  Sabek}, {and} \bibinfo{person}{Federico Spedalieri}.}
  \bibinfo{year}{2024}\natexlab{}.
\newblock \showarticletitle{Constrained quadratic model for optimizing join
  orders}. In \bibinfo{booktitle}{\emph{Proceedings of the 1st Workshop on
  Quantum Computing and Quantum-Inspired Technology for Data-Intensive Systems
  and Applications}}. \bibinfo{pages}{38--44}.
\newblock


\bibitem[\protect\citeauthoryear{Sch{\"o}nberger, Scherzinger, and
  Mauerer}{Sch{\"o}nberger et~al\mbox{.}}{2023a}]%
        {schonberger2023ready}
\bibfield{author}{\bibinfo{person}{Manuel Sch{\"o}nberger},
  \bibinfo{person}{Stefanie Scherzinger}, {and} \bibinfo{person}{Wolfgang
  Mauerer}.} \bibinfo{year}{2023}\natexlab{a}.
\newblock \showarticletitle{Ready to leap (by co-design)? join order
  optimisation on quantum hardware}.
\newblock \bibinfo{journal}{\emph{Proceedings of the ACM on Management of
  Data}} \bibinfo{volume}{1}, \bibinfo{number}{1} (\bibinfo{year}{2023}),
  \bibinfo{pages}{1--27}.
\newblock


\bibitem[\protect\citeauthoryear{Sch{\"o}nberger, Trummer, and
  Mauerer}{Sch{\"o}nberger et~al\mbox{.}}{2023b}]%
        {schonberger2023digitalannealing}
\bibfield{author}{\bibinfo{person}{Manuel Sch{\"o}nberger},
  \bibinfo{person}{Immanuel Trummer}, {and} \bibinfo{person}{Wolfgang
  Mauerer}.} \bibinfo{year}{2023}\natexlab{b}.
\newblock \showarticletitle{Quantum-inspired digital annealing for join
  ordering}.
\newblock \bibinfo{journal}{\emph{Proceedings of the VLDB Endowment}}
  \bibinfo{volume}{17}, \bibinfo{number}{3} (\bibinfo{year}{2023}),
  \bibinfo{pages}{511--524}.
\newblock


\bibitem[\protect\citeauthoryear{Sch{\"o}nberger, Trummer, and
  Mauerer}{Sch{\"o}nberger et~al\mbox{.}}{2023c}]%
        {schonberger2023quantum}
\bibfield{author}{\bibinfo{person}{Manuel Sch{\"o}nberger},
  \bibinfo{person}{Immanuel Trummer}, {and} \bibinfo{person}{Wolfgang
  Mauerer}.} \bibinfo{year}{2023}\natexlab{c}.
\newblock \showarticletitle{Quantum Optimisation of General Join Trees.}. In
  \bibinfo{booktitle}{\emph{VLDB Workshops}}.
\newblock


\bibitem[\protect\citeauthoryear{Serrano, S{\'a}nchez, Santos-Olmo,
  Garc{\'\i}a-Rosado, Blanco, Barletta, Caivano, and
  Fern{\'a}ndez-Medina}{Serrano et~al\mbox{.}}{2024}]%
        {serrano2024minimizing}
\bibfield{author}{\bibinfo{person}{Manuel~A Serrano}, \bibinfo{person}{Luis~E
  S{\'a}nchez}, \bibinfo{person}{Antonio Santos-Olmo}, \bibinfo{person}{David
  Garc{\'\i}a-Rosado}, \bibinfo{person}{Carlos Blanco},
  \bibinfo{person}{Vita~Santa Barletta}, \bibinfo{person}{Danilo Caivano},
  {and} \bibinfo{person}{Eduardo Fern{\'a}ndez-Medina}.}
  \bibinfo{year}{2024}\natexlab{}.
\newblock \showarticletitle{Minimizing incident response time in real-world
  scenarios using quantum computing}.
\newblock \bibinfo{journal}{\emph{Software Quality Journal}}
  \bibinfo{volume}{32}, \bibinfo{number}{1} (\bibinfo{year}{2024}),
  \bibinfo{pages}{163--192}.
\newblock


\bibitem[\protect\citeauthoryear{Shen, Lou, Zhang, and Jiang}{Shen
  et~al\mbox{.}}{2021}]%
        {shen2021failure}
\bibfield{author}{\bibinfo{person}{Qing Shen}, \bibinfo{person}{Jungang Lou},
  \bibinfo{person}{Xiongtao Zhang}, {and} \bibinfo{person}{Yunliang Jiang}.}
  \bibinfo{year}{2021}\natexlab{}.
\newblock \showarticletitle{Failure prediction by regularized fuzzy learning
  with intelligent parameters selection}.
\newblock \bibinfo{journal}{\emph{Applied Soft Computing}}
  \bibinfo{volume}{100} (\bibinfo{year}{2021}), \bibinfo{pages}{106952}.
\newblock


\bibitem[\protect\citeauthoryear{Shi, Gong, Ma, and Jiao}{Shi
  et~al\mbox{.}}{2014}]%
        {shi2014multipopulation}
\bibfield{author}{\bibinfo{person}{Jiao Shi}, \bibinfo{person}{Maoguo Gong},
  \bibinfo{person}{Wenping Ma}, {and} \bibinfo{person}{Licheng Jiao}.}
  \bibinfo{year}{2014}\natexlab{}.
\newblock \showarticletitle{A multipopulation coevolutionary strategy for
  multiobjective immune algorithm}.
\newblock \bibinfo{journal}{\emph{The Scientific World Journal}}
  \bibinfo{volume}{2014}, \bibinfo{number}{1} (\bibinfo{year}{2014}),
  \bibinfo{pages}{539128}.
\newblock


\bibitem[\protect\citeauthoryear{Sun, Feng, and Xu}{Sun et~al\mbox{.}}{2004}]%
        {sun2004particle}
\bibfield{author}{\bibinfo{person}{Jun Sun}, \bibinfo{person}{Bin Feng}, {and}
  \bibinfo{person}{Wenbo Xu}.} \bibinfo{year}{2004}\natexlab{}.
\newblock \showarticletitle{Particle swarm optimization with particles having
  quantum behavior}. In \bibinfo{booktitle}{\emph{Proceedings of the 2004
  congress on evolutionary computation (IEEE Cat. No. 04TH8753)}},
  Vol.~\bibinfo{volume}{1}. IEEE, \bibinfo{pages}{325--331}.
\newblock


\bibitem[\protect\citeauthoryear{Tong, Yang, Zhao, and Liu}{Tong
  et~al\mbox{.}}{2023}]%
        {tong2023can}
\bibfield{author}{\bibinfo{person}{Lian Tong}, \bibinfo{person}{Lan Yang},
  \bibinfo{person}{Xin Zhao}, {and} \bibinfo{person}{Li Liu}.}
  \bibinfo{year}{2023}\natexlab{}.
\newblock \showarticletitle{How can a hybrid quantum-inspired gravitational
  search algorithm decrease energy consumption in IoT-based software-defined
  networks?}
\newblock \bibinfo{journal}{\emph{Sustainable Computing: Informatics and
  Systems}}  \bibinfo{volume}{40} (\bibinfo{year}{2023}),
  \bibinfo{pages}{100920}.
\newblock


\bibitem[\protect\citeauthoryear{Trovato, De~Stefano, Pecorelli, Di~Nucci, and
  De~Lucia}{Trovato et~al\mbox{.}}{2025}]%
        {trovato2025reformulating}
\bibfield{author}{\bibinfo{person}{Antonio Trovato}, \bibinfo{person}{Manuel
  De~Stefano}, \bibinfo{person}{Fabiano Pecorelli}, \bibinfo{person}{Dario
  Di~Nucci}, {and} \bibinfo{person}{Andrea De~Lucia}.}
  \bibinfo{year}{2025}\natexlab{}.
\newblock \showarticletitle{Reformulating regression test suite optimization
  using quantum annealing-an empirical study}.
\newblock \bibinfo{journal}{\emph{International Journal on Software Tools for
  Technology Transfer}} (\bibinfo{year}{2025}), \bibinfo{pages}{1--14}.
\newblock


\bibitem[\protect\citeauthoryear{Trummer and Koch}{Trummer and Koch}{2015}]%
        {trummer2015multiple}
\bibfield{author}{\bibinfo{person}{Immanuel Trummer} {and}
  \bibinfo{person}{Christoph Koch}.} \bibinfo{year}{2015}\natexlab{}.
\newblock \showarticletitle{Multiple query optimization on the D-Wave 2X
  adiabatic quantum computer}.
\newblock \bibinfo{journal}{\emph{arXiv preprint arXiv:1510.06437}}
  (\bibinfo{year}{2015}).
\newblock


\bibitem[\protect\citeauthoryear{Trummer and Venturelli}{Trummer and
  Venturelli}{2024}]%
        {trummer2024leveraging}
\bibfield{author}{\bibinfo{person}{Immanuel Trummer} {and}
  \bibinfo{person}{Davide Venturelli}.} \bibinfo{year}{2024}\natexlab{}.
\newblock \showarticletitle{Leveraging quantum computing for database index
  selection}. In \bibinfo{booktitle}{\emph{Proceedings of the 1st Workshop on
  Quantum Computing and Quantum-Inspired Technology for Data-Intensive Systems
  and Applications}}. \bibinfo{pages}{14--26}.
\newblock


\bibitem[\protect\citeauthoryear{Uotila}{Uotila}{2025}]%
        {uotila2025left}
\bibfield{author}{\bibinfo{person}{Valter Uotila}.}
  \bibinfo{year}{2025}\natexlab{}.
\newblock \showarticletitle{Left-Deep Join Order Selection with Higher-Order
  Unconstrained Binary Optimization on Quantum Computers}.
\newblock \bibinfo{journal}{\emph{arXiv preprint arXiv:2502.00362}}
  (\bibinfo{year}{2025}).
\newblock


\bibitem[\protect\citeauthoryear{Urgelles, Picazo-Martinez, Garcia-Roger, and
  Monserrat}{Urgelles et~al\mbox{.}}{2022}]%
        {urgelles2022multi}
\bibfield{author}{\bibinfo{person}{Helen Urgelles}, \bibinfo{person}{Pablo
  Picazo-Martinez}, \bibinfo{person}{David Garcia-Roger}, {and}
  \bibinfo{person}{Jose~F Monserrat}.} \bibinfo{year}{2022}\natexlab{}.
\newblock \showarticletitle{Multi-objective routing optimization for 6G
  communication networks using a quantum approximate optimization algorithm}.
\newblock \bibinfo{journal}{\emph{Sensors}} \bibinfo{volume}{22},
  \bibinfo{number}{19} (\bibinfo{year}{2022}), \bibinfo{pages}{7570}.
\newblock


\bibitem[\protect\citeauthoryear{Wagner, Kampel, and Simos}{Wagner
  et~al\mbox{.}}{2019}]%
        {wagner2019quantum}
\bibfield{author}{\bibinfo{person}{Michael Wagner}, \bibinfo{person}{Ludwig
  Kampel}, {and} \bibinfo{person}{Dimitris~E Simos}.}
  \bibinfo{year}{2019}\natexlab{}.
\newblock \showarticletitle{Quantum-inspired evolutionary algorithms for
  covering arrays of arbitrary strength}. In \bibinfo{booktitle}{\emph{Analysis
  of Experimental Algorithms: Special Event, SEA$^2$ 2019, Kalamata, Greece,
  June 24-29, 2019, Revised Selected Papers}}. Springer,
  \bibinfo{pages}{300--316}.
\newblock


\bibitem[\protect\citeauthoryear{Wang, Ali, Yue, and Arcaini}{Wang
  et~al\mbox{.}}{2024a}]%
        {wang2024quantum}
\bibfield{author}{\bibinfo{person}{Xinyi Wang}, \bibinfo{person}{Shaukat Ali},
  \bibinfo{person}{Tao Yue}, {and} \bibinfo{person}{Paolo Arcaini}.}
  \bibinfo{year}{2024}\natexlab{a}.
\newblock \showarticletitle{Quantum approximate optimization algorithm for test
  case optimization}.
\newblock \bibinfo{journal}{\emph{IEEE Transactions on Software Engineering}}
  (\bibinfo{year}{2024}).
\newblock


\bibitem[\protect\citeauthoryear{Wang, Muqeet, Yue, Ali, and Arcaini}{Wang
  et~al\mbox{.}}{2024b}]%
        {wang2024test}
\bibfield{author}{\bibinfo{person}{Xinyi Wang}, \bibinfo{person}{Asmar Muqeet},
  \bibinfo{person}{Tao Yue}, \bibinfo{person}{Shaukat Ali}, {and}
  \bibinfo{person}{Paolo Arcaini}.} \bibinfo{year}{2024}\natexlab{b}.
\newblock \showarticletitle{Test case minimization with quantum annealers}.
\newblock \bibinfo{journal}{\emph{ACM Transactions on Software Engineering and
  Methodology}} \bibinfo{volume}{34}, \bibinfo{number}{1}
  (\bibinfo{year}{2024}), \bibinfo{pages}{1--24}.
\newblock


\bibitem[\protect\citeauthoryear{Willsch, Willsch, Jin, De~Raedt, and
  Michielsen}{Willsch et~al\mbox{.}}{2020}]%
        {willsch2020benchmarking}
\bibfield{author}{\bibinfo{person}{Madita Willsch}, \bibinfo{person}{Dennis
  Willsch}, \bibinfo{person}{Fengping Jin}, \bibinfo{person}{Hans De~Raedt},
  {and} \bibinfo{person}{Kristel Michielsen}.} \bibinfo{year}{2020}\natexlab{}.
\newblock \showarticletitle{Benchmarking the quantum approximate optimization
  algorithm}.
\newblock \bibinfo{journal}{\emph{Quantum Information Processing}}
  \bibinfo{volume}{19}, \bibinfo{number}{7} (\bibinfo{year}{2020}),
  \bibinfo{pages}{197}.
\newblock


\bibitem[\protect\citeauthoryear{Wu, Lin, Shi, Ren, and Wang}{Wu
  et~al\mbox{.}}{2020}]%
        {wu2020hybrid}
\bibfield{author}{\bibinfo{person}{Maochuan Wu}, \bibinfo{person}{Junyu Lin},
  \bibinfo{person}{Shouchuang Shi}, \bibinfo{person}{Long Ren}, {and}
  \bibinfo{person}{Zhiwen Wang}.} \bibinfo{year}{2020}\natexlab{}.
\newblock \showarticletitle{Hybrid optimization-based GRU neural network for
  software reliability prediction}. In \bibinfo{booktitle}{\emph{International
  conference of pioneering computer scientists, engineers and educators}}.
  Springer, \bibinfo{pages}{369--383}.
\newblock


\bibitem[\protect\citeauthoryear{Xiong, Hu, Tian, Lan, Li, and Zhou}{Xiong
  et~al\mbox{.}}{2016a}]%
        {xiong2016virtual}
\bibfield{author}{\bibinfo{person}{Gang Xiong}, \bibinfo{person}{Yu-xiang Hu},
  \bibinfo{person}{Le Tian}, \bibinfo{person}{Ju-long Lan},
  \bibinfo{person}{Jun-fei Li}, {and} \bibinfo{person}{Qiao Zhou}.}
  \bibinfo{year}{2016}\natexlab{a}.
\newblock \showarticletitle{A virtual service placement approach based on
  improved quantum genetic algorithm}.
\newblock \bibinfo{journal}{\emph{Frontiers of Information Technology \&
  Electronic Engineering}} \bibinfo{volume}{17}, \bibinfo{number}{7}
  (\bibinfo{year}{2016}), \bibinfo{pages}{661--671}.
\newblock


\bibitem[\protect\citeauthoryear{Xiong, Sun, Hu, Lan, and Li}{Xiong
  et~al\mbox{.}}{2016b}]%
        {xiong2016optimized}
\bibfield{author}{\bibinfo{person}{Gang Xiong}, \bibinfo{person}{Penghao Sun},
  \bibinfo{person}{Yuxiang Hu}, \bibinfo{person}{Julong Lan}, {and}
  \bibinfo{person}{Kan Li}.} \bibinfo{year}{2016}\natexlab{b}.
\newblock \showarticletitle{An optimized deployment mechanism for virtual
  middleboxes in NFV-and SDN-enabling network}.
\newblock \bibinfo{journal}{\emph{KSII Transactions on Internet and Information
  Systems (TIIS)}} \bibinfo{volume}{10}, \bibinfo{number}{8}
  (\bibinfo{year}{2016}), \bibinfo{pages}{3474--3497}.
\newblock


\bibitem[\protect\citeauthoryear{Yue, Ali, and Arcaini}{Yue
  et~al\mbox{.}}{2023}]%
        {yue2023QRE}
\bibfield{author}{\bibinfo{person}{Tao Yue}, \bibinfo{person}{Shaukat Ali},
  {and} \bibinfo{person}{Paolo Arcaini}.} \bibinfo{year}{2023}\natexlab{}.
\newblock \showarticletitle{Towards quantum software requirements engineering}.
  In \bibinfo{booktitle}{\emph{2023 IEEE International Conference on Quantum
  Computing and Engineering (QCE)}}, Vol.~\bibinfo{volume}{2}. IEEE,
  \bibinfo{pages}{161--164}.
\newblock


\bibitem[\protect\citeauthoryear{Zhang}{Zhang}{2020}]%
        {zhang2020intelligent}
\bibfield{author}{\bibinfo{person}{Feng Zhang}.}
  \bibinfo{year}{2020}\natexlab{}.
\newblock \showarticletitle{Intelligent task allocation method based on
  improved QPSO in multi-agent system}.
\newblock \bibinfo{journal}{\emph{Journal of ambient intelligence and humanized
  computing}} \bibinfo{volume}{11}, \bibinfo{number}{2} (\bibinfo{year}{2020}),
  \bibinfo{pages}{655--662}.
\newblock


\bibitem[\protect\citeauthoryear{Zhang, Zhang, Yue, Ali, and Li}{Zhang
  et~al\mbox{.}}{2020}]%
        {zhang2020uncertainty}
\bibfield{author}{\bibinfo{person}{Huihui Zhang}, \bibinfo{person}{Man Zhang},
  \bibinfo{person}{Tao Yue}, \bibinfo{person}{Shaukat Ali}, {and}
  \bibinfo{person}{Yan Li}.} \bibinfo{year}{2020}\natexlab{}.
\newblock \showarticletitle{Uncertainty-wise requirements prioritization with
  search}.
\newblock \bibinfo{journal}{\emph{ACM Transactions on Software Engineering and
  Methodology (TOSEM)}} \bibinfo{volume}{30}, \bibinfo{number}{1}
  (\bibinfo{year}{2020}), \bibinfo{pages}{1--54}.
\newblock


\bibitem[\protect\citeauthoryear{Zhang, Ali, and Yue}{Zhang
  et~al\mbox{.}}{2019}]%
        {zhang2019uncertaintyTesting}
\bibfield{author}{\bibinfo{person}{Man Zhang}, \bibinfo{person}{Shaukat Ali},
  {and} \bibinfo{person}{Tao Yue}.} \bibinfo{year}{2019}\natexlab{}.
\newblock \showarticletitle{Uncertainty-wise test case generation and
  minimization for cyber-physical systems}.
\newblock \bibinfo{journal}{\emph{Journal of Systems and Software}}
  \bibinfo{volume}{153} (\bibinfo{year}{2019}), \bibinfo{pages}{1--21}.
\newblock


\bibitem[\protect\citeauthoryear{Zhang and Arcuri}{Zhang and Arcuri}{2021}]%
        {zhang2021adaptive}
\bibfield{author}{\bibinfo{person}{Man Zhang} {and} \bibinfo{person}{Andrea
  Arcuri}.} \bibinfo{year}{2021}\natexlab{}.
\newblock \showarticletitle{Adaptive hypermutation for search-based system test
  generation: A study on REST APIs with EvoMaster}.
\newblock \bibinfo{journal}{\emph{ACM Transactions on Software Engineering and
  Methodology (TOSEM)}} \bibinfo{volume}{31}, \bibinfo{number}{1}
  (\bibinfo{year}{2021}), \bibinfo{pages}{1--52}.
\newblock


\bibitem[\protect\citeauthoryear{Zhang, Li, Yue, and Cai}{Zhang
  et~al\mbox{.}}{2025}]%
        {zhang2025quantum}
\bibfield{author}{\bibinfo{person}{Man Zhang}, \bibinfo{person}{Yuechen Li},
  \bibinfo{person}{Tao Yue}, {and} \bibinfo{person}{Kai-Yuan Cai}.}
  \bibinfo{year}{2025}\natexlab{}.
\newblock \showarticletitle{Quantum Optimization for Software Engineering: A
  Survey}.
\newblock \bibinfo{journal}{\emph{arXiv preprint arXiv:2506.16878}}
  (\bibinfo{year}{2025}).
\newblock


\bibitem[\protect\citeauthoryear{Zhang, Li, Liu, Ouyang, Fang, Mu, and
  Gao}{Zhang et~al\mbox{.}}{2021}]%
        {zhang2021new}
\bibfield{author}{\bibinfo{person}{Quanyuan Zhang}, \bibinfo{person}{Haolun
  Li}, \bibinfo{person}{Yanli Liu}, \bibinfo{person}{Shangrong Ouyang},
  \bibinfo{person}{Caiting Fang}, \bibinfo{person}{Wentao Mu}, {and}
  \bibinfo{person}{Hao Gao}.} \bibinfo{year}{2021}\natexlab{}.
\newblock \showarticletitle{A new quantum particle swarm optimization algorithm
  for controller placement problem in software-defined networking}.
\newblock \bibinfo{journal}{\emph{Computers and Electrical Engineering}}
  \bibinfo{volume}{95} (\bibinfo{year}{2021}), \bibinfo{pages}{107456}.
\newblock


\bibitem[\protect\citeauthoryear{Zhang, Yang, Lin, Dai, and Li}{Zhang
  et~al\mbox{.}}{2017}]%
        {zhang2017test}
\bibfield{author}{\bibinfo{person}{Ya-nan Zhang}, \bibinfo{person}{Hong Yang},
  \bibinfo{person}{Zheng-kui Lin}, \bibinfo{person}{Qing Dai}, {and}
  \bibinfo{person}{Yu-feng Li}.} \bibinfo{year}{2017}\natexlab{}.
\newblock \showarticletitle{A test suite reduction method based on novel
  quantum ant colony algorithm}. In \bibinfo{booktitle}{\emph{2017 4th
  International Conference on Information Science and Control Engineering
  (ICISCE)}}. IEEE, \bibinfo{pages}{825--829}.
\newblock


\end{thebibliography}
